\documentclass[aps,pra,onecolumn,superscriptaddress,preprintnumbers,secnumroman,nofootinbib]{revtex4-2}
\usepackage[T1]{fontenc}
\usepackage{epsfig}
\usepackage{graphicx}
\usepackage[english]{babel}
\usepackage{hyphenat}
\usepackage{amsmath}
\usepackage{amssymb}
\usepackage{bbold}
\usepackage{mathtools}
\usepackage{mathrsfs}
\usepackage{slashed}
\usepackage{epstopdf}
\usepackage{xcolor}
\usepackage{braket}
\usepackage{booktabs}
\definecolor{lcolor}{rgb}{0.,0.0,0.}
\definecolor{citcolor}{rgb}{0,0.,0.5}
\usepackage[breaklinks,colorlinks,urlcolor=blue,citecolor=blue,linkcolor=blue]{hyperref}
\usepackage{multirow}
\usepackage{ltablex}
\usepackage[normalem]{ulem} %makes underlining possible: \uline \uwave \sout

\usepackage[a4paper,margin=1.2in]{geometry}
\usepackage{setspace}
\usepackage[e]{esvect}

\usepackage{media9}

\newcommand{\secn}[1]{Section~1}
\newcommand{\appn}[1]{Appendix~1}

\long\def\comment#1{ }

\def\and{\quad\text{and}\quad}

\def\0{{\boldsymbol 0}}
\def\1{{\boldsymbol 1}}

\def\0{{\boldsymbol 0}}

\renewcommand{\part}{{\rm part}}
\newcommand{\be}{\begin{equation}}
\newcommand{\ee}{\end{equation}}
\newcommand{\bes}{\begin{subequations}}
\newcommand{\ees}{\end{subequations}}
\newcommand{\bea}{\begin{eqnarray}}
\newcommand{\eea}{\end{eqnarray}}

\def\bea#1\eea{\begin{align}#1\end{align}}
\newcommand{\bef}{\begin{figure}[h!tb]\centering}
\newcommand{\eef}{\end{figure}}

\allowdisplaybreaks

\begin{document}

\title{Out-of-Equilibrium Dynamics in a U(1) Lattice Gauge Theory \\ via Local Information Flows:  
Scattering and String Breaking}

\author{Claudia Artiaco}
\email{artiaco@kth.se}
\affiliation{Department of Physics, KTH Royal Institute of Technology, Stockholm, 106 91 Sweden}

\author{Jo\~{a}o Barata}
\email{joao.lourenco.henriques@cern.ch}
\affiliation{European Organization for Nuclear Research (CERN),  Theoretical Physics Department, CH-1211 Geneva, Switzerland}

\author{Enrique Rico}
\email{enrique.rico.ortega@cern.ch}
\affiliation{European Organization for Nuclear Research (CERN),  Theoretical Physics Department, CH-1211 Geneva, Switzerland}
\affiliation{EHU Quantum Center and Department of Physical Chemistry, University of the Basque Country UPV/EHU, P.O. Box 644, 48080 Bilbao, Spain}
\affiliation{DIPC - Donostia International Physics Center, Paseo Manuel de Lardizabal 4, 20018 San Sebastián, Spain}
\affiliation{IKERBASQUE, Basque Foundation for Science, Plaza Euskadi 5, 48009 Bilbao, Spain}

\preprint{CERN-TH-2025-191}

\begin{abstract}
We introduce local information flows as a diagnostic tool for characterizing out-of-equilibrium quantum dynamics in lattice gauge theories. We employ the information lattice framework, a local decomposition of total information into spatial- and scale-resolved contributions, to characterize the propagation and buildup of quantum correlations in real-time processes. Focusing on the Schwinger model, a canonical $(1+1)$-dimensional U(1) lattice gauge theory, we apply this framework to two scenarios. First, in the near-threshold scattering of two vector mesons, we demonstrate that the emergence of correlations at a longer length scale in the information lattice marks the production of heavier scalar mesons. Second, in the dynamics of electric field strings, we clearly distinguish between the confining regime, which evolves towards a steady state with a static correlation profile, and the string-breaking sector. The latter is characterized by dynamic correlation patterns that reflect the sequential formation and annihilation of strings. This information-centric approach provides a direct, quantitative, and interpretable visualization of complex many-body phenomena, offering a promising tool for analyzing dynamics in higher-dimensional gauge theories and experiments on quantum hardware.
\end{abstract}

\maketitle

\clearpage
\tableofcontents

\clearpage

\section{Introduction}\label{sec:introduction}

The ability to simulate the real-time, out-of-equilibrium dynamics of quantum gauge theories is a primary objective in modern physics, promising to unlock insights into phenomena from high-energy elementary particle scattering to the properties of the early Universe. Recent advances in quantum technologies, including analog quantum simulators~\cite{Banuls:2019bmf}, tensor network methods~\cite{Banuls:2018jag}, and digital quantum computers~\cite{Jordan:2012xnu}, have opened a direct experimental and computational window into these challenging domains. These tools enable the exploration of quantum field theory in regimes previously inaccessible to traditional methods, such as perturbation theory or Euclidean lattice simulations. However, a significant bottleneck remains: the development of effective diagnostics to extract and interpret the complex correlations that govern non-equilibrium many-body systems.

Characterizing a system's evolution typically begins by tracking local observables. Yet, for systems far from equilibrium, where extensive and non-local correlations dominate, such simple probes are insufficient. Although higher-order diagnostics such as multi-point correlators can, in principle, capture this complexity, they are computationally and experimentally demanding to measure and often challenging to interpret physically. Higher-order diagnostics usually require elaborate post-processing, thus obscuring the direct connection to the underlying physics \cite{PhysRevD.89.074011}.

An alternative and more robust approach, which is the main focus of this work, lies in the context of quantum information theory. By abstracting away microscopic details, quantum information measures, such as the von Neumann entropy, distill the universal features of a state's correlation structure~\cite{Pichler:2015yqa} and allow us to map out its properties. Building on this idea, the recently developed information lattice framework provides a powerful tool to characterize states solely from the distribution of local information and local information flows~\cite{klein2022time,artiaco2025universal}. The information lattice decomposes the total information of a quantum state into local contributions that are uniquely assigned to a spatial location and scale. During unitary evolution, this local information behaves like a perfect fluid, showing well-defined currents and flows.\footnote{This holds provided the underlying theory is unitary and local.} This hydrodynamic picture offers a universal and intuitive way of presenting and analyzing real-time quantum dynamics, as shown for three exemplary quench dynamics in noninteracting fermionic chains in Ref.~\cite{bauer2025local}. Moreover, this provides a framework for performing the efficient approximate time evolution of local observables in large-scale quantum systems~\cite{klein2022time,artiaco2024efficient,harkins2025nanoscale}. In this work, we expand this program and employ the information lattice to characterize the non-equilibrium dynamics of the Schwinger model, the paradigmatic U(1) gauge theory in (1+1) dimensions. To that end, we consider two quench experiments in this model.

In a particle scattering simulation, we study the collision of two vector mesons near the threshold to produce a heavier scalar meson, following the setup of Ref.~\cite{Papaefstathiou:2024zsu}. Our analysis reveals that production of a scalar meson is directly signaled by the emergence of local information at a new, larger length scale, corresponding to the characteristic information scale of the heavier meson particle. In its absence, the information flow remains confined to the scales characteristic of the initial vector mesons. We then follow to characterize string-breaking dynamics in the Schwinger model, analyzing the evolution of an electric-field string, a problem first explored in Ref.~\cite{Casher:1974vf}. The information lattice provides a clean distinction between two dynamical regimes. In the confining regime, where the string is stable, the system relaxes to a non-thermal steady state with a static finite-range correlation profile. Conversely, in the string-breaking regime, we observe a clear cyclical buildup and destruction of correlations, a direct signature of the repeated creation of particle-antiparticle pairs that screen the electric field.

This paper is organized as follows. Section~\ref{sec:Information_lattice} introduces the theoretical construction of the information lattice as a diagnostic for quantum states in $(1+1)$D. Section~\ref{sec:schwinger_model} provides the necessary background on the Schwinger model. In Section~\ref{sec:numerical_results}, we present our main results on mapping the real-time dynamics in terms of local information flows for the two quench scenarios. Finally, in Section~\ref{sec:conclusion}, we summarize our findings and discuss the broader potential of local information flows as a diagnostic tool for lattice gauge theories.

\section{The information lattice}\label{sec:Information_lattice}

\begin{figure}[tb]
    \centering
    \includegraphics[width=\linewidth]{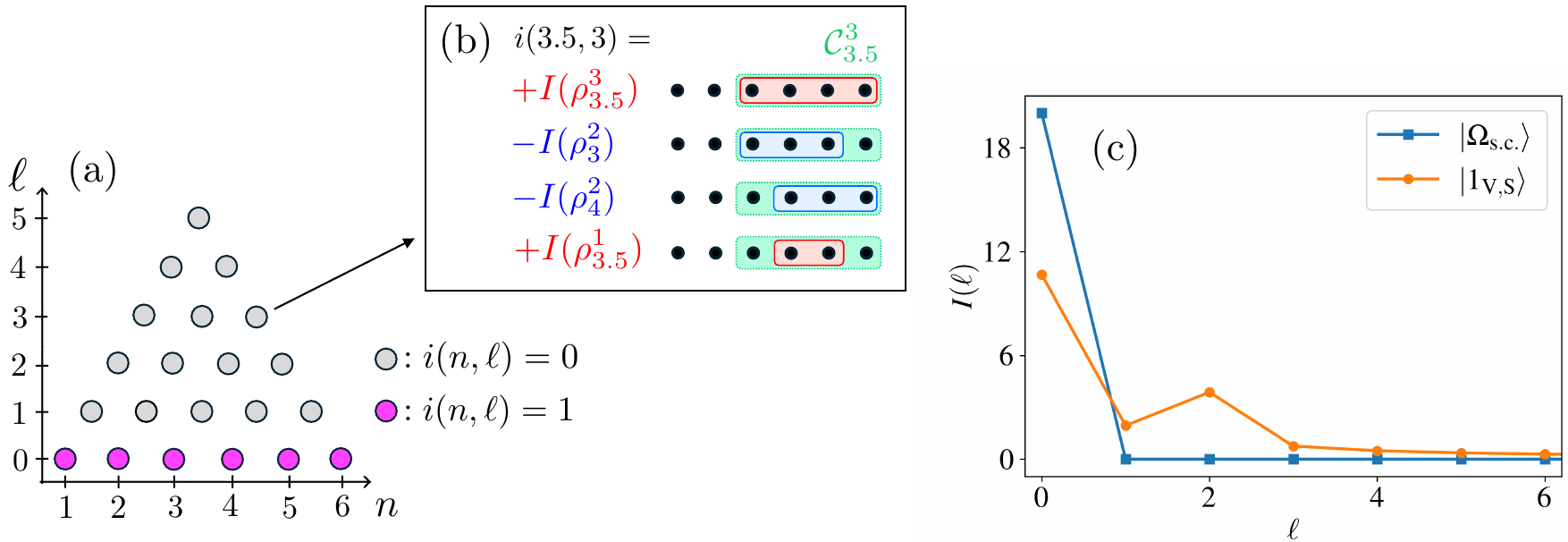}
    \caption{(a) Information lattice for a product state of qubits, e.g., $\ket{\Omega_\mathrm{s.c.}}$ in Eq.~\eqref{eq:vac_SC}. (b) Illustration of the formula for local information in Eq.~\eqref{eq:local_info_formula} for $i(3.5,3)$; red denotes positive contributions and blue negative ones. The underlying green area shows the subsytem $\mathcal{C}^{3}_{3.5}$. (c) Information per scale for the ground state $\ket{\Omega_{\mathrm{s.c.}}}$ and first excited states $\ket{1_{\mathrm{V,S}}}$ of the Schwinger model in the strong coupling limit in Eqs.~\eqref{eq:vac_SC} and \eqref{eq:vector_meson_strong_coupling}, respectively, with $N=20$. Only scales $\ell < 7$ are shown.}
    \label{fig:info_lattice_picture}
\end{figure}

The information lattice is the hierarchical triangular structure shown in Fig.~\ref{fig:info_lattice_picture} (a).
Each information lattice site is associated with local information, quantified by the color of the circle (gray or pink), which corresponds to the total amount of correlations in a spatial region on a given scale~\cite{artiaco2025universal}.
The information lattice decomposes the total information in a quantum state into local contributions (that is, local information), making the information akin to a hydrodynamic quantity that, during time evolution, locally flows within the information lattice through well-defined local currents~\cite{klein2022time,artiaco2024efficient,bauer2025local}. 

The information lattice can be given a precise mathematical definition in (1+1)D; generalizations to higher spatial dimensions can also be constructed~\cite{flor2025workinprogress}.
To define it mathematically, let $\rho$ be the density matrix that describes the state of a quantum system with total Hilbert space dimension $\dim(\rho)$.
The total information (or von Neumann information) in the state, $I(\rho)$, is the difference between the von Neumann entropy of the state, $S(\rho)$, and its maximum value,
\begin{align}
    I(\rho) &= \log_2[\dim(\rho)] - S(\rho) = \log_2[\dim(\rho)] + \mathrm{Tr}[\rho \log_2(\rho)].
    \label{eq:total_information}
\end{align}
 This quantity corresponds to the additional average number of bits that can be predicted about measurement outcomes from knowledge of $\rho$ compared to the maximally mixed state.
Analogously, $I(\rho_A)$ is the information stored in the reduced density matrix $\rho_A = \mathrm{Tr}_{\bar{A}}(\rho)$ of the subsystem $A$, where $\bar{A}$ is the complement of $A$.

For a quantum system in one spatial dimension, length $N$, and under open boundary conditions, the information lattice is constructed as follows.
Consider the subsystem $\mathcal{C}_n^{\ell}$ made up of $\ell + 1$ neighboring physical sites centered around position $n$.
Subsystems composed of one physical site are labeled $\ell=0$, subsystems made of two physical sites have $\ell=1$, etc.
For $\ell$ even, $n$ is integer; for $\ell$ odd, $n$ is half-integer.
Each subsystem is uniquely defined by the pair of labels $(n, \ell)$.
The reduced density matrix of the subsystem $\mathcal{C}_n^{\ell}$ is $\rho^\ell_n = \mathrm{Tr}_{\bar{\mathcal{C}}_n^{\ell}}(\rho)$, with $\bar{\mathcal{C}}_n^{\ell}$ the complement of $\mathcal{C}_n^{\ell}$.
We define local information by imposing that it provides the decomposition of the total information in the reduced density matrix of the subsystem $\rho^\ell_n$ for any $(n, \ell)$,
\begin{equation}
\label{eq:local_info_decomposition}
    I(\rho^\ell_n) = \sum_{(n', \ell') \in \mathcal{D}^\ell_n} i(n',\ell'),
\end{equation}
where $\mathcal{D}^\ell_n = \{(n', \ell') \,|\, \mathcal{C}^{\ell'}_{n'} \subseteq \mathcal{C}^\ell_n \} $. 
This gives
\begin{equation}
\label{eq:local_info_formula}
    i(n,\ell) = I(\rho^\ell_n) - I(\rho^{\ell-1}_{n - 1/2}) - I(\rho^{\ell-1}_{n + 1/2}) + I(\rho^{\ell-2}_n),
\end{equation}
with the convention that $I(\rho^\ell_n) = 0$ for empty subsystems.
Local information $i(n,\ell)$ is the quantum conditional mutual information between the subsystems $\mathcal{C}^{\ell-1}_{n-1/2}$ and $\mathcal{C}^{\ell-1}_{n+1/2}$~\cite{wilde2013quantum}. 
The summation and subtraction formula in Eq.~\eqref{eq:local_info_formula} is schematically represented in Fig.~\ref{fig:info_lattice_picture} (b).
As a result, $i(n,\ell)$ quantifies how much more information is stored in $\rho^\ell_n$ rather than in the density matrices $\rho^{\ell-1}_{n-1/2}$ and $\rho^{\ell-1}_{n+1/2}$ of the smaller subsystems contained in $\mathcal{C}^\ell_n$.
It follows that $i(n,\ell) \geq 0$. The labels $(n, \ell)$ can be organized in a hierarchical structure with $n$ increasing from left to right and $\ell$ increasing from bottom to top, and associating each site $(n,\ell)$ with local information $i(n,\ell)$.
This defines the information lattice shown in Fig.~\ref{fig:info_lattice_picture} (a).

As an example, consider the maximally entangled state between two sites, $\ket{\beta} = \frac{1}{\sqrt{d}} \sum_{n=1}^{d} \ket{n}_1 \otimes \ket{n}_2$, where $d = d_1 = d_2$, with $d_i$ the local Hilbert-space dimension of site $i$, and
$\{\ket{n}_i\}_{n=1}^{d}$ an orthonormal basis for site $i$.
The single-site reduced density matrices are maximally mixed, $\rho_1 = \rho_2 = \frac{1}{d} \mathbb{1}_N$, hence $i^0_1 = i^0_2 = 0$.
This means that knowing that the system is in a maximally entangled pair state does not provide any predictive power about the outcome of measurements performed on single sites.
In contrast, the two-site state contains all information of the system, $i(3/2,1) =2 \log_2{d}$.

As a second example, consider the vacuum of a spin-$1/2$ antiferromagnetic Ising spin chain in the broken phase, 
\begin{align}
\label{eq:vac_SC}
   \ket{\Omega_{\rm s.c.}} =  \ket{\uparrow \downarrow\uparrow \downarrow \uparrow \downarrow \cdots \uparrow \downarrow} \, .
\end{align}
Since this is a product state, one has $i(n,\ell) = \delta_{\ell,0}$ for all $n$; this is exemplified in Fig.~\ref{fig:info_lattice_picture} (a).
In fact, in a product state such as $\ket{\Omega_{\rm s.c.}}$ all the information is encoded in the single-site reduced density matrices $\rho^0_n$; from each $\rho^0_n$ one can predict the result of a single-site optimal measurement with certainty.
Excitations of $\ket{\Omega_{\rm s.c.}}$ could, for instance, be constructed by flipping pairs of neighboring spins: 
\begin{equation}
\label{eq:vector_meson_strong_coupling}
   \ket{1_{\rm{V,S}}} = \frac{1}{\sqrt{N-1}}\sum_{n=1}^{N-1} (\sigma^+_{n+1} \sigma^-_n \mp {\rm h.c.}) \ket{\Omega_{\rm s.c.}} \, ,
\end{equation}
where parity by translation over one lattice site distinguishes the V and S states.
The states $\ket{1_{\rm{V,S}}}$ are characterized by finite correlations centered around $\ell = 2$, as it is shown in Fig.~\ref{fig:info_lattice_picture} (c) through the information per scale~\cite{artiaco2025universal}, defined as
\begin{align}
\label{eq:information_per_scale}
    I(\ell) = \sum_{n} i(n,\ell) \, ,
\end{align}
which captures the distribution of the total amount of correlations across different scales. The equivalent distribution for the ground state is also shown, making clear the distinction between this and the above excited states. The examples in Eqs.~\eqref{eq:vac_SC} and \eqref{eq:vector_meson_strong_coupling} correspond to the vacuum and first excited states, a vector (V) and scalar (S) meson state, of the lattice Schwinger model in the strong-coupling limit~\cite{Coleman:1976uz}, respectively, which we will discuss in Sec.~\ref{sec:scattering}. 

More generally, the behavior of $I(\ell)$ provides a clear signature for different physical phases, which can be given a compelling interpretation in terms of holographic duality~\cite{ryu2006holographic,nozaki2013holographic}.
For a more detailed discussion on this and a visual representation of these profiles, we refer the reader to Ref.~\cite{artiaco2025universal}.
Localized states, such as the ground states of gapped local Hamiltonians and many-body localized states, are defined by an extensive amount of information concentrated at short scales, $\ell \ll N$. This local information distribution results in an area law for entanglement entropy and corresponds holographically to a “capped off” spacetime where the bulk is shallow, confining correlations near the boundary. In sharp contrast, ergodic states feature extensive information on macroscopic scales. For infinite-temperature states like those drawn from the Haar-random ensemble, this information peaks at the half-system-size scale, $\ell \sim N/2$. More complex are the finite-temperature thermal states, which exhibit extensive information on both short scales ($\ell \ll N$) and half-system size scales ($\ell \sim N/2$). The short-scale component ensures that local observables thermalize according to the eigenstate thermalization hypothesis~\cite{Deutsch:1991msp,Srednicki:1994mfb}, while the long-scale component signifies system-spanning correlations that produce a volume-law for entanglement entropy. Holographically, this dual structure is represented by a black hole in the bulk spacetime, where the region near the boundary governs local thermal properties and the deep interior's horizon generates the volume-law entanglement. Crucially, the presence of this long-range information is the defining feature of a true thermal (pure) state; an eigenstate with only a short-scale information peak remains localized, not thermal. Finally, critical states sit between these limits. At a quantum critical point, the state's scale invariance is reflected in a power-law decay of $I(\ell)$ with scale $\ell$, which corresponds holographically to a pure, uncapped Anti-de Sitter spacetime whose symmetries mirror the state's scale-free nature.

The von Neumann information \( I(\rho) \) is conserved under unitary time evolution. Consequently, the total sum of local information in the information lattice is also conserved,
\begin{equation}
    I[\rho(t)] = \sum_{(n, \ell)} i(n,\ell,t) = \text{const}.
\end{equation}
This makes $i(n,\ell,t)$ analogous to a conserved local density in hydrodynamics, with well-defined local currents. Under time evolution governed by a local Hamiltonian, information flows through the lattice along diagonals in the \( (n, \ell) \) space.
This current of information can be derived from the von Neumann equation of motion for subsystem density matrices, and they obey continuity equations similar to those in conventional conservation laws~\cite{klein2022time,artiaco2024efficient}.
The information lattice thus provides a natural and general framework to study quantum dynamics, enabling the tracking of how correlations emerge, propagate, and equilibrate following a quench~\cite{bauer2025local} and during thermalization~\cite{klein2022time,artiaco2024efficient,harkins2025nanoscale}.

\section{The Schwinger model}
\label{sec:schwinger_model}

The Schwinger model is a $(1+1)$-dimensional $U(1)$ gauge field theory coupled to fermionic matter~\cite{Schwinger:1951nm}, naturally related to quantum electrodynamics (QED) in ($3+1$)D. It shares some physical features, e.g., the existence of a chiral condensate, finite mass gap, bound states, and a linear confining potential, with higher-dimensional theories relevant for high-energy physics~\cite{Schwinger:1951nm, Coleman:1976uz, Mandelstam:1975hb, Banks:1975gq, Coleman:1975pw}, such as quantum chromodynamics (QCD). Although the Schwinger model cannot quantitatively describe high-energy physics phenomena, it has served over the decades as an ideal theoretical laboratory to explore complex processes, such as the production of particles from the vacuum (Schwinger effect), the hadronization transition, and string dynamics. In more recent decades, with the technical developments in quantum information science, there has been a renewed interest in exploring the real-time properties of the theory, which so far had only been studied through strong coupling, perturbative, lattice quantum field theory, or statistical methods, see,  e.g., Refs.~\cite{Banks:1975gq, Coleman:1976uz,Hebenstreit:2014rha}.

In the continuum, the theory's Hamiltonian in the temporal $A^0=0$ gauge reads~\cite{Schwinger:1951nm}
\begin{align}\label{eq:H_Schwinger}
	H&=  \int dx \, \frac{g^2}{2 } L^2(x) + \psi^\dagger(x) \gamma^0 (-i \gamma^1 \partial_1 +g \gamma^1 A_1(x) +m ) \psi(x)\, ,
\end{align}
where $L(x)$ is the (reduced) electric field, and $\psi(x)$ is a two-component fermionic field, with mass $m$ and with a coupling to the gauge field $g$. The Dirac matrices are denoted by $\gamma^\mu$. Physical states in the theory must be invariant under local gauge transformations; in ($1+1$)D QED, this results in physical states that must satisfy Gauss's law: $\partial_x L = \psi^\dagger(x) \psi (x)$. 

The continuum theory can be mapped to a variety of lattice theories with a common ultraviolet fixed point~\cite{Banuls:2019bmf,Banuls:2018jag}; here we employ the Kogut-Susskind construction~\cite{Kogut:1974ag,Susskind:1976jm}, commonly considered in the context of quantum simulation of this theory. In this formulation,  Eq.~\eqref{eq:H_Schwinger} is mapped to the spin chain
\begin{align}\label{eq:Spin_Hamiltonian}
	H_{\rm latt} &= 	\frac{g^2a}{2} \sum_{n=1}^{N-1} \left[\frac{1}{2}\sum_{k=1}^n (\sigma^z_k+(-1)^k)  \right]^2  +\sum_{n=1}^{N} m  (-1)^n \frac{\sigma^z_n}{2} +\frac{1}{2a} \sum_{n=1}^{N-1}  \sigma^+_n \sigma^-_{n+1} + \, {\rm h.c.} \,, 
\end{align}
where $a$ is the lattice spacing; $m$ and $g$ have mass dimension $[a^{-1}]$. Note the first term on the right-hand side of Eq.~\eqref{eq:Spin_Hamiltonian}, where we have explicitly integrated out the gauge fields using Gauss's law and used open boundary conditions with vanishing fields at the edge of the lattice. We refer the reader to Ref.~\cite{Hamer:1997dx} for details on this mapping, which we shall not further discuss, and we rather take Eq.~\eqref{eq:Spin_Hamiltonian} as a starting point of our study.  

The Schwinger model's spectrum is purely bosonic, with the lowest energy excitation being the so-called Schwinger boson~\cite{Coleman:1976uz,Schwinger:1951nm}, which, in the strong coupling limit, can be understood as a vector bound state of fermions. These statements can be understood via a bosonization map of Eq.~\eqref{eq:H_Schwinger} to a Sine-Gordon theory~\cite{Coleman:1976uz,Mandelstam:1975hb}. In the strong coupling limit, i.e., $m/g \ll 1$ (and $g a \gg 1$ on the lattice), the dual bosonic theory is noninteracting, and the mass of the vector boson reads $m_{V}=g/\sqrt{\pi}$~\cite{Coleman:1976uz}. Moreover, in this limit of the theory, the higher energy bound states are formed by clustering vector mesons; indeed, the second excited state is a scalar meson which can be thought of as a bound state of two Schwinger bosons, with twice the mass of the vector meson. Although the wavefunctions of these two states are not known in general, they can be systematically computed at strong coupling. It has been found that, on the lattice, the vector meson is dominated by two-point correlations, while the scalar state is more sensitive to higher point correlators of fermionic operators~\cite{Mo:1992sv,Harada:1995tv}. These observations are natural since the vector state directly couples to the electromagnetic current $J^1(x) = \psi^\dagger(x) \gamma^0 \gamma^1 \psi(x)$, while the scalar would couple mainly to $(J^1)^2$. More recent studies have confirmed this by testing the overlap of the states' wavefunctions with appropriate fermionic operators~\cite{Papaefstathiou:2024zsu}, and the coupling to the vector current~\cite{Barata:2025jhd}. In what follows, we shall show that the information lattice gives a new perspective into particle formation at the threshold for the formation of the scalar meson from two scattering vector states.

Another key phenomenon investigated in the Schwinger model is the formation and subsequent evolution of electric flux strings in real-time. In its simplest formulation, in (1+1) dimensions, there is a linear confining Coulomb potential between oppositely charged fermion–antifermion pairs. This potential arises from the fact that, in the absence of transverse dimensions, the electric field generated by the charges remains constant along the spatial interval separating them. The result is the formation of a one-dimensional “string” of uniform electric flux connecting the two charges, with the energy stored in the string growing linearly with separation. This mechanism is directly analogous to the confinement of quarks in QCD, where a color flux tube plays the role of the string. Note that while the latter arises due to the non-Abelian character of the theory, the former is due to the dimensionality of spacetime.

If the strength of the electric field exceeds a certain critical value, the vacuum becomes unstable against spontaneous particle–antiparticle creation via the Schwinger effect. In this regime, the energy density stored in the electric field over a unit length can be converted into the rest mass of new fermion–antifermion pairs. The critical field threshold for string breaking is obtained by equating the energy stored in the field over the distance separating the charges to the rest energy of a pair, yielding the semi-classical critical field value $L_c = m^2 / g^2$. Once this threshold is reached, the initial flux string connecting the original charges can break, with the liberated ends of the new pairs screening the original charges.

In scenarios where additional work is supplied, either through external driving or from kinetic energy imparted to the charges, the system can undergo multiple successive string breakings, where the endpoint charges recede from each other. This leads to the formation of a “multi-string” state in which several fermion–antifermion pairs emerge from the vacuum and propagate apart. The repeated breaking of flux strings produces a striking dynamical effect: the spatially uniform electric field undergoes large-amplitude oscillations, with the sign of the field inverting after each pair-creation event. Such oscillations reflect the repeated reversal of field polarization due to the alternating arrangement of charges.

As the system evolves, energy is stored in the rest masses and kinetic energies of the produced particles. Once the field amplitude drops below the critical value $L_c$, further Schwinger pair production becomes energetically forbidden. At this stage, the system relaxes into a configuration where the residual electric field is localized between the last-produced fermion–antifermion pair, forming a static flux tube. This final configuration closely parallels the phenomenology of hadronization in QCD: when the invariant mass of a color flux tube falls below the threshold for quark–antiquark pair creation, no further string breaking occurs, and the quarks at the tube endpoints become bound into a color-neutral hadron. This picture is closely followed by phenomenological models for hadronization of parton showers of high-energy partonic cascades~\cite{Ferreres-Sole:2018vgo,Andersson:1997xwk,Winter:2003tt}. Thus, the Schwinger model not only provides an analytically tractable framework for understanding real-time confinement and string-breaking but also serves as a valuable low-dimensional analog for exploring non-perturbative aspects of strong interactions.

In recent years, advances in numerical and experimental techniques have enabled detailed studies of real-time dynamics in the Schwinger model, which are well beyond the reach of conventional perturbative field theory~\cite{Banuls:2019bmf}. On the computational side, tensor network methods, particularly matrix product states (MPS) and matrix product operators (MPO), have emerged as powerful tools for simulating the out-of-equilibrium evolution of strongly correlated gauge theories in one spatial dimension. These approaches exploit the limited growth of entanglement in such systems to efficiently represent the quantum state, enabling high-precision simulations~\cite{banuls2013mass}.
In particular, tensor network studies have resolved the oscillatory electric field dynamics following quench protocols and have quantified the interplay between string breaking timescales, fermion mass, and gauge coupling~\cite{Pichler:2015yqa,PhysRevX.6.041040}; they have enabled the first observations of inelastic scattering processes in several theories~\cite{Papaefstathiou:2024zsu,PhysRevD.96.114501,PhysRevD.101.054507}, among other remarkable achievements.

Complementing these numerical efforts, analog quantum simulators based on ultracold atoms in optical lattices, trapped ions, and superconducting qubit arrays have demonstrated the ability to engineer gauge-invariant Hamiltonians closely related to the Schwinger model. Such platforms have enabled the real-time observation of particle–antiparticle pair creation and subsequent string-breaking events~\cite{KASPER2016742,yang2020observation}, allowing for the direct measurement of the electric field and particle density evolution. Trapped-ion experiments have leveraged long-range interactions to simulate gauge-invariant couplings, achieving strong control over system parameters and providing access to regimes where classical simulations become intractable~\cite{martinez2016real,kokail2019self,PRXQuantum.3.020324}.

These developments underscore the Schwinger model’s dual role: as a benchmark for testing and validating advanced numerical algorithms for real-time dynamics in gauge theories, and as a testbed for emerging quantum simulation platforms aimed at exploring confinement and other non-perturbative phenomena in regimes relevant to high-energy physics.

\section{Out-of-equilibrium dynamics in the Schwinger model through the information lattice}\label{sec:numerical_results}
 
In this section, we study how the information lattice can give new insights into the out-of-equilibrium dynamics in the Schwinger model. As mentioned above, we consider two quench protocols. In the first, closely following Ref.~\cite{Papaefstathiou:2024zsu}, we scatter wave packets of vector mesons, with a momentum vector $k$. By adjusting the momentum, at fixed $m/g$, one can go above the particle threshold to generate slower scalar meson states. In the second quench experiment, we study the dynamics of electric flux strings by inserting an expanding electric field on the lattice, extending between two external charges of absolute value $Q$. Depending on the value of $m/g$ and $Q$, we can study the transition between the no-breaking and string-breaking regimes. We discuss below in more detail each one of these simulation protocols.

The following numerical simulations were performed using the tensor network software package {\tt{iTensor}}~\cite{itensor}, which allows to conveniently cast $H_{\rm latt}$ in Eq.~\eqref{eq:Spin_Hamiltonian} in terms of a matrix product operator, while the quantum state is represented by a matrix product state (MPS) \textit{ansatz}. More, using the {\tt{iTensor}}'s native implementations of the density matrix renormalization group algorithm (DMRG)~\cite{PhysRevB.48.10345,PhysRevLett.69.2863}, and the time-dependent variational principle algorithm (TDVP)~\cite{Haegeman:2011zz,PhysRevB.94.165116}, we can implement the complete simulation protocols. The information lattice can be constructed by contracting the tensors and then using singular value decomposition to extract the entanglement entropy for the different density matrices entering Eq.~\eqref{eq:local_info_formula}, see Refs.~\cite{artiaco2025universal,aceituno2024thermalization}. 

Although we use a range of different values for the model parameters in our results, we constrain the simulations to small values of maximal bond dimension $D\leq 30$. The reason for this is the numerical complexity inherently tied to the exact calculation of each $i(n,\ell)$. Although the calculation of local information can be made more efficient~\cite{artiaco2025universal,aceituno2024thermalization}, one should bear in mind that computational time should scale exponentially with the relevant bond dimensions defining the reduced density matrix of interest. Nonetheless, we have checked that in both quench experiments, considering a reduced bond dimension of $D=20$ does not qualitatively change the results for local observables, entanglement entropies, and the information lattice. For the scattering protocol, we have tested that taking $D=40$ also does not lead to any qualitatively new features. Finally, we also studied these protocols in a smaller lattice, where one can explore slightly larger $D$; again, we found no new qualitative features compared to the results below. Of course, the following results cannot be extrapolated to infinite bond dimension, and one should take this into account in the ensuing discussion and conclusions.

\subsection{Near-threshold particle scattering}
\label{sec:scattering}

We first consider the scattering of two wave packets formed from the vector meson state. In what follows, we take the coupling $ga= \{1,2\}$, which are close to the strong coupling regime, and $ma=10^{-5}$. The meson momentum is varied between $ka=0.7$ to $ka=1.3$, following the strong-coupling estimates for the threshold momentum; we have numerically verified that $k_{\rm thresh} a \approx 1.12$ for the parameters used, in accordance with the results reported in Ref.~\cite{Papaefstathiou:2024zsu}. 

\begin{figure}[h]
    \centering
    \includegraphics[width=0.45\linewidth]{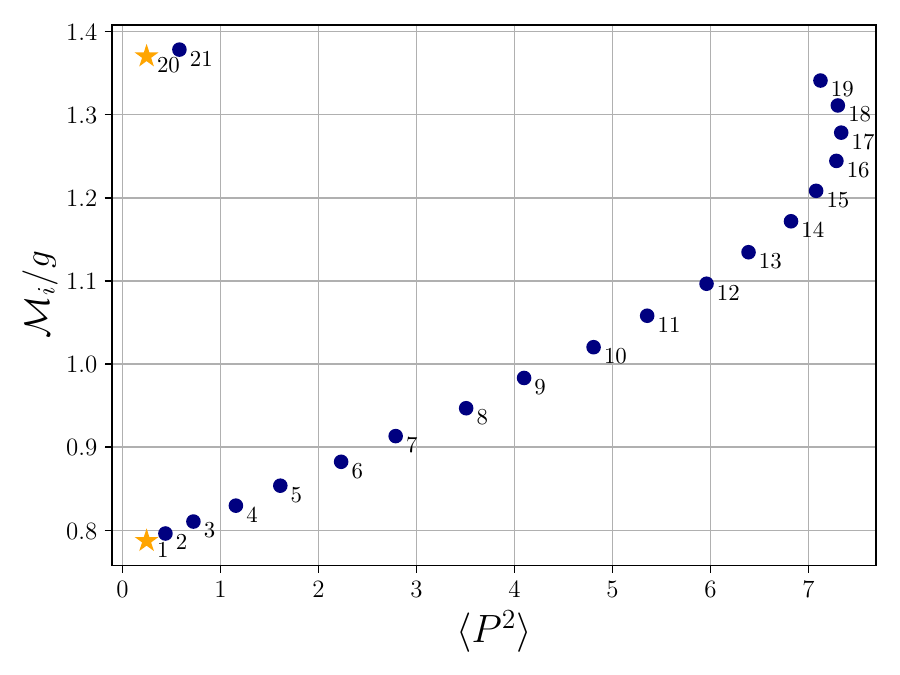}
    \includegraphics[width=0.5\linewidth]{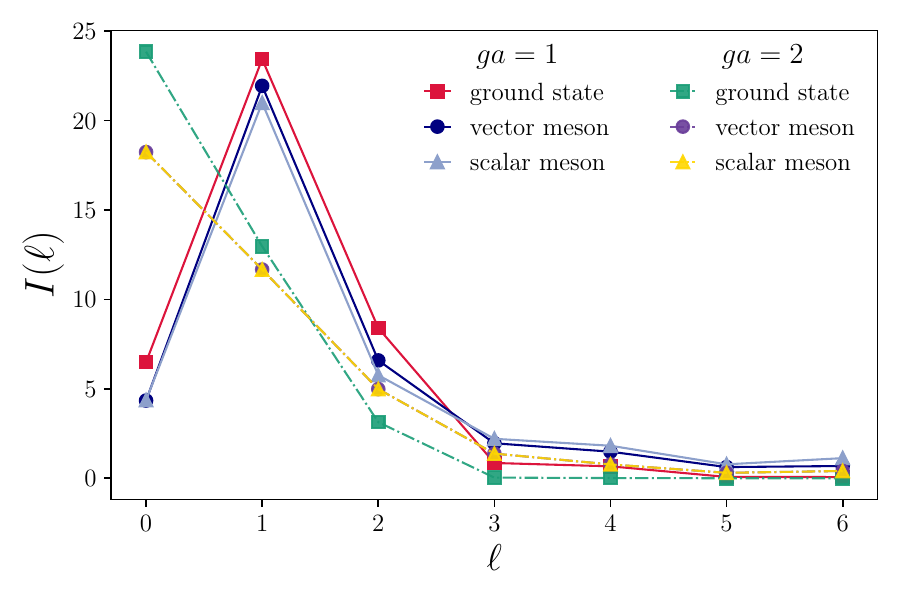}
    \caption{\textbf{Left:} Mass spectrum as a function of the expectation value of the squared pseudo momentum operator $P =-i \sum_n \left( \sigma_{n}^- \sigma^z_{n+1} \sigma^+_{n+2}- {\rm h.c.} \right)$ for $ga=1$. Here, $\mathcal{M}_i= E_i-E_{\rm vac}$ is the energy gap to the vacuum of the $i$-th state. The vector ($i=1$) and scalar ($i=20$) states (gold stars) are identified by having the minimal momentum and exhibiting a mass gap. Their identification was further confirmed by checking their parity. Blue circular markers denote finite momentum excitations of the vector meson, which appear only in the lattice theory. The results for mass gaps of the identified states agree quantitatively with those reported in Ref.~\cite{Papaefstathiou:2024zsu}.
    \textbf{Right:} $I(\ell)$ distribution for the vacuum, vector, and scalar meson states identified in the left panel. $I(\ell)$ is also shown for the same states at $ga=2$, which are identified through the same DMRG procedure. Note that for $ga=2$ the curves of the vector and scalar meson states overlap.
    We set $ma=10^{-5}$ and $N=40$ in both panels.}
    \label{fig:spectrum}
\end{figure}

We begin by characterizing the lowest lying states, i.e., vector and scalar mesons, using the information lattice. In Fig.~\ref{fig:spectrum} (left), we show the energy gap $\mathcal{M}_i= E_i-E_{\rm vac}$ of the lowest lying states as a function of their squared momentum. The states are obtained by performing consecutive runs of DMRG, changing the seed state, and removing the lower energy states by raising their energy in the spectrum. For $ga=1$, we identify two bound states with no net momentum marked as gold stars on the figure.\footnote{In fact, they have a finite pseudo momentum due to lattice discretization and open boundary conditions being used.} The distinction between the two identified states can be straightforwardly checked by, e.g., computing the states' properties under translation by a lattice site~\cite{Papaefstathiou:2024zsu}, or their coupling to the vector current~\cite{Barata:2025jhd}; this allows us to identify these excitations as the vector and scalar mesons, respectively.
The intermediate states marked as blue dots in Fig.~\ref{fig:spectrum} (left) correspond to finite momentum excitation states of the vector meson. 

In the right panel, we show the information per scale $I(\ell)$ defined in Eq.~\eqref{eq:information_per_scale} for the vacuum and the identified meson states, including the results for the equivalent simulation with $ga=2$. For $ga=1$, we observe that all the states are characterized by a peak at the $\ell \approx 1$ level. For the excited states, correlations at $\ell=\{0,1\}$ diminish, leading to enhanced correlations at larger $\ell$ due to the decomposition property of local information in Eq.~\eqref{eq:local_info_decomposition}, which implies that $\sum_\ell I(\ell) = N$. The correlations of the vector and scalar meson states are similar at larger scales. Nonetheless, the vector has a larger amount of information at $\ell \lesssim 2$, while the scalar exhibits stronger correlations for $\ell \gtrsim 3$.

These observations seem in contradiction with what is expected from a strong coupling analysis. In that limit, one expects the vacuum to be a product state with the form in Eq.~\eqref{eq:vac_SC}; thus, information should be localized at $\ell \approx 0$, rather than peaking at $\ell \approx 1$. Furthermore, at strong coupling the vector meson state is $\ket{1_{\rm{V}}}$ in Eq.~\eqref{eq:vector_meson_strong_coupling}, which has the information per scale distribution shown in Fig.~\ref{fig:info_lattice_picture} (c) with a peak at $\ell \approx 2$ and a decaying tail at larger scales. The discrepancy between the information per scale distributions expected in the strong coupling limit in Fig.~\ref{fig:info_lattice_picture} and those obtained at $ga=1$ in Fig.~\ref{fig:spectrum} (right) indicates that the Hamiltonian parameters at $ga=1$ remain far from the asymptotic strong coupling limit. To make this more evident, we complement Fig.~\ref{fig:spectrum} with the information per scale of the vacuum, vector, and scalar meson states for $ga=2$, identified with the same DMRG procedure described above. For this parameter choice, the vacuum state exhibits dominant correlations at the $\ell \approx 0$ level. Moreover, the vector and scalar mesons show nearly identical information per scale distribution. The latter feature is also present at strong coupling, where the vector and scalar meson states, given by Eq.~\eqref{eq:vector_meson_strong_coupling}, have coinciding information per scale distributions. Nevertheless, the information per scale distributions for the meson states in Fig.~\ref{fig:spectrum} are qualitatively distinct from those in Fig.~\ref{fig:info_lattice_picture} (c), indicating that the lattice model at $ga=2$ is also far from the asymptotic strong coupling limit. Some features of the strong coupling limit, namely the overlap of the $I(\ell)$ distributions of the meson states, are, however, already present.

\begin{figure}[tb]
    \centering
    \includegraphics[width=0.45\textwidth]{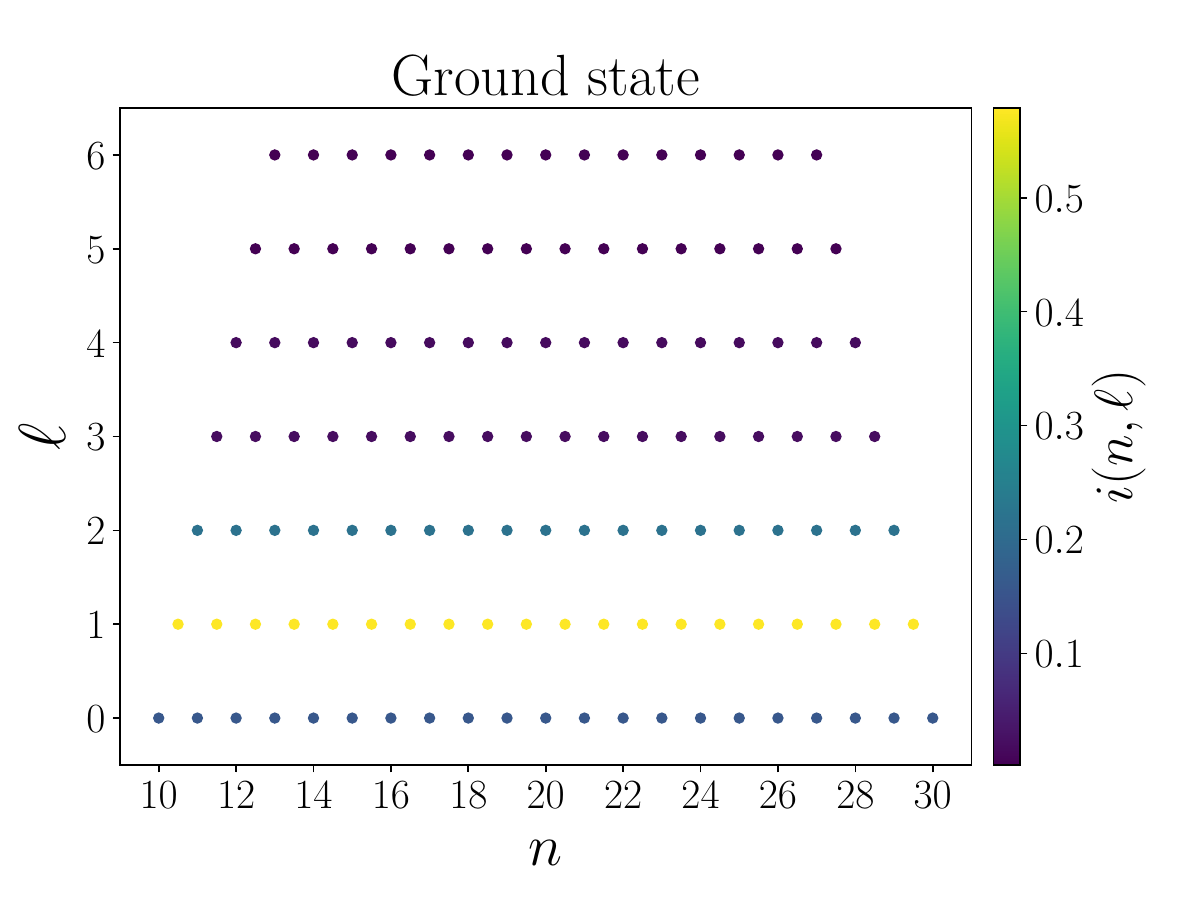}
    \includegraphics[width=0.45\textwidth]{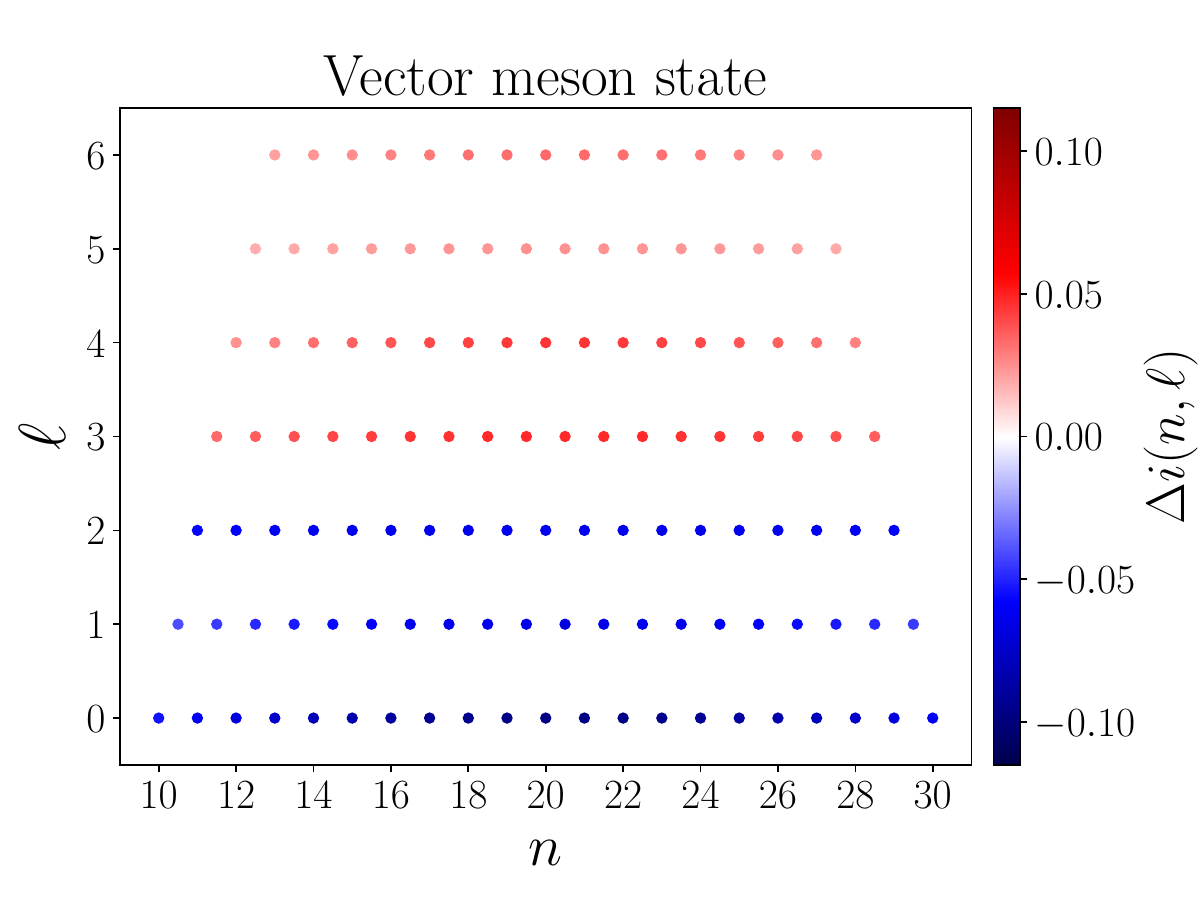}
    \includegraphics[width=0.45\textwidth]{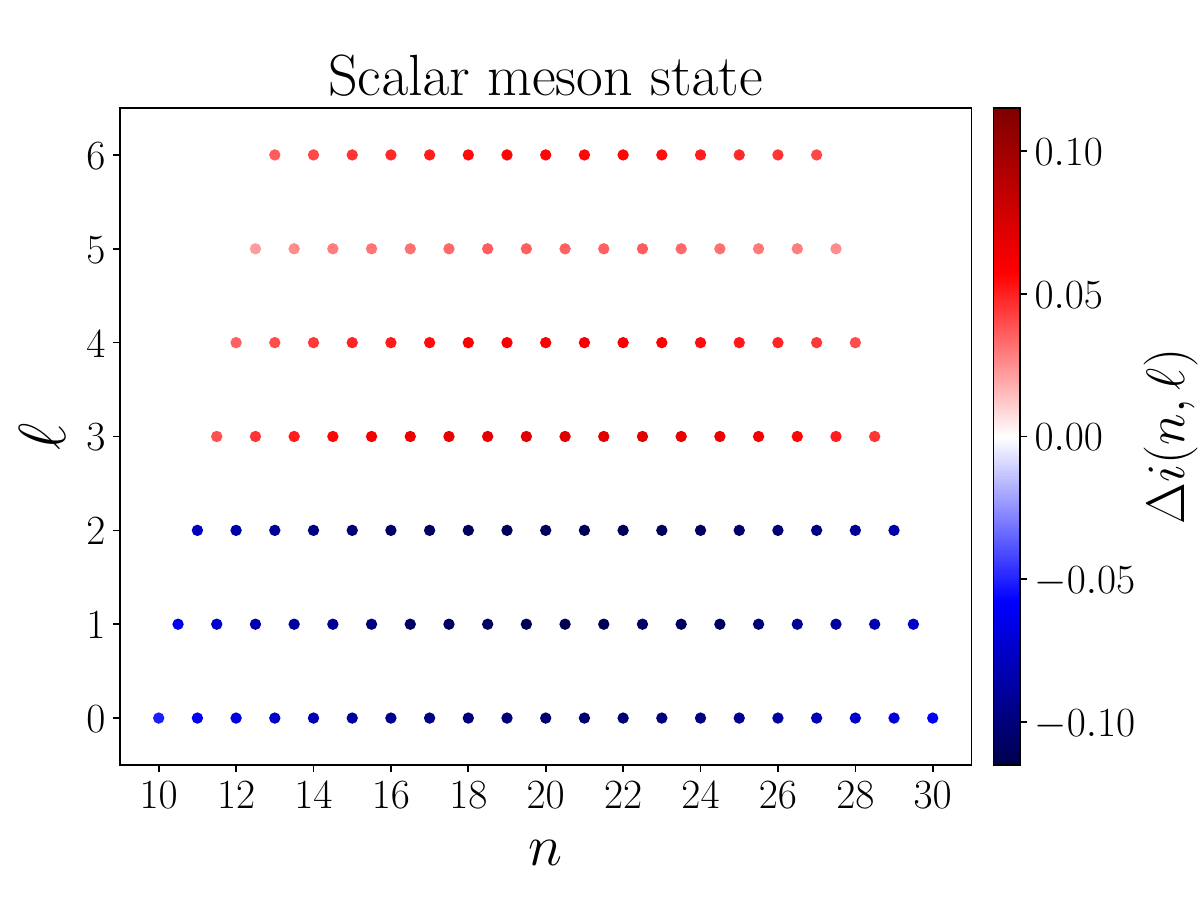}
    \caption{Information lattice for $\ell \leq 6$ for the ground, vector meson, and scalar meson states in the Schwinger model for $ga=1$, $ma=10^{-5}$, and $N=40$. \textbf{Left}: The ground state's $i(n,\ell)$ distribution is dominated by correlations at $\ell \approx 1$ with a decaying tail. Note that the exact strong coupling vacuum is a product state, and thus, there one would have correlations only at $\ell=0$. \textbf{Right}: Difference between the $i(n,\ell)$ distributions for the first excited state and the ground state. The vector meson state is dominated by higher-level correlations for $\ell \gtrsim 3$. \textbf{Center}: The difference between the ground state and the scalar meson $i(n,\ell)$ distributions, which shows that this excited state has a strong enhancement of correlations at $\ell \approx \{3,4\}$.}
    \label{fig:iln_spectrum}
\end{figure}

Finally, to better distinguish the properties of the different states, in Fig.~\ref{fig:iln_spectrum} we show the local information distribution $i(n,\ell)$ for $\ell<7$ for the vacuum (left) and the differences of the meson's local information to the ground state, $\Delta i(n,\ell) = i(n,\ell)\big|_{\rm{meson}} - i(n,\ell)\big|_{\rm{ground\;state}}$, (center and right). With the information lattice, one can better observe the enhancement of information in the excited states with respect to the ground state at the $\ell \gtrsim 3$ levels. In particular, the scalar meson shows a more prominent enhancement for $\ell \gtrsim 4$.

\begin{figure}[tb]
    \centering
    \includegraphics[width=0.48\linewidth]{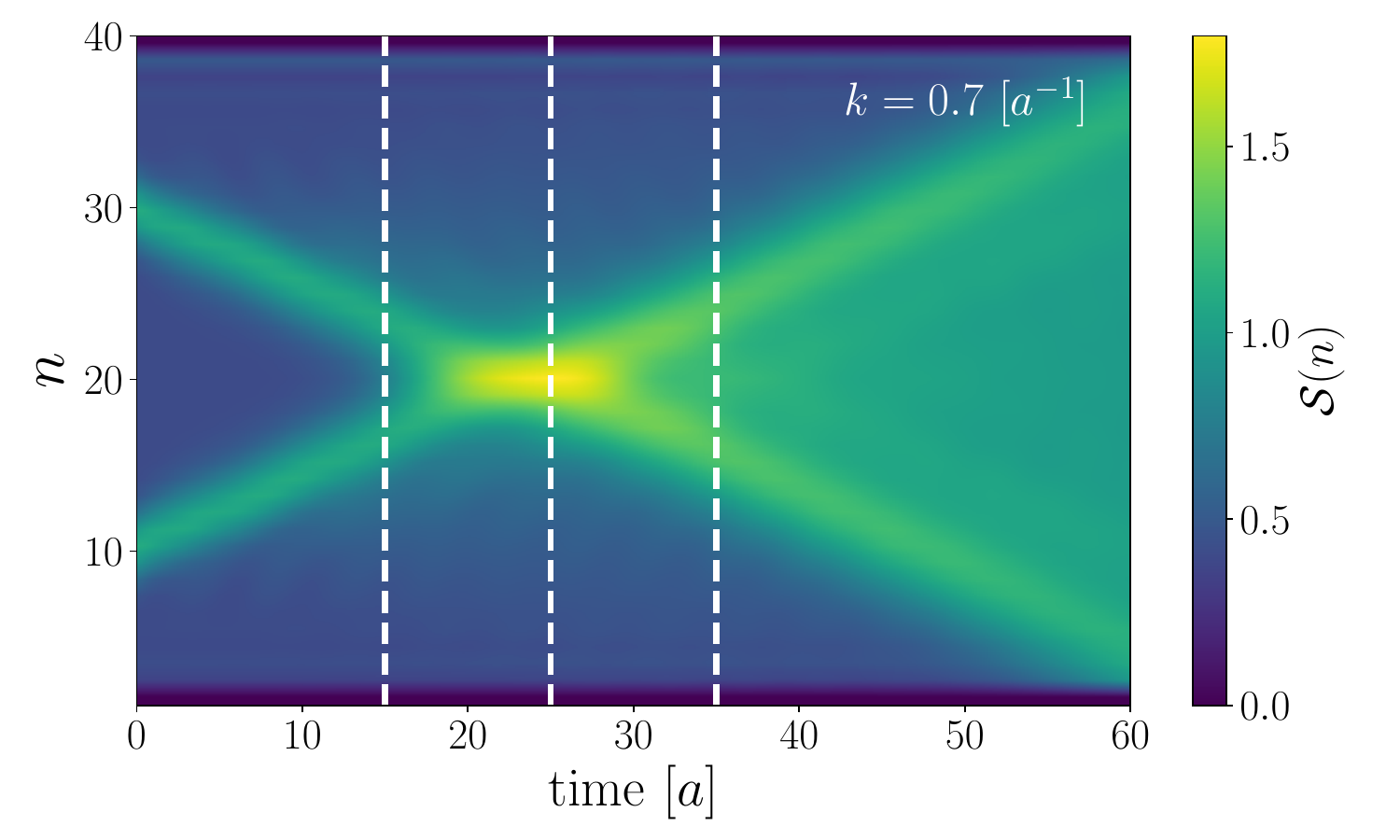}
    \includegraphics[width=0.48\linewidth]{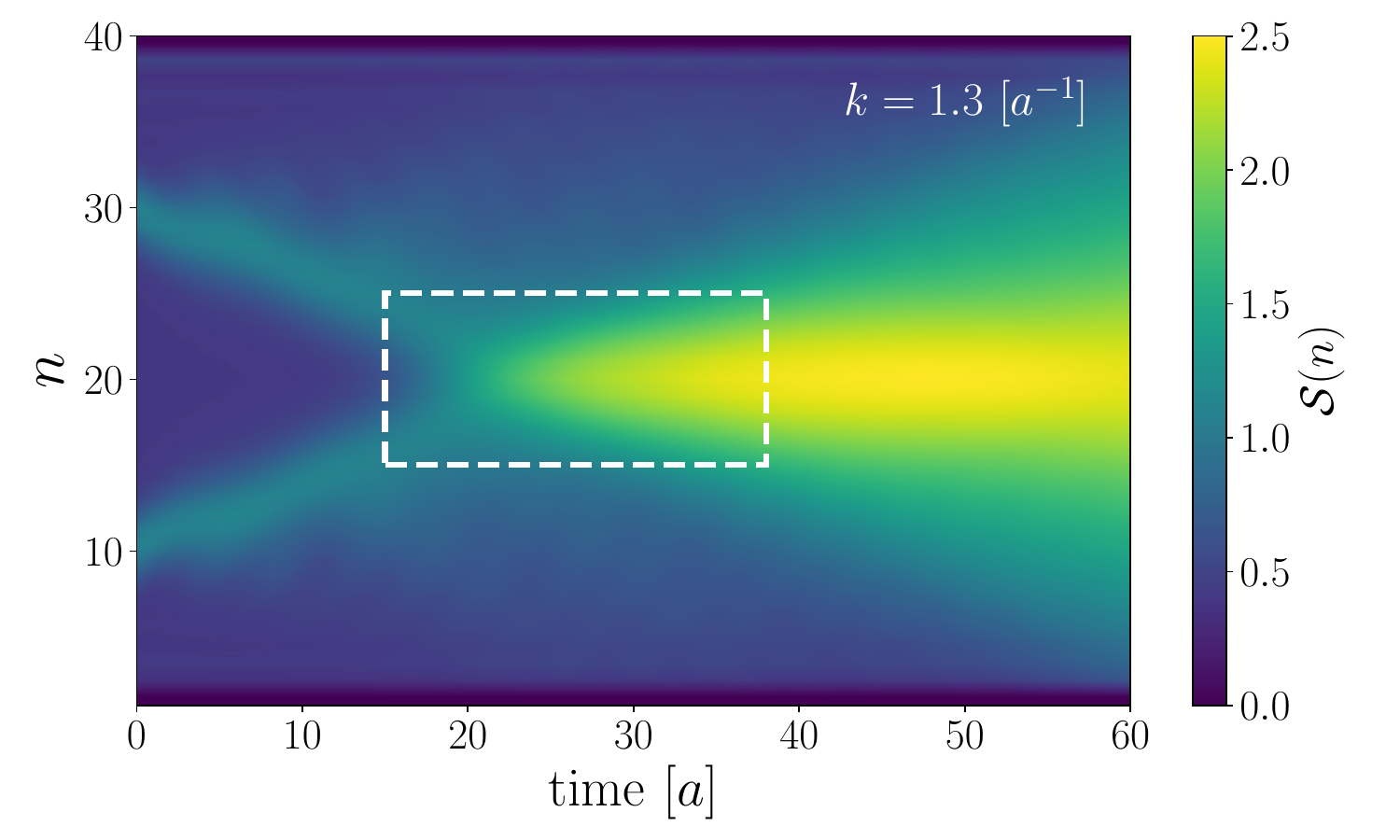}
    \caption{Bipartite entanglement entropy across a cut between sites $n$ and $n+1$, i.e., for the region from site 1 to $n$.  We set $ga=1$, $ma=10^{-5}$, and $N=40$. \textbf{Left}: Scattering below particle threshold ($ka=0.7$). Vertical white lines indicate the time slices used in Figs.~\ref{fig:scattering_k07} and \ref{fig:scattering_k13}. \textbf{Right}: Scattering above threshold ($ka=1.3$). The white box highlights the region used to compute the distributions in Fig.~\ref{fig:cut_Iln_scattering}.}
    \label{fig:entropy}
\end{figure}

Having discussed the information properties of the states in the theory, we now use these insights to interpret the results of scattering processes between two initial wave packets made of the vector meson. We closely follow the protocol used in Ref.~\cite{Papaefstathiou:2024zsu}, except we consider a smaller lattice.\footnote{We also do not implement the truncation in the local electric field Hilbert spaces, and take a maximum bond dimensions of $D=30$. We checked that using $D=40$ does not lead to any significant changes in the results.} We first prepare two wave packets of the form
\begin{align}\label{eq:jet_init}
\sum_{n} \phi(n,j) e^{-i n k}  \left( \sigma^+_{n}\sigma^-_{n+1}- \sigma^+_{n+1}\sigma^-_{n}\right) \ket{\Omega} \, ,
\end{align}
where $\phi(n,j)$ is a Gaussian centered around the lattice site $j$ with dispersion $\sigma=a$, $k$ is the momentum of the wave packet, and $\ket{\Omega}$ is the vacuum state which is obtained from the DMRG routine. Notice that in the asymptotic strong coupling limit, where $\ket{\Omega}=\ket{\Omega_{\rm s.c.}}$, this is nothing but a wave-packet constructed from the vector meson state $\ket{1_{\rm{V}}}$ in Eq.~\eqref{eq:vector_meson_strong_coupling}. We then time-evolve the system, allowing for the wave packets to scatter. In Fig.~\ref{fig:entropy}, we illustrate the time evolution of the system by considering the bipartite entanglement entropy 
\begin{equation}
\label{eq:bipartite_entropy}
    \mathcal{S}(n) = - \mathrm{Tr} \, \rho(n) \log_2[\rho(n)] \, ,
\end{equation}
where $\rho(n)$ is the density matrix of the subsystem containing the lattice sites from $1$ to $n$.\footnote{Notice that this entropy is distinct from the one studied in Ref.~\cite{Papaefstathiou:2024zsu}, which was computed for the two-site subsystem made of the lattice sites $n$ and $n+1$. Both these entropy measures are incorporated in the information lattice. In fact, $\mathcal{S}(n) = \log_2[\text{dim}(\rho^{n-1}_{n+1/2})] - I(\rho^{n-1}_{n+1/2})$ where $I(\rho^{n-1}_{n+1/2})$ can be computed by using the decomposition property in Eq.~\eqref{eq:local_info_decomposition}.
} Recalling that $k_{\rm thersh}a\approx 1.12$, we consider the case $k\ll k_{\rm thersh}$ on the left panel and $k\gg k_{\rm thersh}$ on the right panel. The transition between the elastic and inelastic regimes is evident from the profile of the entropy distribution in space. If no scalar meson is produced, then after the scattering, the entropy peak follows the light-cone structure set by the incoming wave packets; in contrast, when above threshold, the new particles produced at low momentum lead to an extensive peak at intermediate $n$ in the entropy distribution.

In Ref.~\cite{Papaefstathiou:2024zsu}, the production of scalar mesons was determined by computing the projection of the wavefunction on a four-body fermionic operator, which is expected to have significant overlap with the vector but not the scalar state. Note that this method can only cleanly separate the two states at strong coupling; away from this limit, the vector state also couples to higher-body excitations. Therefore, this method has certain limitations in the characterization of the states of the theory for parameter regimes such as $ga=\{1,2\}$. Here, we discuss how the information lattice can be used to map out the properties of the states produced after the scattering event. Importantly, the information lattice does not require any \textit{a priori} knowledge of the correct operator one should consider to distinguish the states, and thus could also be used to study the structure of scattering processes in more complex theories.\footnote{We note, however, that establishing whether the state connects to an eigenstate of the theory still requires a more detailed analysis.}

\begin{figure*}[tb]
    \centering
    \includegraphics[width=0.48\linewidth]{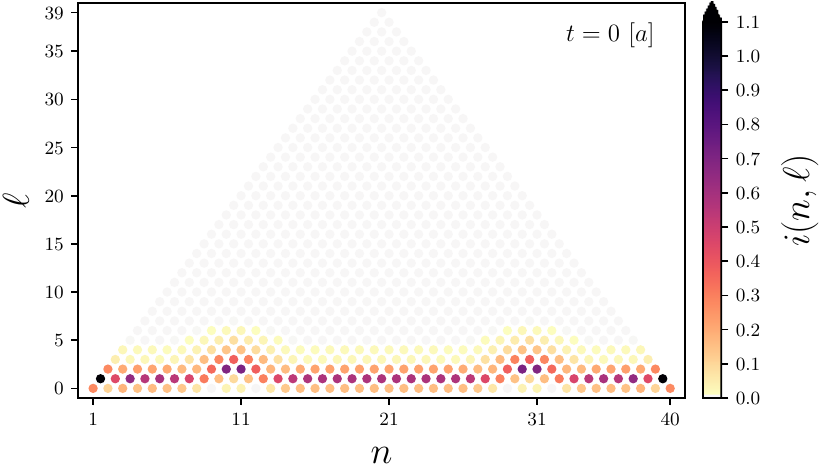}
    \includegraphics[width=0.48\linewidth]{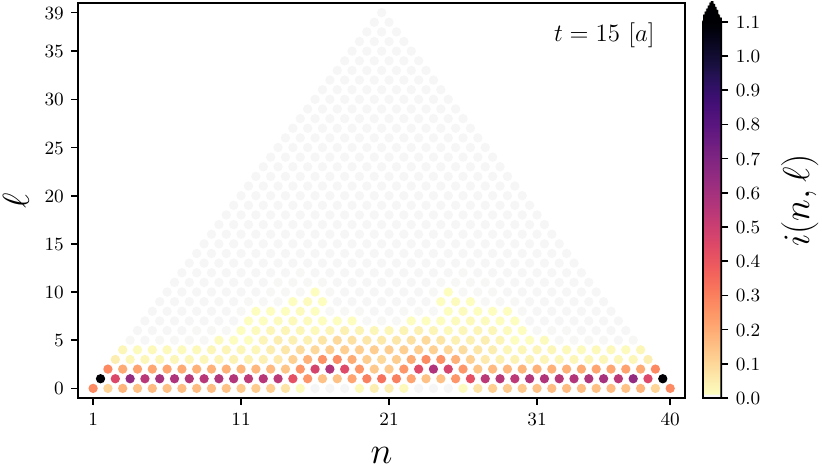}
    \includegraphics[width=0.48\linewidth]{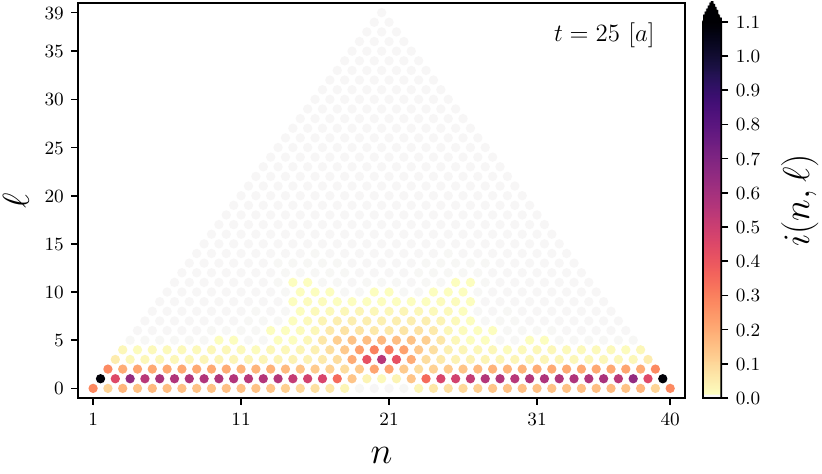}
    \includegraphics[width=0.48\linewidth]{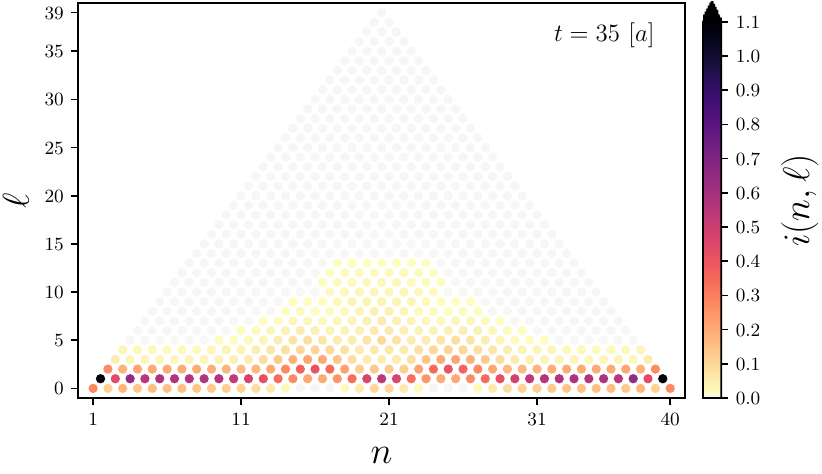}
    \caption{Snapshots of the information lattice for the scattering of two wave packets as in Eq.~\eqref{eq:jet_init} for $ka=0.7$. The selected times correspond to the dashed vertical lines shown in Fig.~\ref{fig:entropy}. We set $ga=1$, $ma=10^{-5}$, and $N=40$.}
    \label{fig:scattering_k07}
\end{figure*}

\begin{figure*}[tb]
    \centering
    \includegraphics[width=0.48\linewidth]{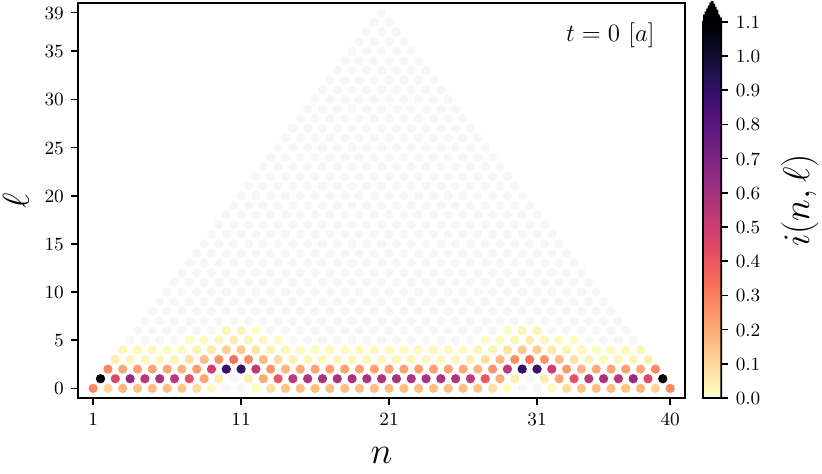}
    \includegraphics[width=0.48\linewidth]{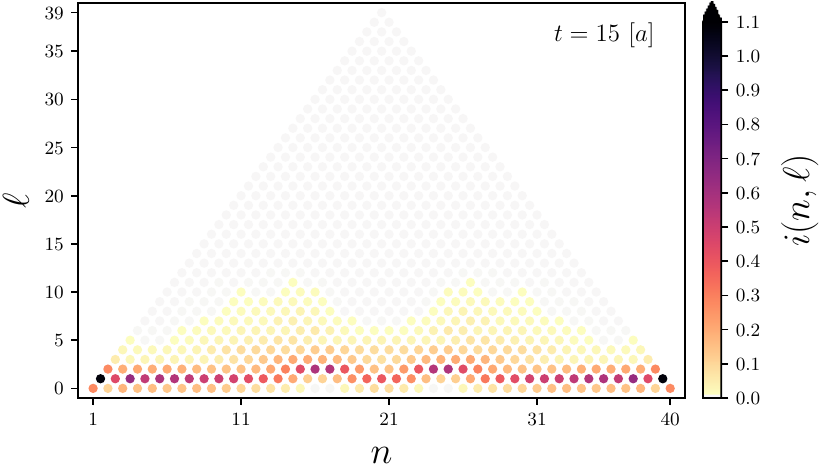}
    \includegraphics[width=0.48\linewidth]{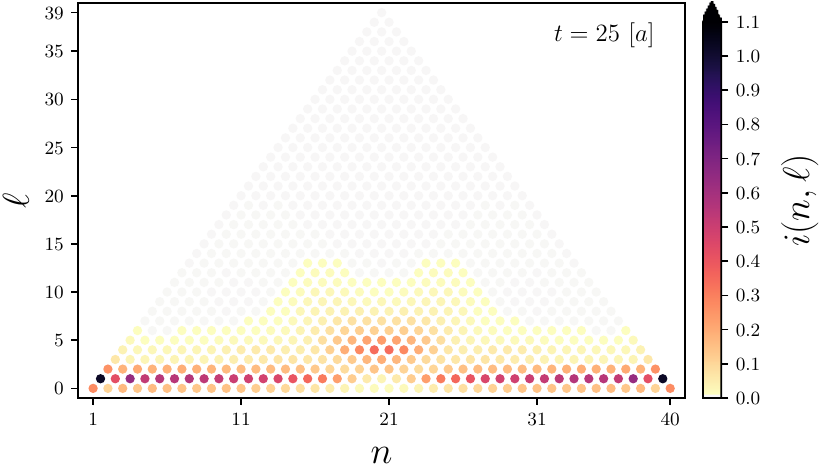}
    \includegraphics[width=0.48\linewidth]{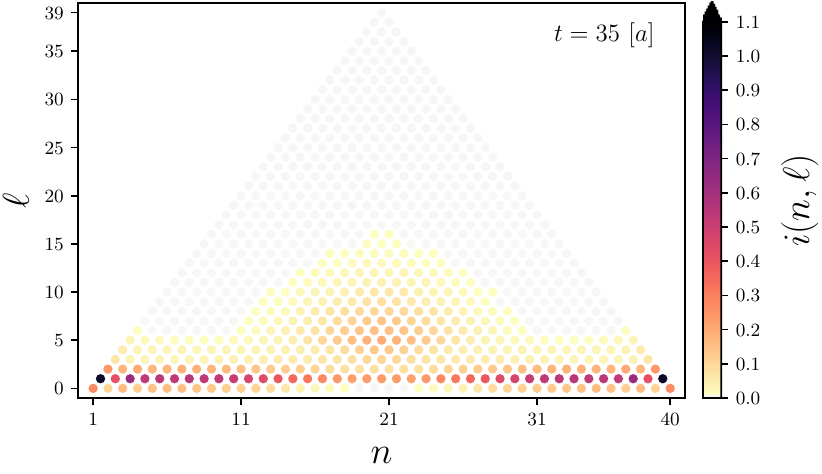}
    \caption{Plots analogous to Fig.~\ref{fig:scattering_k07}, now for $ka=1.3$. We set $ga=1$, $ma=10^{-5}$, and $N=40$.}
    \label{fig:scattering_k13}
\end{figure*}

In Figs.~\ref{fig:scattering_k07} and \ref{fig:scattering_k13}, we analyze the time evolution of the scattering through the information lattice for $ka=0.7$ and $ka=1.3$, respectively, setting $ga=1$. Before the scattering, the local information distribution inside the wave packets has the characteristic features of the vector meson's distribution at strong coupling inherited from $\left( \sigma^+_{n}\sigma^-_{n+1}- \sigma^+_{n+1}\sigma^-_{n}\right) \ket{\Omega}$, thus similar to that of $\ket{1_{\rm{V}}}$. After scattering ($t > 25 \,a$), the central region of the information lattice is populated by an intermediate state with correlations peaked at $\ell\approx3$ for $ka=0.7$ and $\ell\approx4$ for $ka=1.3$. This is consistent with the entropy peak observed in both scenarios. At later times ($t = 35\, a$), we observe that, in the elastic scenario, the local information distribution relaxes back to smaller scales, is centered at the receding wave packets, and becomes broader than in the initial state.
In contrast, for the $ka=1.3$ scenario, we observe that, after the collision, there is the formation of a state localized at the center of the information lattice with characteristic $\ell\approx5$ dominant correlations. Although in Figs.~\ref{fig:spectrum} and ~\ref{fig:iln_spectrum} we observe a significant overlap between the two meson states, we emphasize that the scalar meson has a larger amount of information at large scales $\ell \gtrsim 4$.

\begin{figure}[tb]
    \centering
    \includegraphics[width=.48\columnwidth]{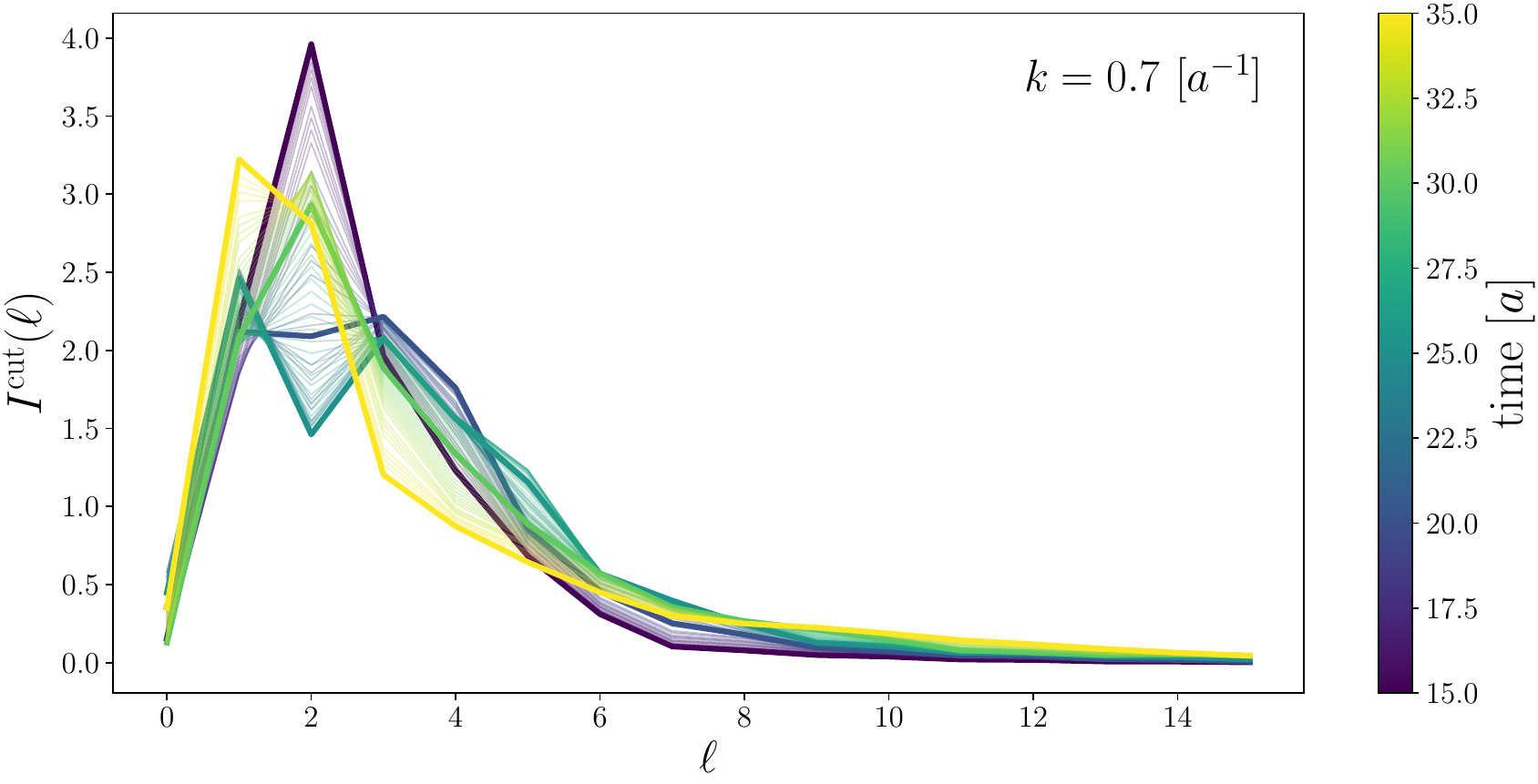}
    \includegraphics[width=.48\columnwidth]{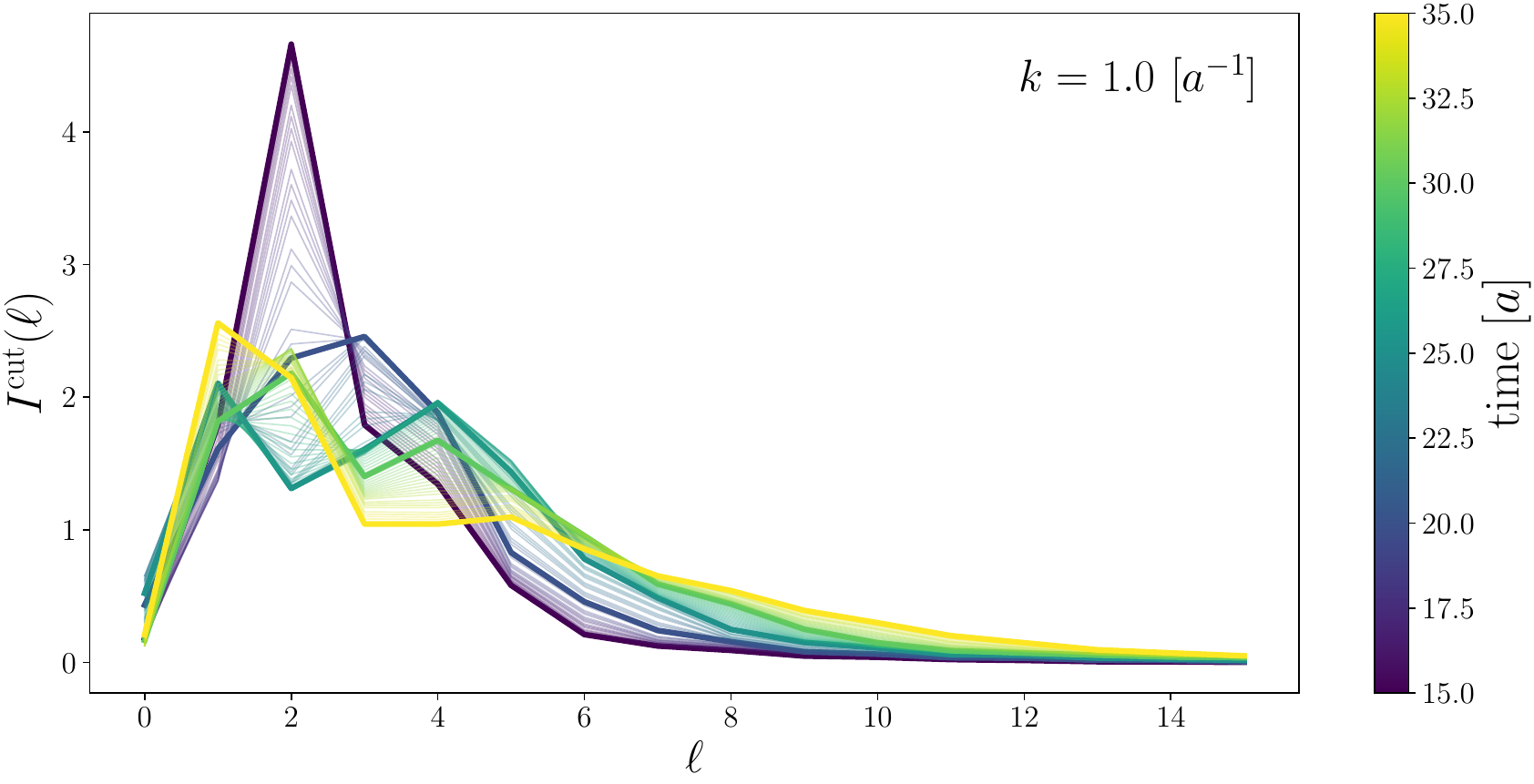}
    \includegraphics[width=.48\columnwidth]{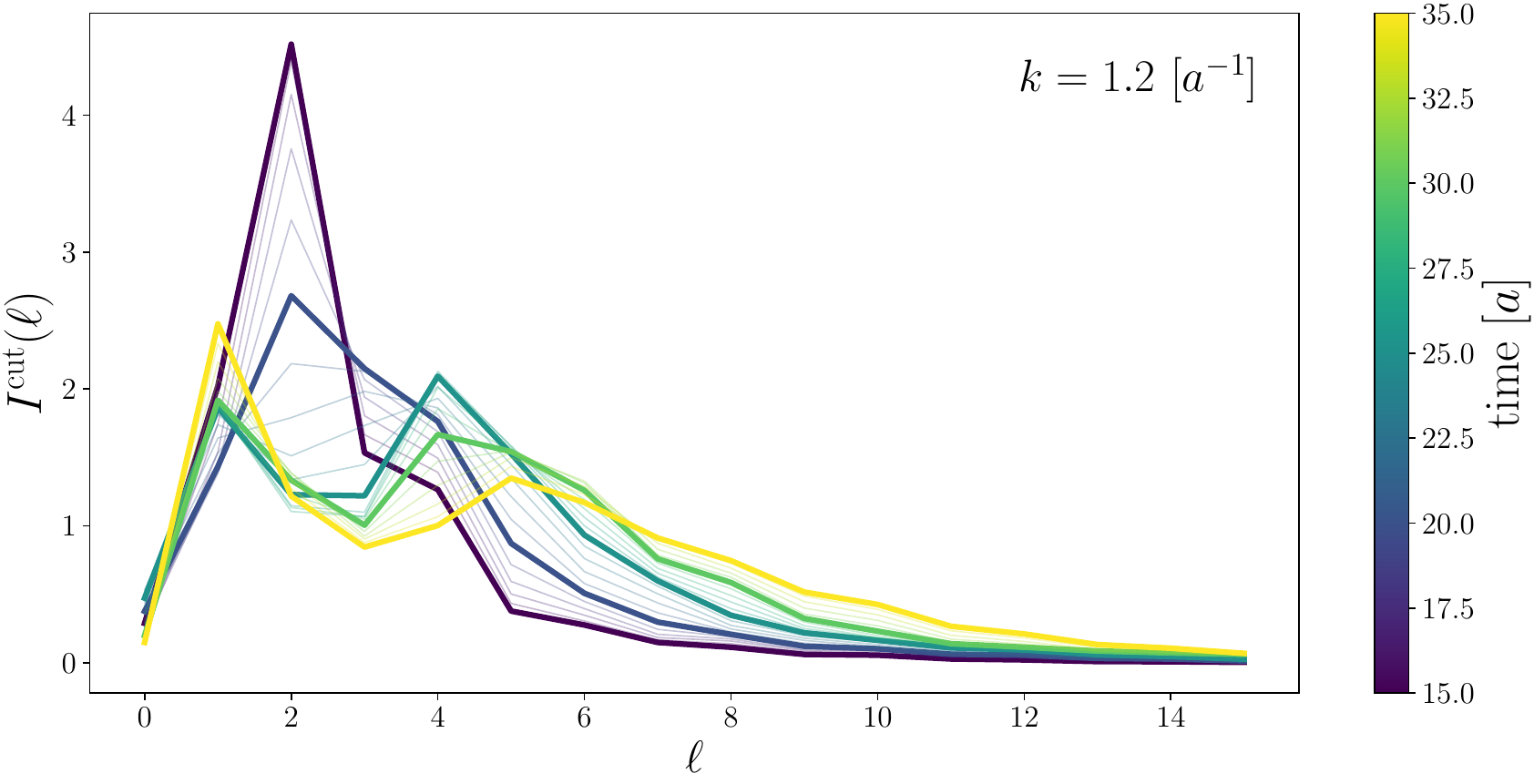}
    \includegraphics[width=.48\columnwidth]{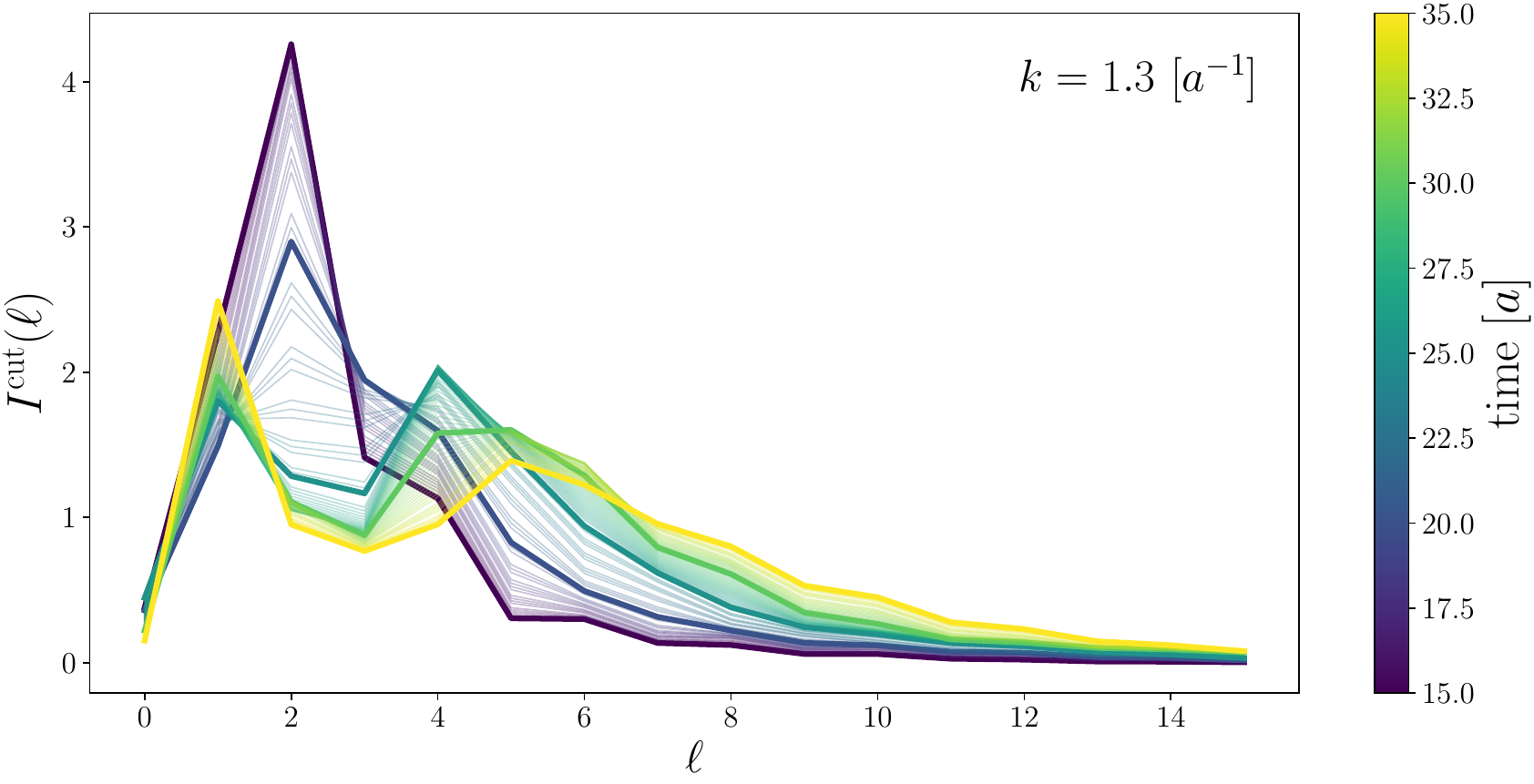}
    \caption{Integrated information distribution for the central region of the scattering process shown in Fig.~\ref{fig:entropy} (right) for $ka=\{0.7,1,1.2,1.3\}$. Thicker lines highlight times multiples of $5a$.}
    \label{fig:cut_Iln_scattering}
\end{figure}

We furthermore investigate the properties of this scattering process by computing the integrated distribution of local information for the central sites of the information lattice and corresponding to the time region shown on the right plot of Fig.~\ref{fig:entropy}:
\begin{align}
    I^{\rm cut}(\ell) = \sum_{\substack{15\leq n<25}} i(\ell,n) \, ,\quad 15 a<t<35 a\, .
\end{align}
The results for $I^{\rm cut}(\ell)$ are shown in Fig.~\ref{fig:cut_Iln_scattering} for $ka=\{0.7,1,1.2,1.3\}$.
Note that the sum over $\ell$ of $I^{\rm cut}(\ell)$ is not a conserved quantity under unitary time evolution [as it is instead $I(\ell)$ in Eq.~\eqref{eq:information_per_scale}], and thus one should only qualitatively compare the shapes of the distributions of $I^{\rm cut}(\ell)$. In all cases, the initial configuration is peaked at $\ell \approx 2$, characteristic of the incoming states prepared according to Eq.~\eqref{eq:jet_init}. During the collision process, the peak of $I(\ell)$ increases up to $\ell \approx 5$. For the smaller momenta, $I(\ell)$ relaxes back to a configuration dominated by a peak at $\ell \approx 1$, characteristic of the lowest energy state at $ga=1$, see Fig.~\ref{fig:spectrum}, and a finite decaying tail. As the initial momentum increases, in addition to the $\ell \approx 1$ peak, a second distribution emerges centered around $\ell \approx 4\text{--}5$. This is an indicator for the production of a state that has a large overlap with the scalar meson state, complementing the above observations.

\subsection{Electric string dynamics}\label{sec:strings}

In this section, we study the out-of-equilibrium dynamics following a local quench where two external charges are injected at the center of the physical chain and move apart along their respective light cones. The electric flux string expanding between the external charges excites the vacuum of the theory and can lead to particle production via the Schwinger mechanism, potentially leading to the dynamical breaking of the string. For this to happen, the local electric field has to become larger than the critical field, see Section~\ref{sec:schwinger_model}. Conversely, if the field is too weak to dynamically generate and accelerate new charges, then the original string is never broken, and it only deposits energy into the bulk state. This quench experiment was first proposed by Casher \textit{et al.}~\cite{Casher:1974vf} as a simple model for the hadronization transition in QCD. More recently, spurred by the seminal work of Calabrese and Cardy~\cite{Calabrese:2006rx,Alba:2017lvc,Cheneau_2012}, there has been a large interest in understanding how related dynamics take place in integrable models deformed by a confining potential~\cite{Chanda:2019fiu,Kormos_2016,James_2019,Robinson_2019}. These models are characterized by modifications to the characteristic light-cone structure of integrable theories, exhibiting richer dynamics that may be shared by lattice gauge theories. In the Schwinger model, the string dynamics have also been widely explored, see, e.g., Refs.~\cite{Liu:2024lut,Buyens:2015tea}, as well as in related theories~\cite{Chanda:2019fiu,Mallick:2024slg}. More recently,  preliminary studies on stringy dynamics in $(2+1)$D models have also appeared in the literature~\cite{Cobos:2025krn,DiMarcantonio:2025cmf,Gonzalez-Cuadra:2024xul,Cataldi:2025cyo,Tian:2025mbv,Xu:2025abo}.

\begin{figure}[h!]
    \centering
    \includegraphics[width=0.48\linewidth]{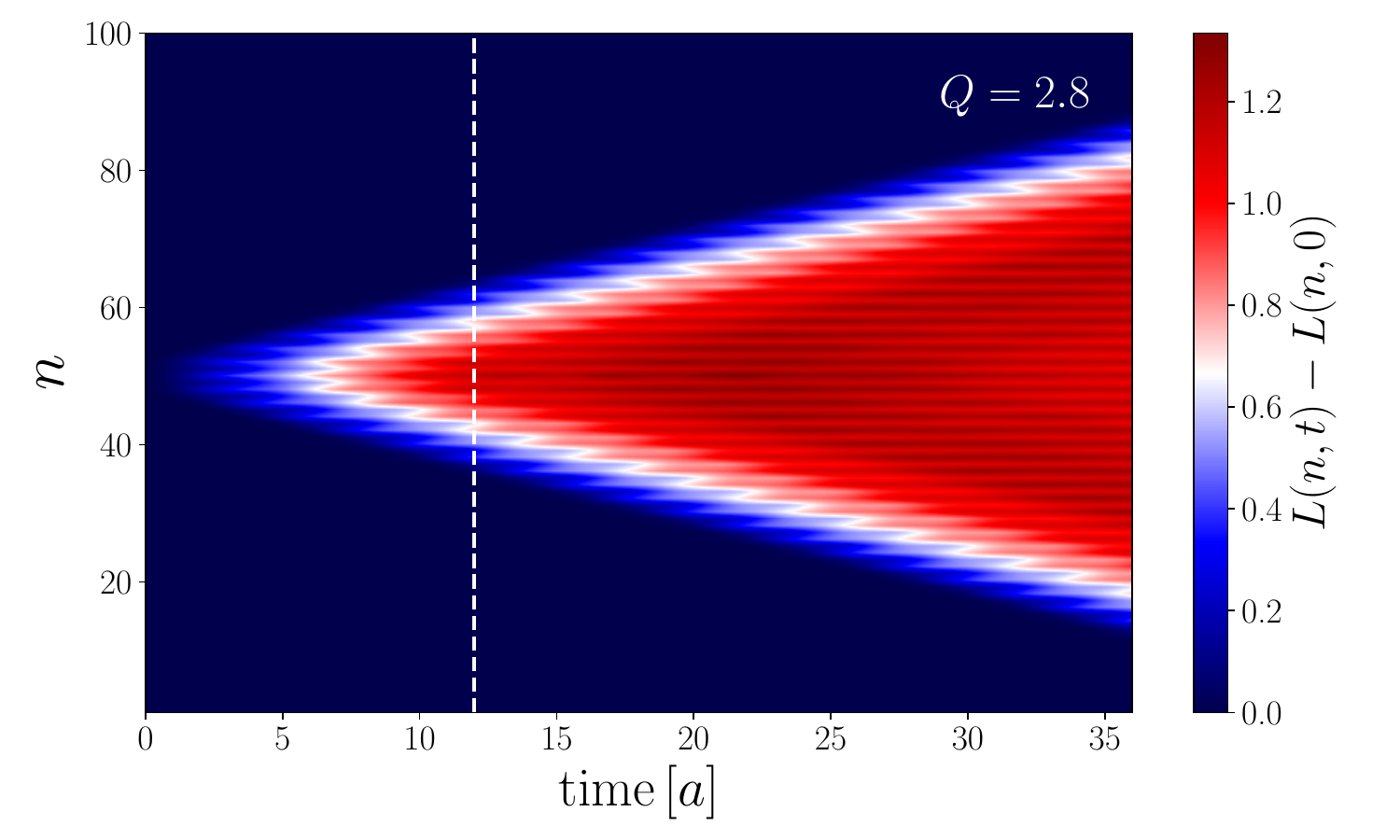}
    \includegraphics[width=0.48\linewidth]{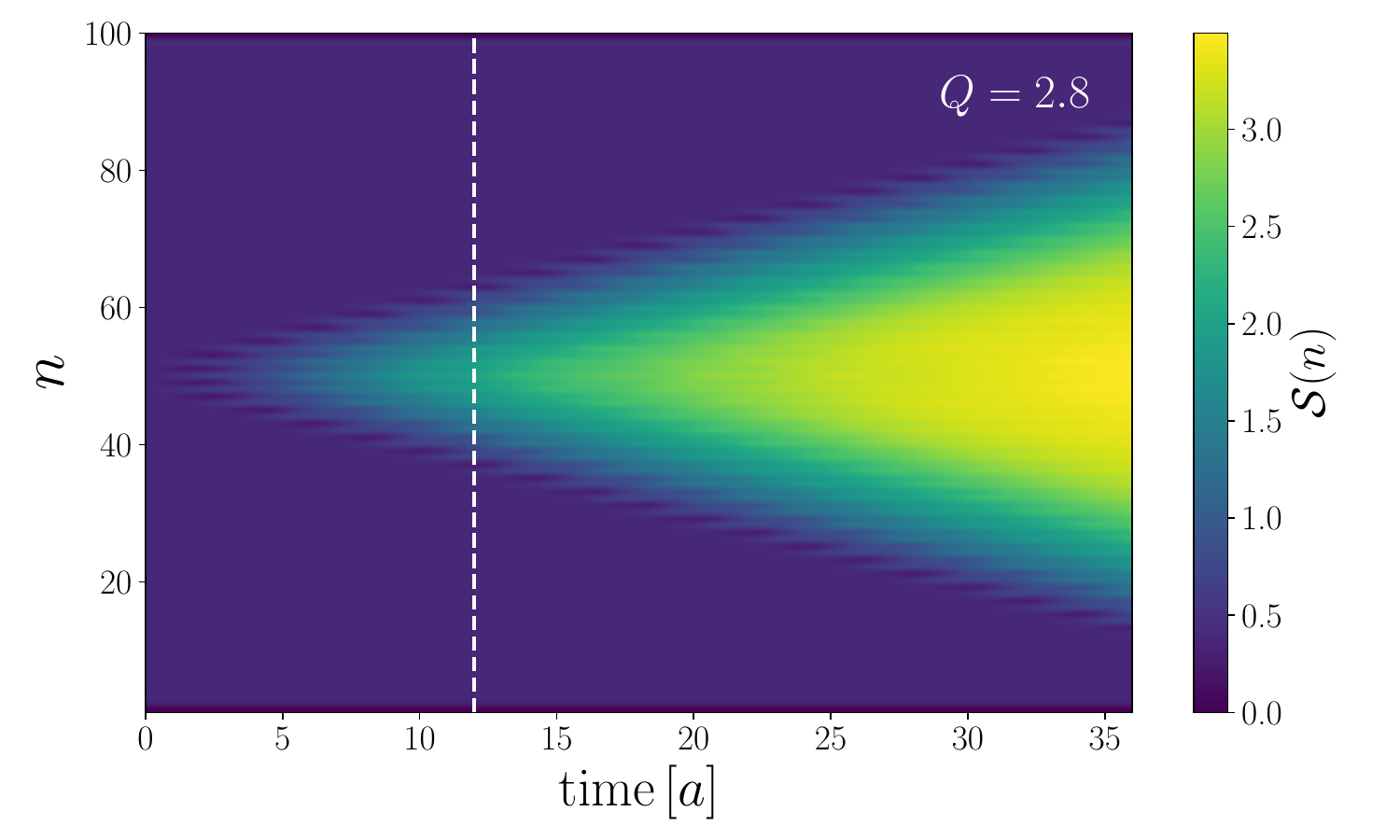}\\
    \includegraphics[width=0.48\linewidth]{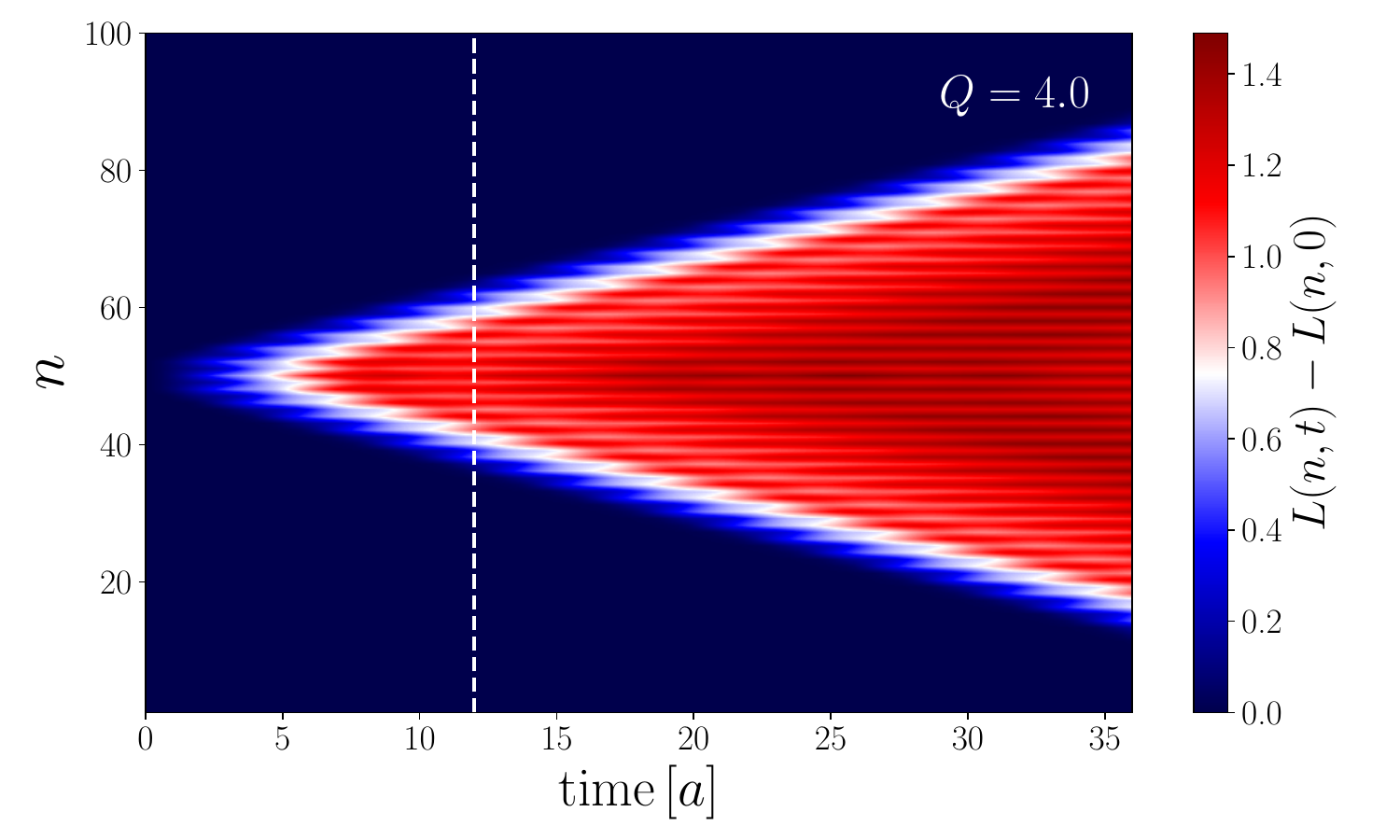}
    \includegraphics[width=0.48\linewidth]{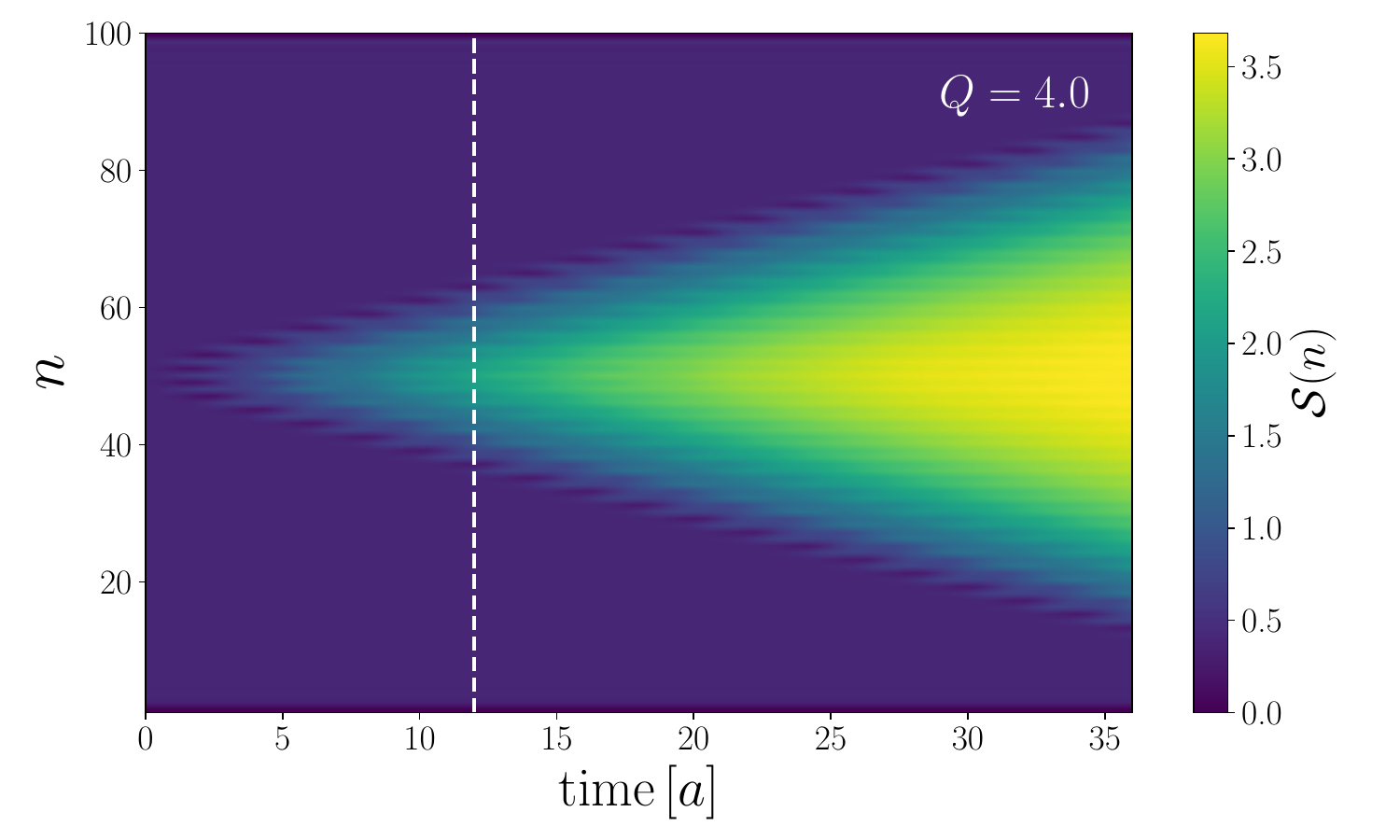}\\
    \caption{Time evolution of the expectation value of the onsite electric field $L(n,t)-L(n,0)$ and the bipartite entanglement entropy $\mathcal{S}(n)$ for quenches with injected external charges using $Q=\{2.8,4\}$, $ga=0.5$, $ma=0.25$, and $N=100$. }
    \label{fig:11_full_charges}
\end{figure}

\begin{figure}[h!]
    \centering
    \includegraphics[width=0.48\linewidth]{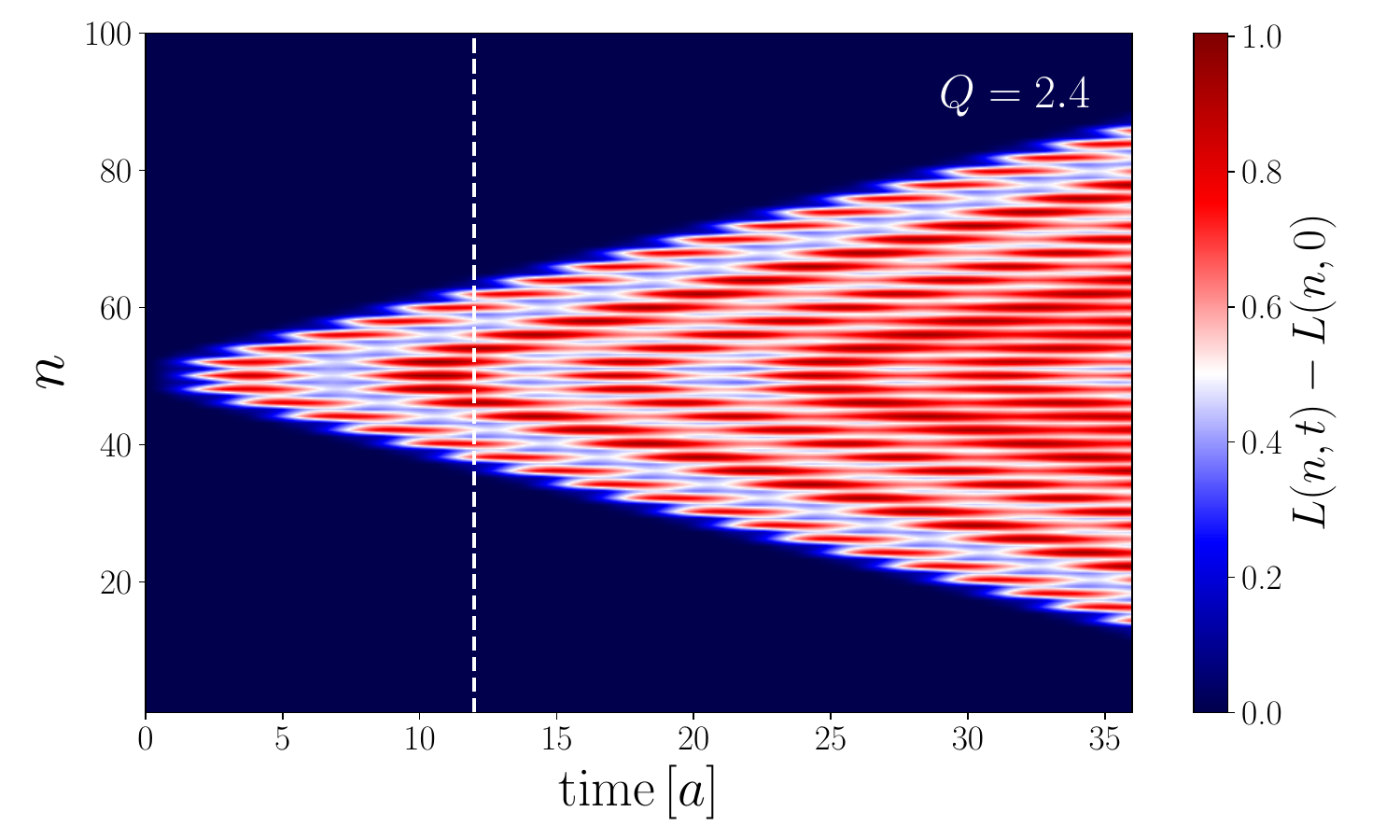}
    \includegraphics[width=0.48\linewidth]{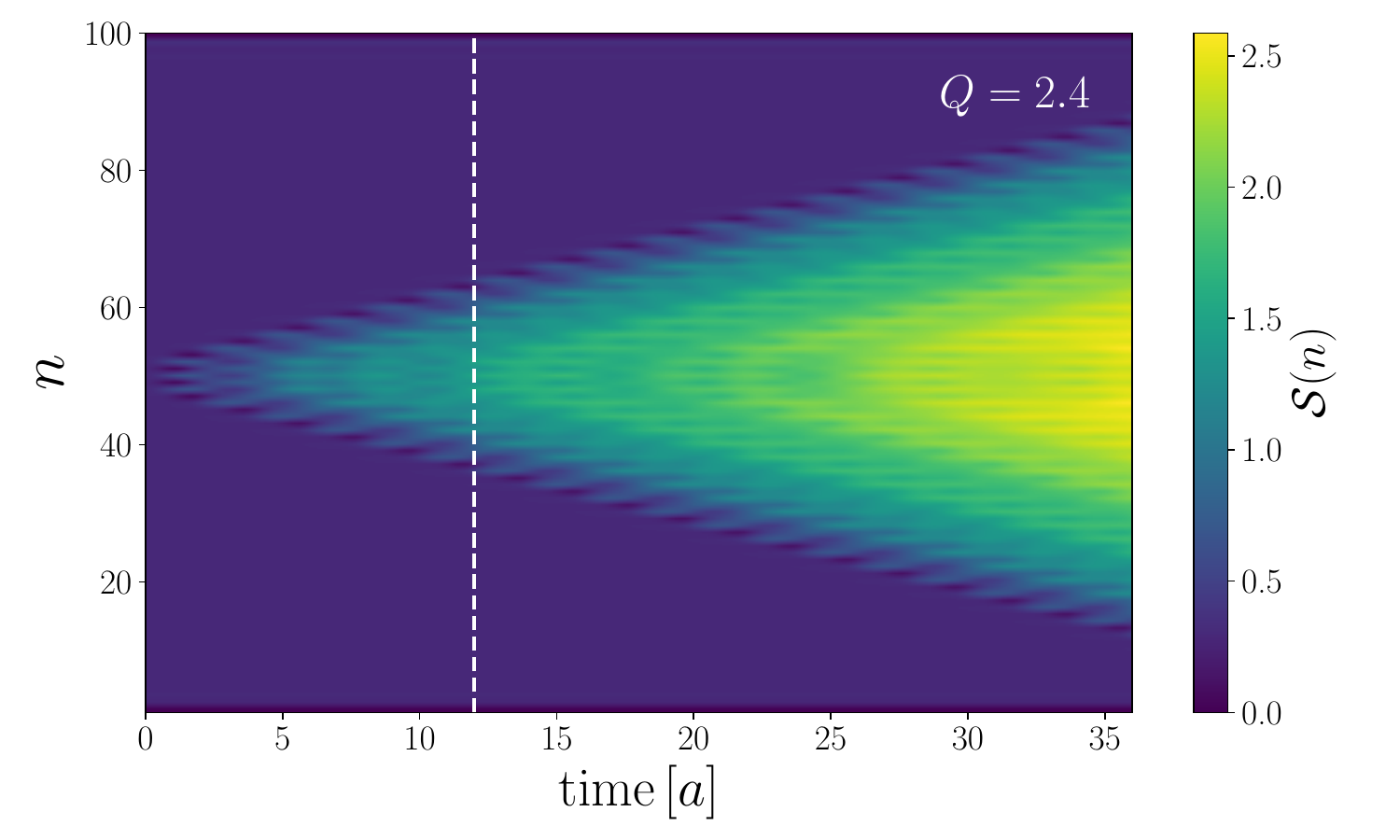}\\
    \includegraphics[width=0.48\linewidth]{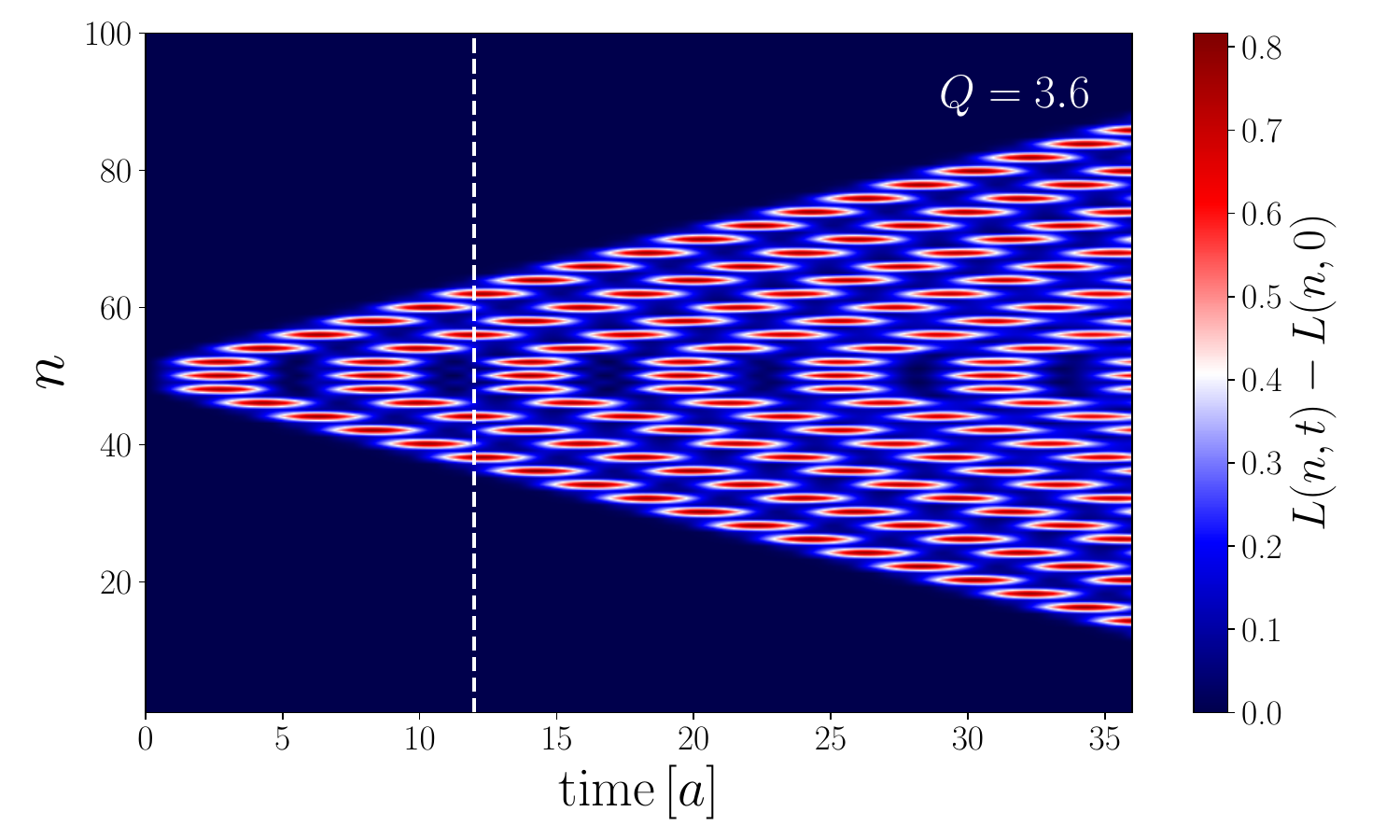}
    \includegraphics[width=0.48\linewidth]{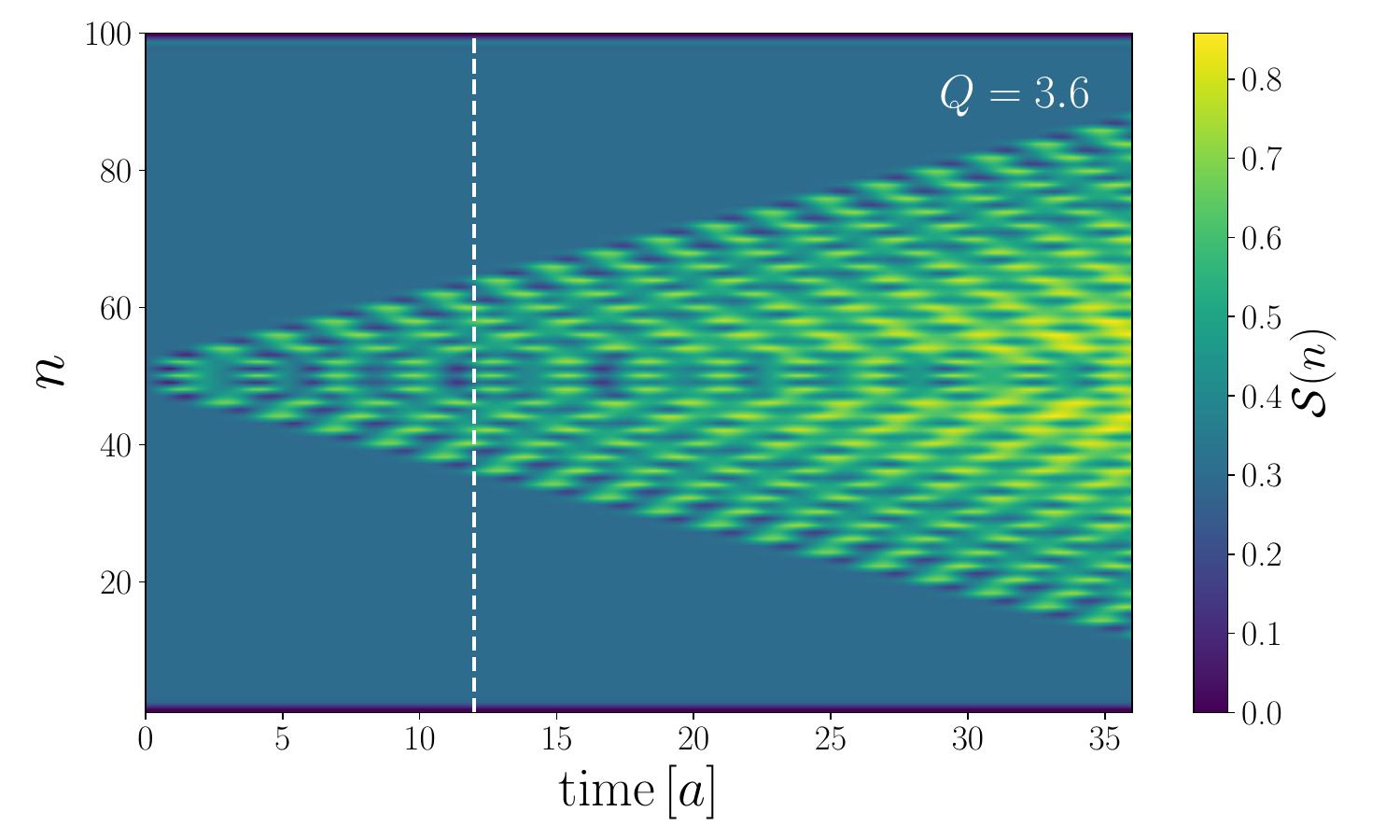}\\

    \caption{Plots analogous to Fig.~\ref{fig:11_full_charges}, now for $ga=1$. We use $Q=\{ 2.4,3.6 \}$, and set $ma=0.25$ and $N=100$.}
    \label{fig:22_full_charges}
\end{figure}

We first consider a quench where two charges are injected with an initial separation of six lattice sites and are then moved along the respective light cones with opposite momenta. Their trajectories can be visualized in Fig.~\ref{fig:11_full_charges}, where we illustrate the expectation value of the onsite electric field $L(n)= 1/2 \sum_{k=1}^n (\sigma^z_k+(-1)^k)$ and the bipartite entanglement entropy $\mathcal{S}(n)$ in Eq.~\eqref{eq:bipartite_entropy} for $ga=0.5$, $ma=0.25$, and $Q = \{2.8, 4\}$. Fig.~\ref{fig:22_full_charges} shows the same observables, now for $ga=1$. The external charges are introduced by adding a local topological term to the lattice Hamiltonian in Eq.~\eqref{eq:Spin_Hamiltonian}, such that $L(x) \to L(x) - Q\, g^{-1} \, \Theta(-ut<x<ut)$, where $u$ defines the light-cone speed. Here, it is set to the speed of light on the lattice.  

The results for $ga=0.5$ agree with the picture of an unbroken string that stretches and excites the vacuum in between the charges, indicating that the electric field for $Q = \{ 2.8,4 \}$ is below the critical value. In this scenario, there is a monotonic increase in the bipartite entanglement entropy at the center of the chain. In contrast, for $ga=1$ the external field can more easily go above the critical value $L_c\sim g^{-2}$, and for sufficiently large $Q$ this leads to the transition into the regime where a succession of string \textit{wedges} is formed. As the coupling increases, oscillations between different strings become nearly instantaneous and decay more slowly.
Multiple string breaking limits the growth of bipartite entanglement entropy, as correlations cannot build up between successive breakings.

\begin{figure}[h!]
    \centering
    \includegraphics[width=0.48\linewidth]{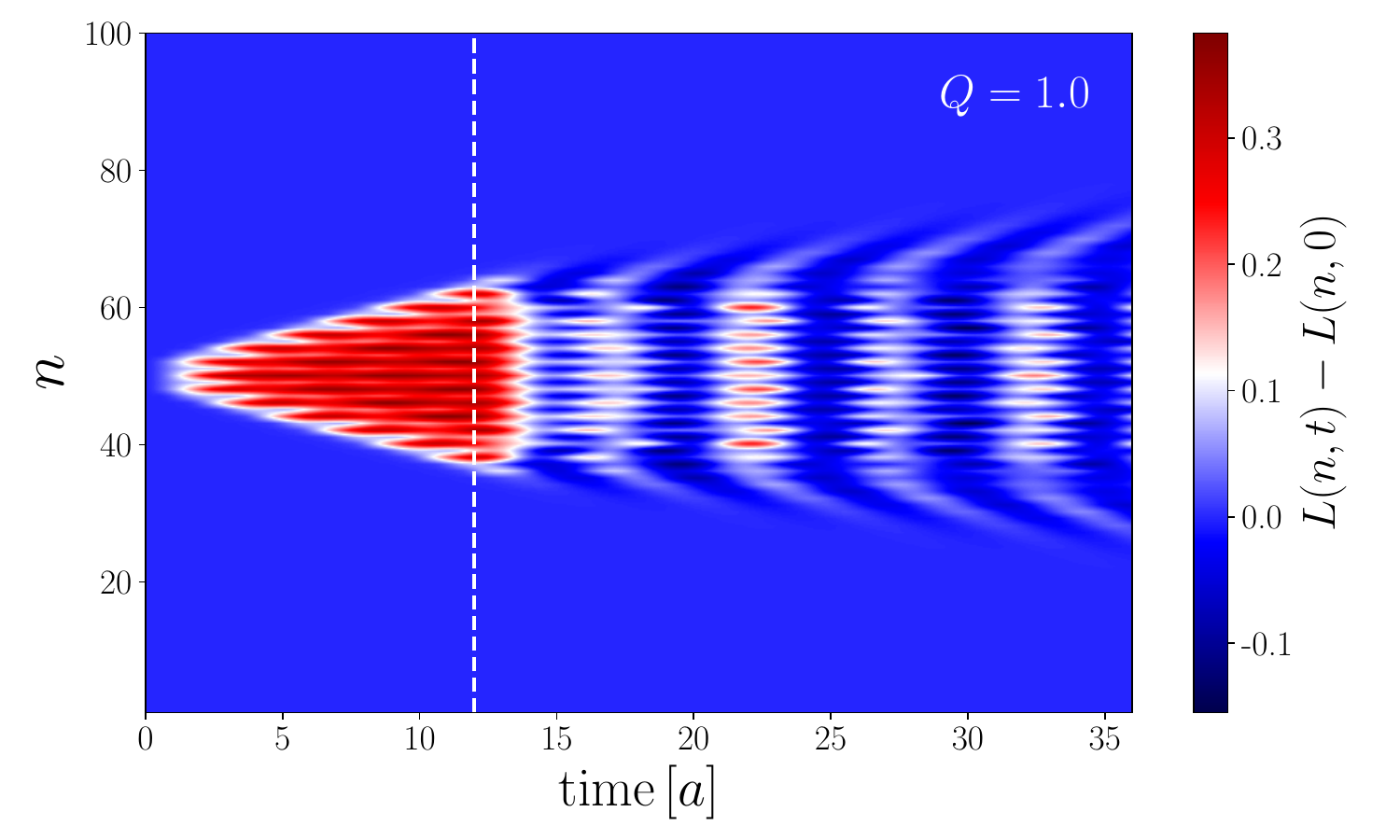}
    \includegraphics[width=0.48\linewidth]{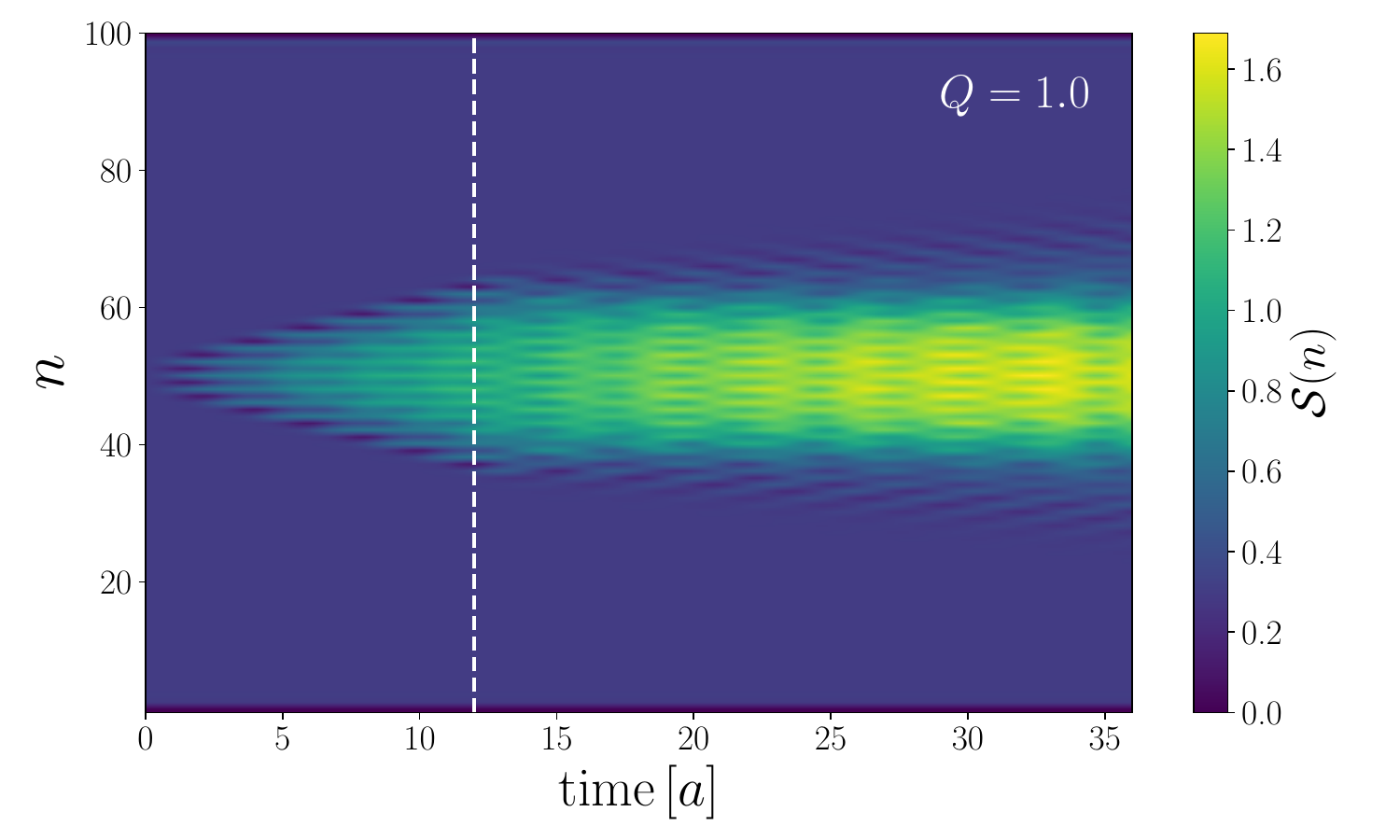}\\
    \includegraphics[width=0.48\linewidth]{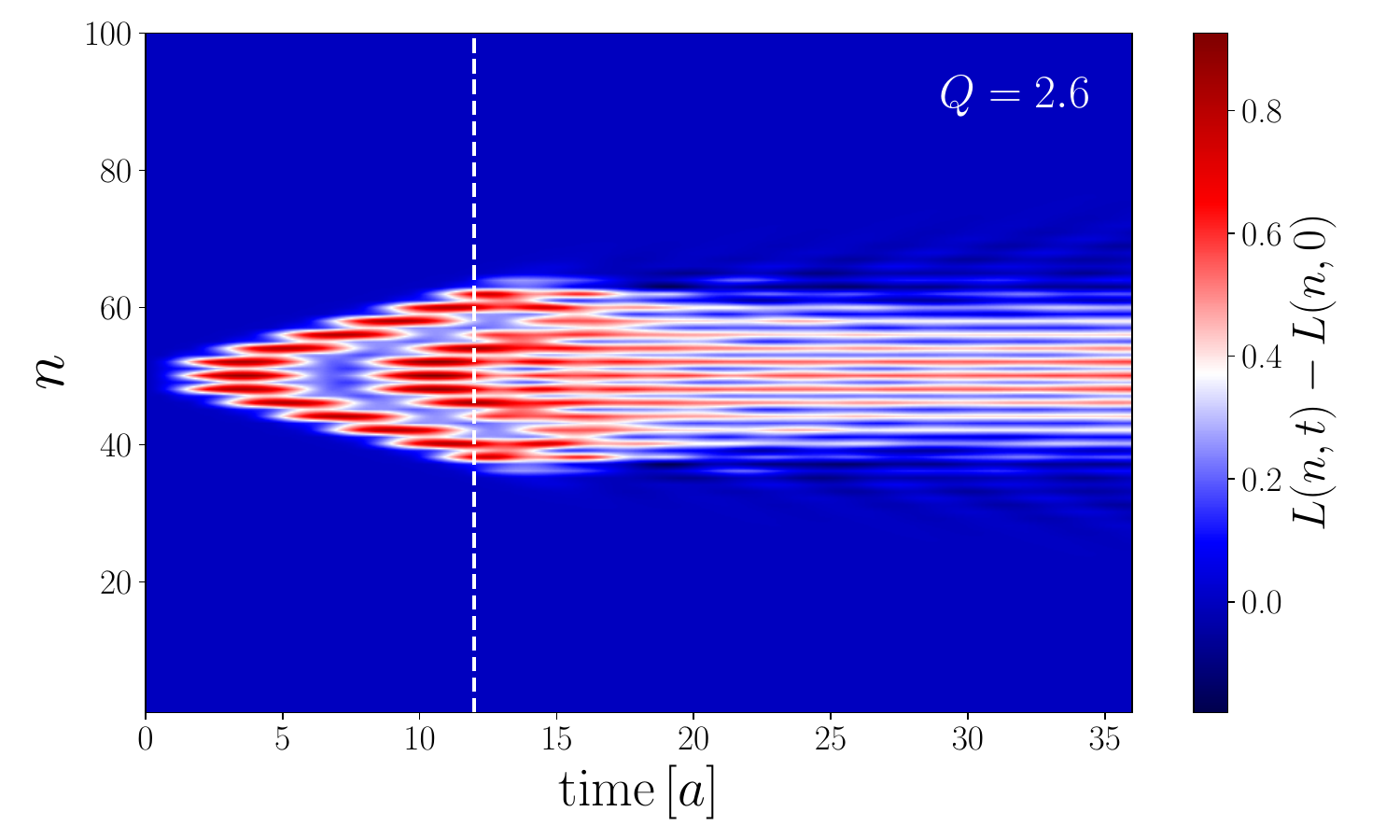}
    \includegraphics[width=0.48\linewidth]{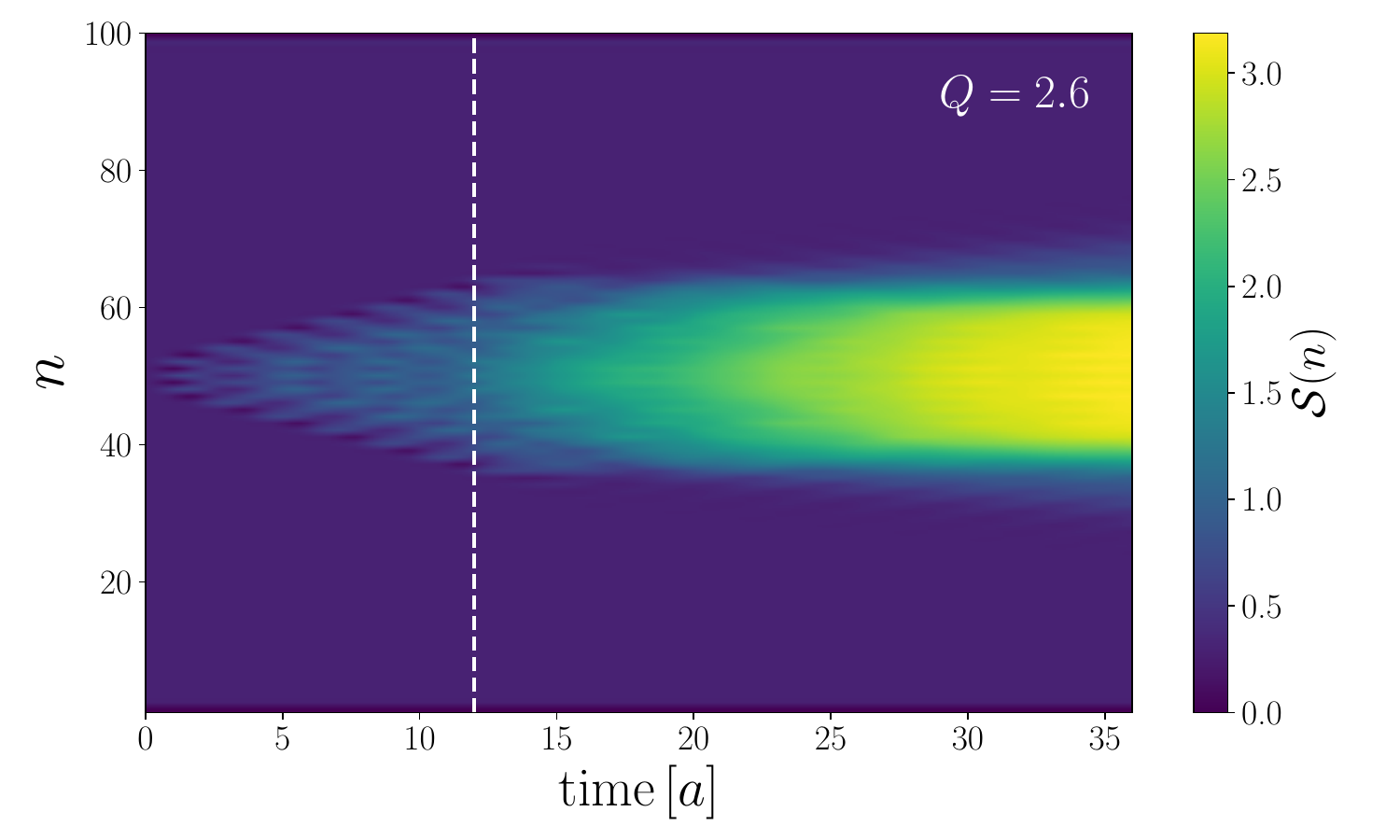}\\ 
    \includegraphics[width=0.48\linewidth]{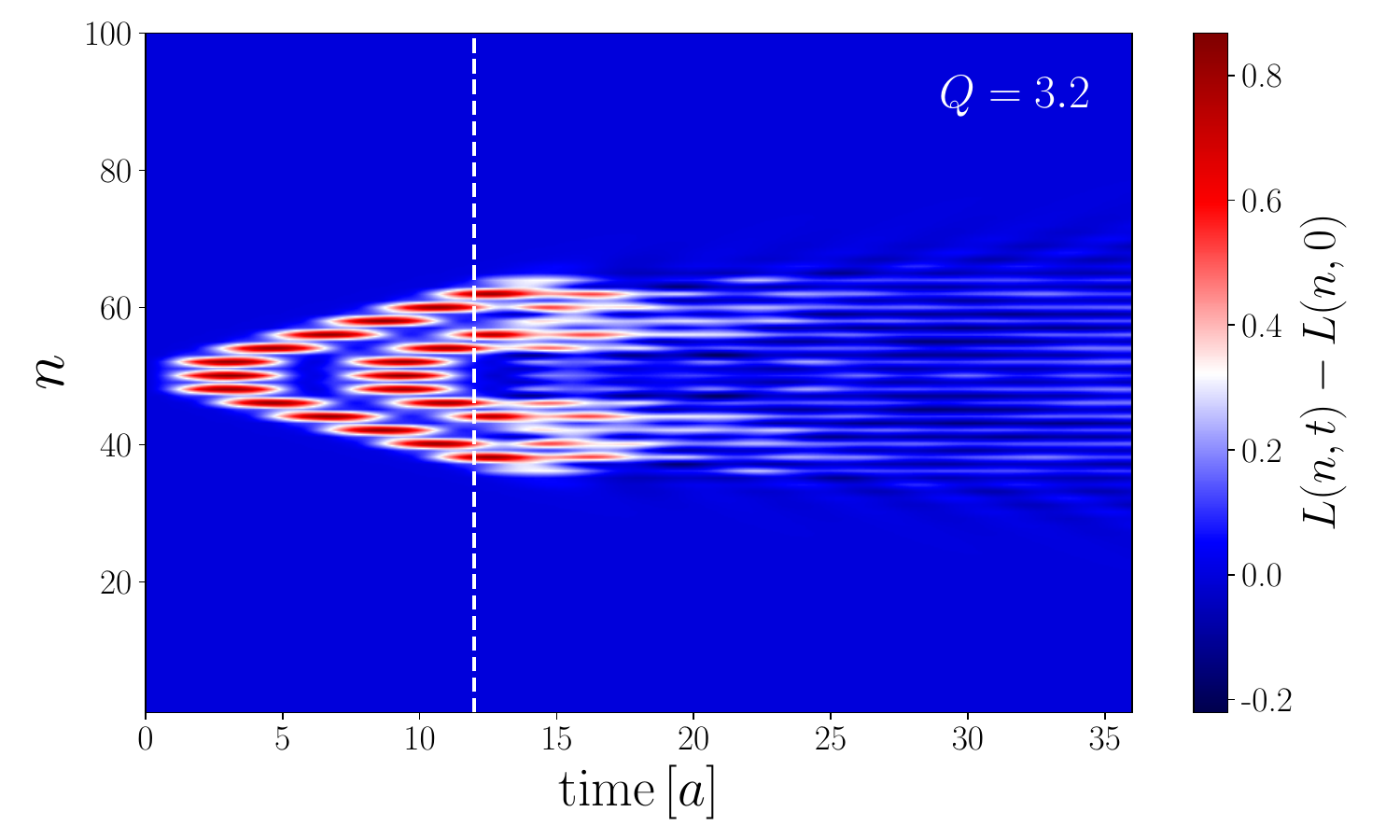}
    \includegraphics[width=0.48\linewidth]{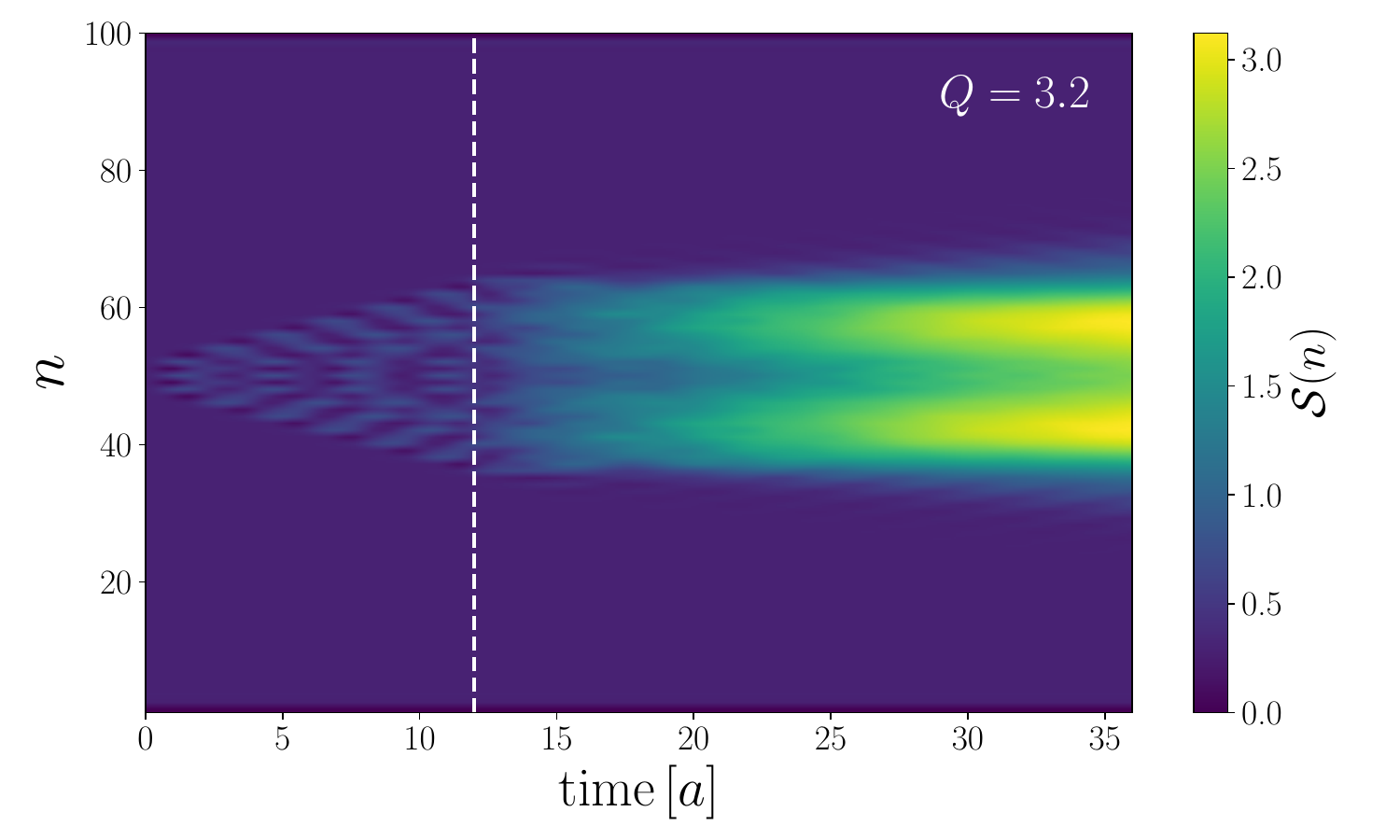}
    \caption{Time evolution of the expectation value of the onsite electric field and bipartite entanglement entropy for $Q=\{1,2.6,3.2\}$ for the quench in which the initially injected external charges are removed at time $t=12 \, a$, indicated by the vertical white dashed line. We set $ga=1$, $ma=0.25$, and $N=100$.}
    \label{fig:E_field_Entropy_stopped}
\end{figure}

The previous quench can be considered to be, in some sense, unphysical since the external charges are continuously injecting energy into the system. This continuous driving of the system obscures some of the dynamical properties of the model. We thus complement these results with the case where the initial external charges are removed from the system at a later time $t=12 \, a$. The results for the electric field and the entropy for three values of $Q=\{1, 2.6, 3.2 \}$ and $ga=1$ are shown in Fig.~\ref{fig:E_field_Entropy_stopped} for the same mass and system size as in Figs.~\ref{fig:11_full_charges} and \ref{fig:22_full_charges}. Note that for $t>12 \, a$ this quench follows into the general class considered by Cardy and Calabrese for conformal field theories~\cite{Calabrese:2006rx}. However, here, the breaking of integrability and the presence of a linear potential do not lead to the characteristic formation of a light-cone structure as in integrable models. Indeed, at strong coupling, one expects that the initial electric strings remain confined, as it is energetically expensive to extend or break them. The results without driving the system show that, after the charges are removed, the sequential string-breaking pattern stops, and the vacuum screens the applied electric field (see top figure). Deeper into the string-breaking regime (middle and bottom figures), we observe a growth of bipartite entropy compared to the top panel.

\begin{figure}[h!]
    \centering
    \includegraphics[width=0.5\columnwidth]{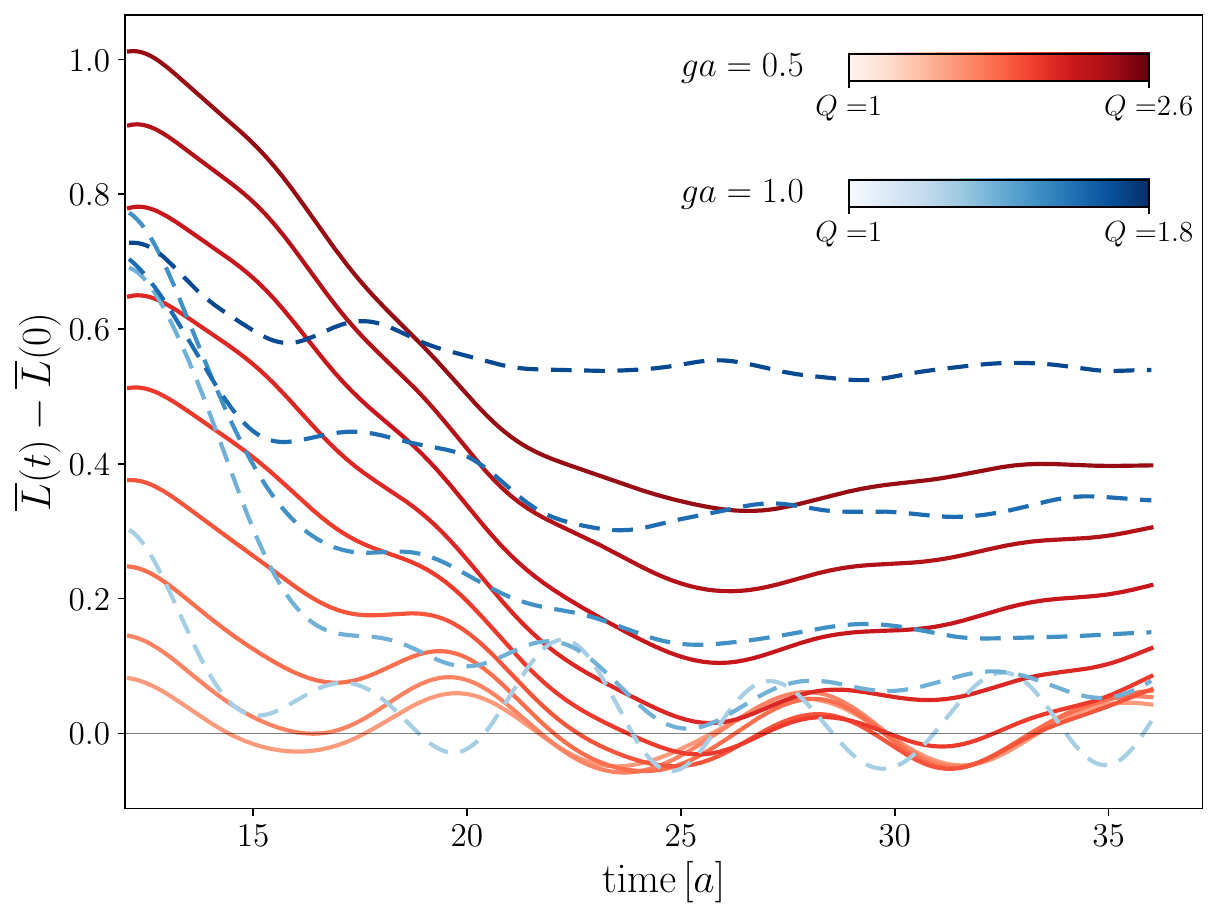}
    \caption{Time evolution of the average electric field $\overline{L}(t)$ at the center of the lattice (see main text) minus its initial value at $t=0$ for the quench in which the external charges are removed at $t=12 \, a$ as in Fig.~\ref{fig:E_field_Entropy_stopped}. Here we use $ma=0.25$ and $ga=\{0.5,1\}$, with varying values of the external charge $Q$. The plot shows the results from the instant at which the external charges are removed from the system.}
    \label{fig:E_field_oscillations_stopped}
\end{figure}

Finally, in Fig.~\ref{fig:E_field_oscillations_stopped}, we complement these results with the time evolution of the electric field values averaged between sites $48\leq n \leq 51$, $\overline{L}(t)$, for the quench in which charges are removed at $t = 12 \, a$, using $ga=\{0.5,1\}$ and the same mass and system size as in Fig.~\ref{fig:E_field_Entropy_stopped}. Here we observe further evidence for the transition between two regimes roughly separated at $Q\lesssim 2$ ($Q\lesssim 1.5$) for $ga=0.5$ ($ga=1$) for the parameter values used. The first regime is characterized by an oscillatory pattern in the sign of the electric field for $t > 12 \, a$, as a back-reaction to the originally injected field, preceded by an evolution without string breaking; in the second regime, there is no field inversion and the strings produced during the initial string-breaking evolution for $t < 12 \, a$ survive at longer times and are slowly attenuated.

\begin{figure}[h!]
    \centering
    \includegraphics[width=0.32\linewidth]{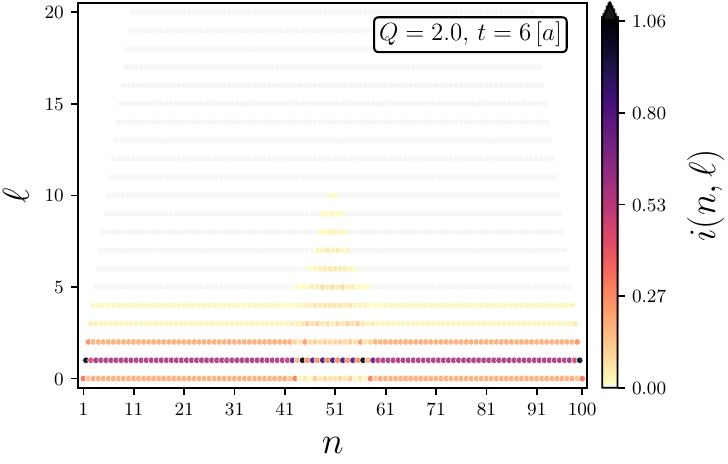}
    \includegraphics[width=0.32\linewidth]{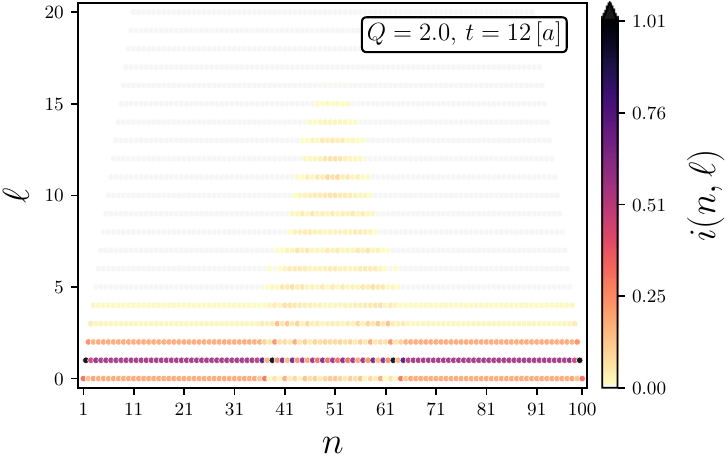}
    \includegraphics[width=0.32\linewidth]{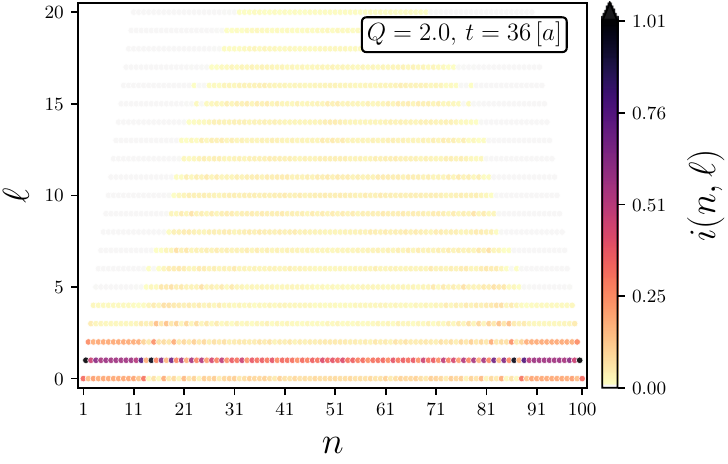}
    \includegraphics[width=0.32\linewidth]{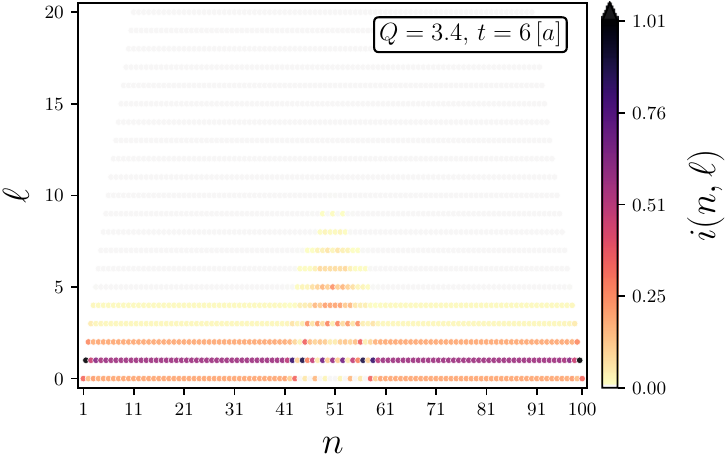}
    \includegraphics[width=0.32\linewidth]{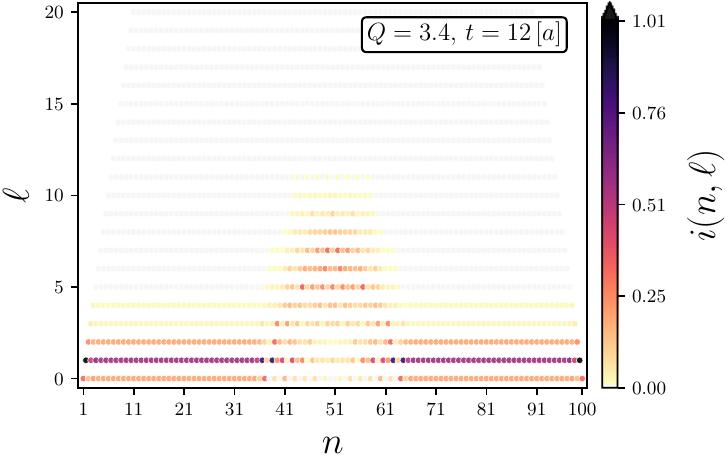}
    \includegraphics[width=0.32\linewidth]{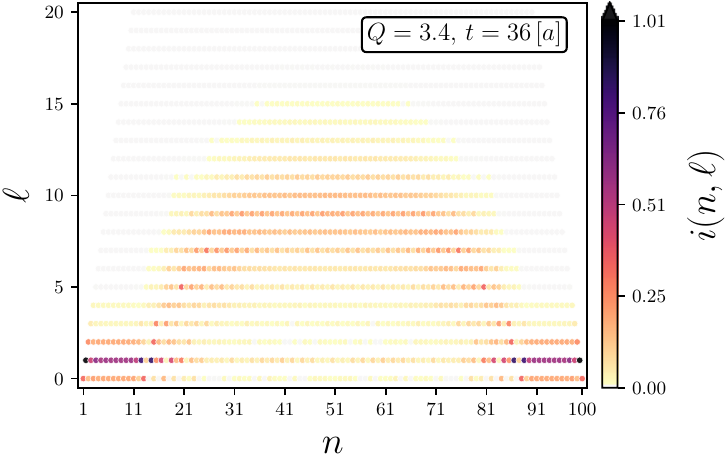}
    \includegraphics[width=0.32\linewidth]{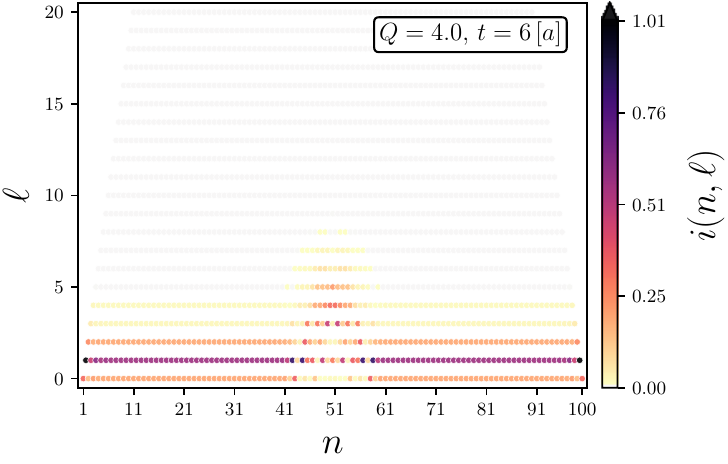}
    \includegraphics[width=0.32\linewidth]{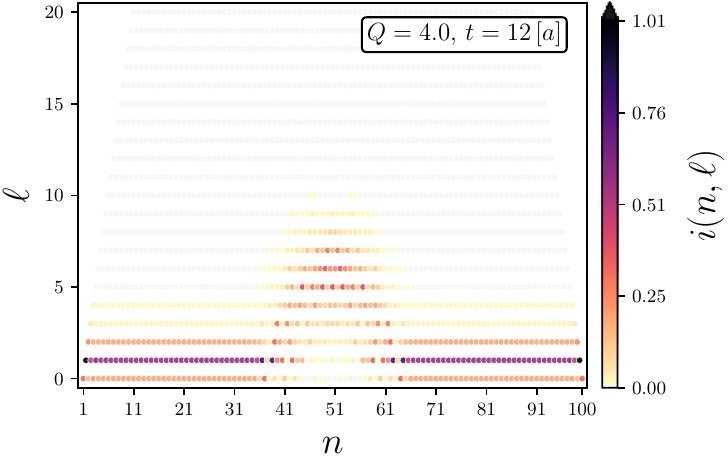}
    \includegraphics[width=0.32\linewidth]{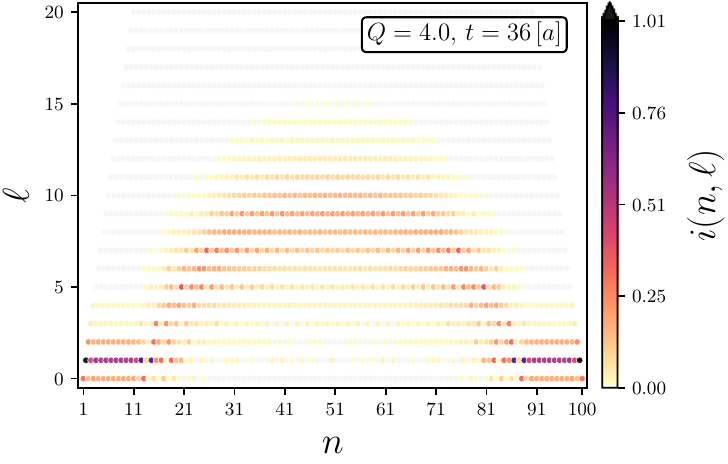}
    \caption{Snapshots of the information lattice for the time evolution with injected external charges for $Q=\{2, 3.4, 4\}$. We set $ga=0.5$, $ma=0.25$, and $N=100$.}
    \label{fig:string_full_iln}
\end{figure}

We now proceed to study these quench protocols through the lens of the information lattice. We first consider the quench where the external charges are never removed. In Fig.~\ref{fig:string_full_iln} we show the evolution of the $i(n,\ell)$ distribution for three values of the external charges $Q=\{2, 3.4, 4\}$, using $ga=0.5$ and $ma = 0.25$; compare with Fig.~\ref{fig:11_full_charges}. It is clear from these results that the information lattice provides a more detailed picture of the buildup of correlations in the system, which is hard to judge using local observables or the bipartite entropy. In this regime, where there is no string-breaking, the evolution of the original string leads to a buildup of correlations to a plateau at $\ell \approx 9$ for sufficiently large $Q$. Notice that the saturation level seems independent of the quench; we verified this to be true for the numerically possible bond dimensions. Still, one should not exclude the possibility that the use of tensor-network methods and truncation effects could play a role. This observed behavior indicates the formation of a nearly translational invariant state at the center of the lattice with larger correlations than those characterizing the vacuum.

Importantly, although for large $Q$ the states display the characteristic features of thermal pure states---namely, an information per scale profile with two peaks, one at finite $\ell$ with decaying tails and another at small $\ell$, which is characteristic of finite temperature states as discussed in Sec.~\ref{sec:Information_lattice} (see also Ref.~\cite{artiaco2025universal})---the information lattice reveals that the state is in fact nonthermal.
A first indication is that the peak at finite $\ell$ of the information per scale does not reach $\ell \sim N/2$ at long times, as expected in thermal states. This implies that the long-time state in this quench protocol does not exhibit a volume law for entanglement entropy as thermal states do.
In turn, this feature is what allows us to reach long simulation times with tensor-network methods, which would not be able to capture genuinely thermalizing dynamics~\cite{artiaco2024efficient,PhysRevB.109.134304}. Nonetheless, in the long-time state, the expectation values of local observables become time independent, as shown by the stabilization of the local information distribution. However, we expect these values not to coincide with those obtained in a thermal ensemble.
We emphasize that our observations may be affected by numerical artifacts due, for instance, to tensor-network approximations and truncation effects, and should therefore be interpreted with due caution.
While nonthermal properties are difficult to infer from local observables, the information lattice provides a straightforward characterization of the states, including both thermal and nonthermal features.

\begin{figure}[h!]
    \centering
    \includegraphics[width=.45\columnwidth]{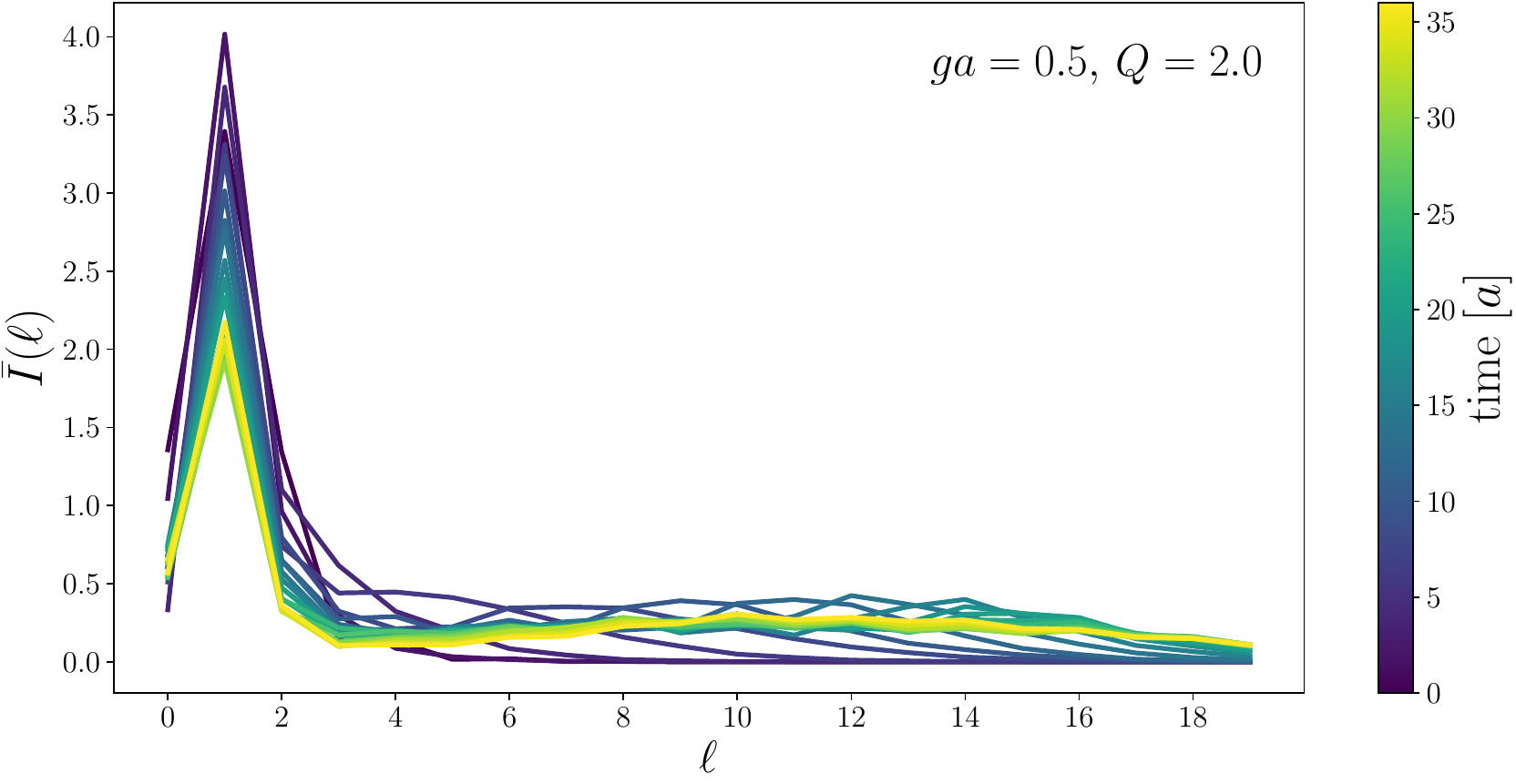}
    \includegraphics[width=.45\columnwidth]{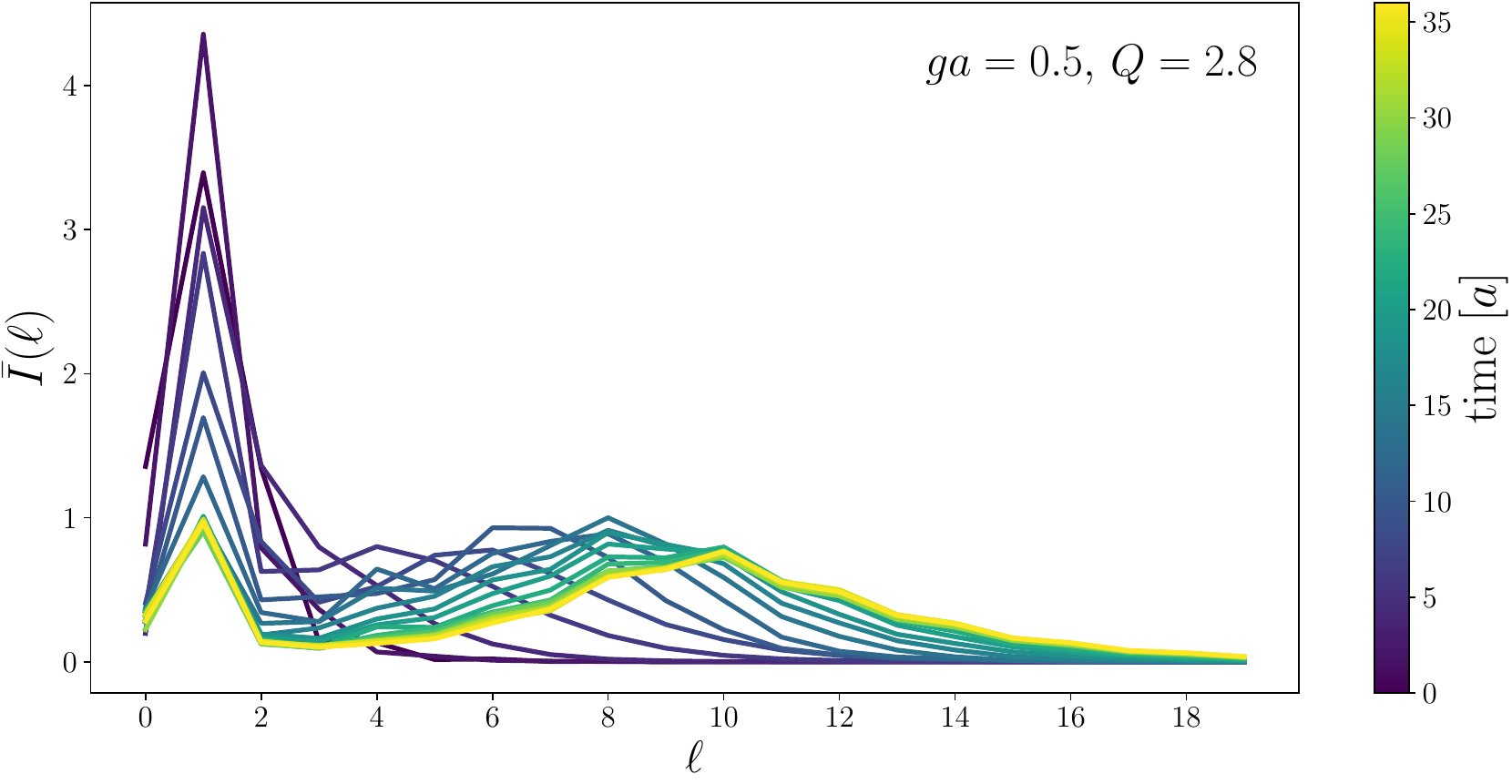}
    \includegraphics[width=.45\columnwidth]{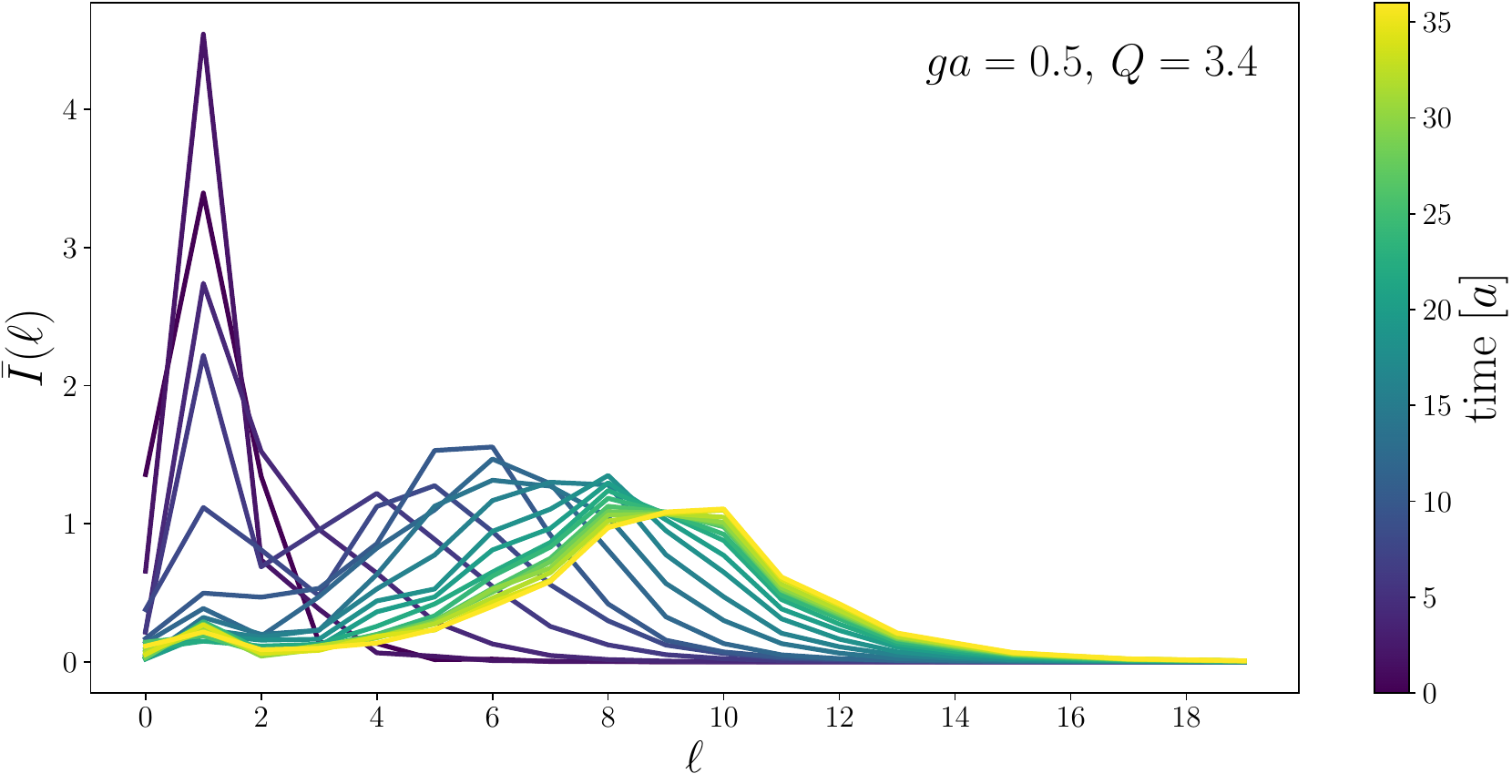}
    \includegraphics[width=.45\columnwidth]{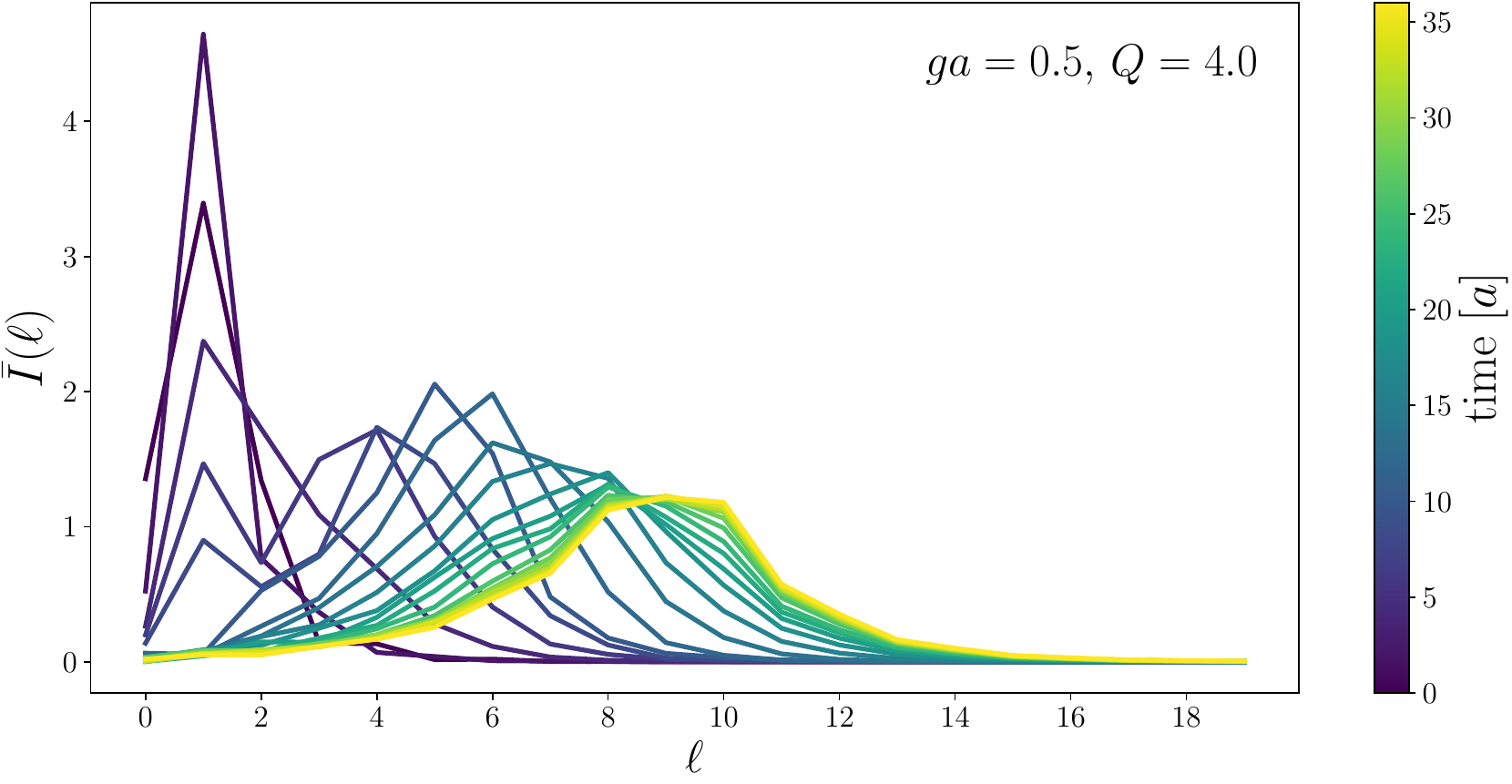}
    \caption{Partially integrated information per scale $\bar{I}(\ell)$ (see main text) for the time evolution illustrated in Fig.~\ref{fig:string_full_iln}.}
    \label{fig:partially_integrated_information}
\end{figure}

We complement these observations by studying the corresponding partially integrated information per scale $\bar{I}(\ell) = \sum_{n=47}^{59} i(n,\ell)$, shown in Fig.~\ref{fig:partially_integrated_information}. Here, one observes that for weak quenches, i.e., smaller $Q$, the correlations are peaked around the typical value for the vacuum, $\ell \approx 1$, while for sufficiently strong excitations, i.e., larger $Q$, this peak disappears giving rise to a distribution centered around $\ell \approx 9$. Notice that this distribution becomes static in time. 

\begin{figure}[tb]
    \centering
    \includegraphics[width=0.32\linewidth]{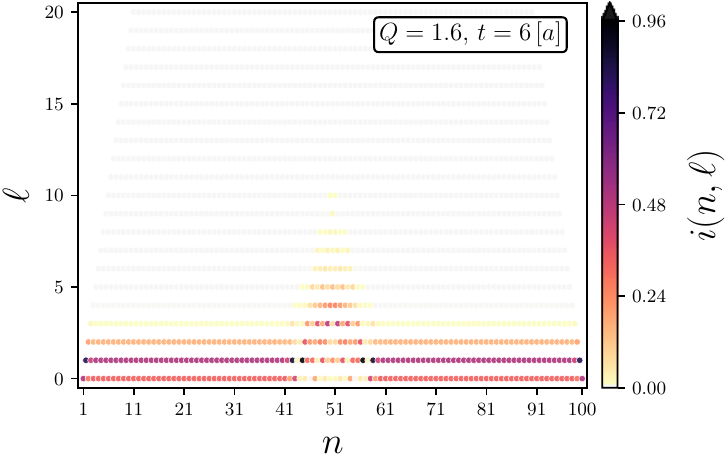}
    \includegraphics[width=0.32\linewidth]{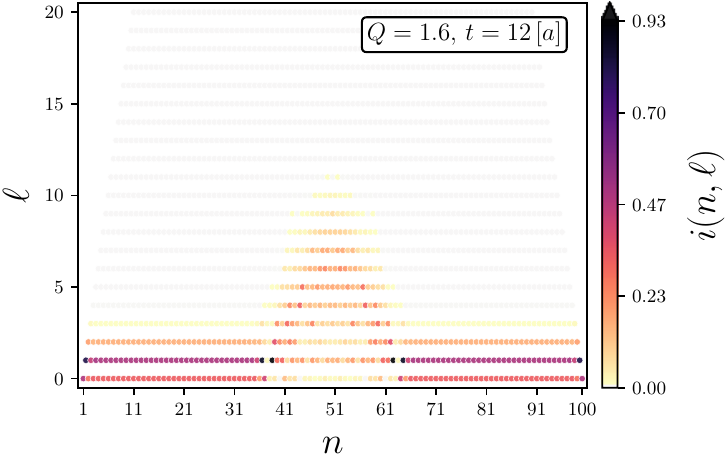}
    \includegraphics[width=0.32\linewidth]{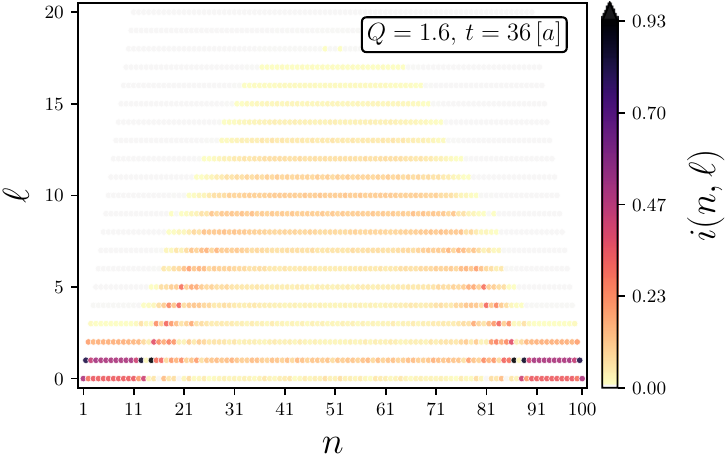}
    \includegraphics[width=0.32\linewidth]{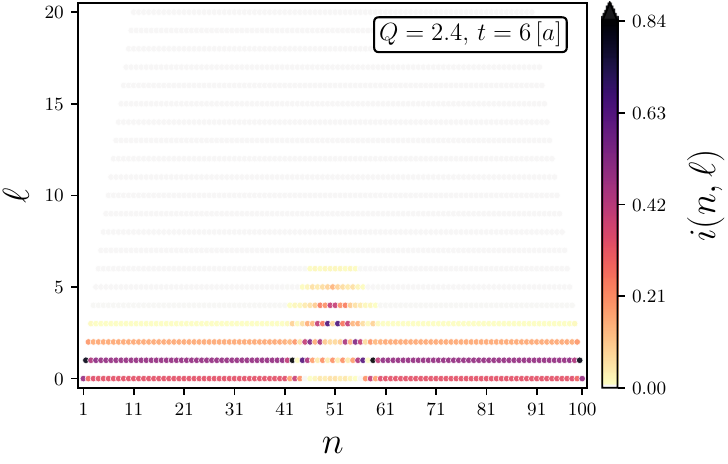}
    \includegraphics[width=0.32\linewidth]{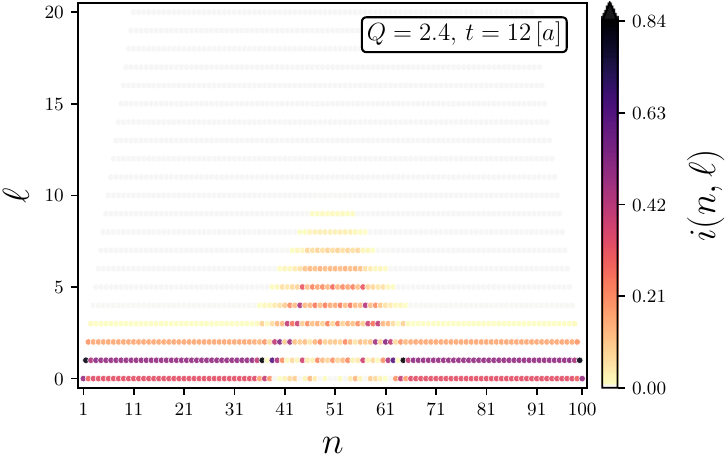}
    \includegraphics[width=0.32\linewidth]{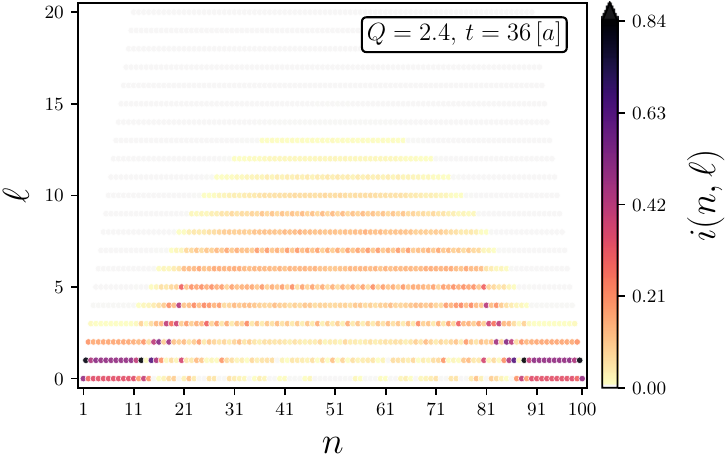}
    \includegraphics[width=0.32\linewidth]{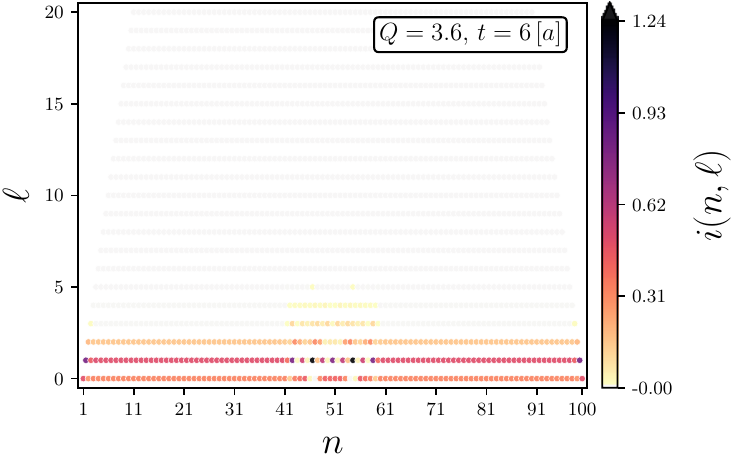}
    \includegraphics[width=0.32\linewidth]{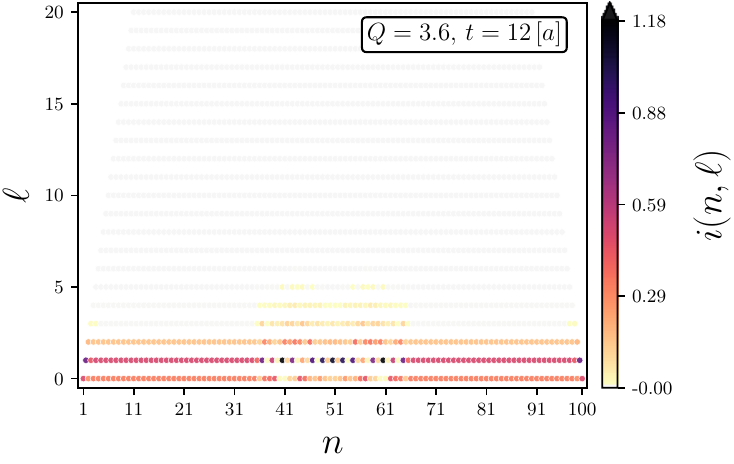}
    \includegraphics[width=0.32\linewidth]{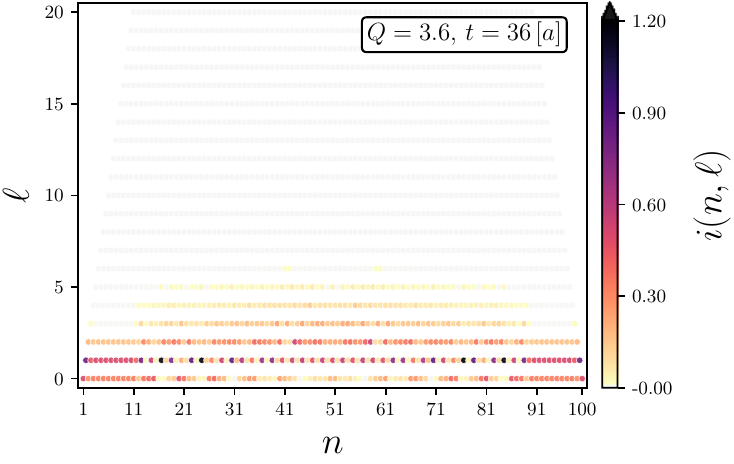}
    \caption{Snapshots of the information lattice for the time evolution with injected external charges as in Fig.~\ref{fig:string_full_iln}, now for $Q=\{1.6,2.4,3,6\}$, $ga=1$, $ma=0.25$, and $N=100$.}
    \label{fig:string_full_iln_2}
\end{figure}

\begin{figure}[tb]
    \centering
    \includegraphics[width=.48\columnwidth]{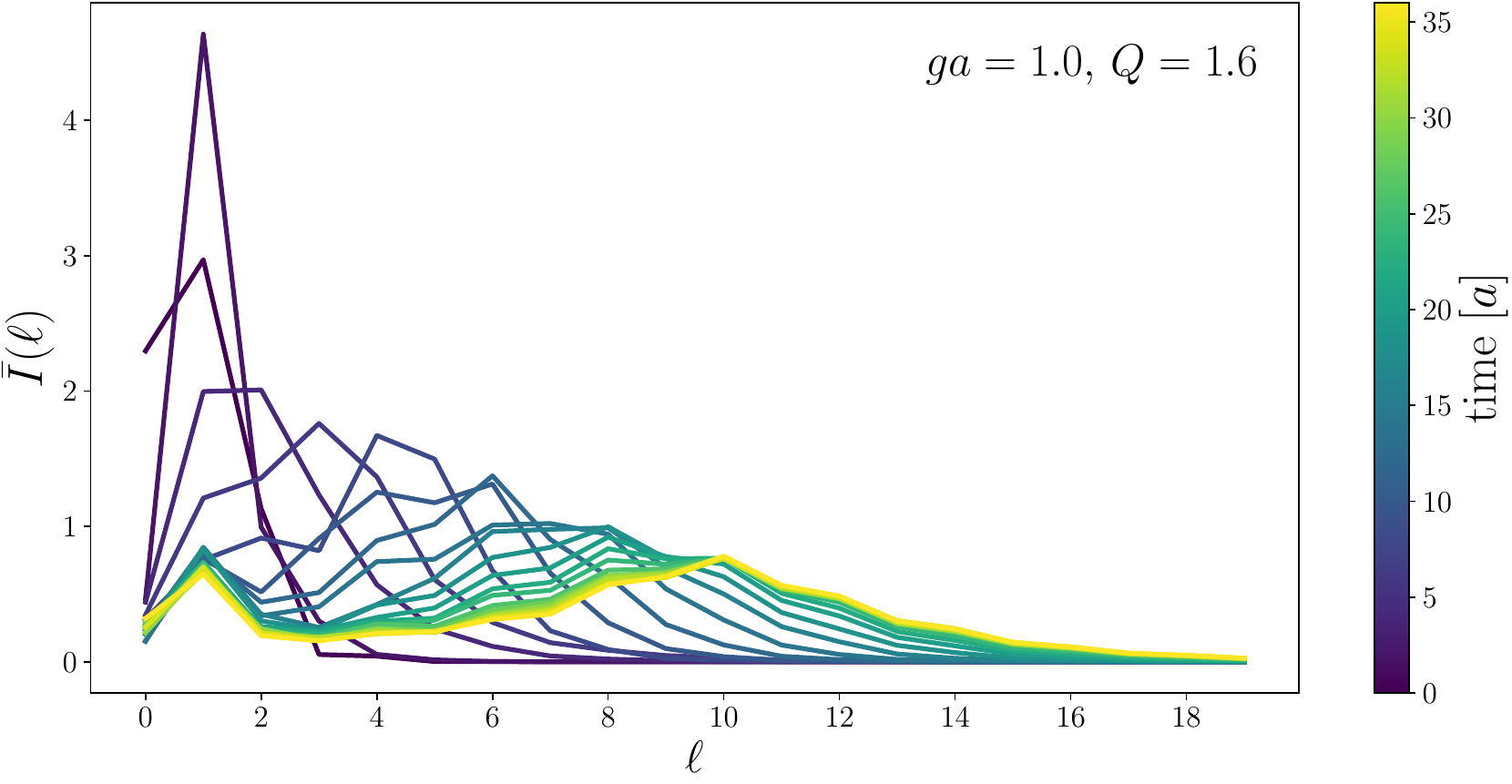}
    \includegraphics[width=.48\columnwidth]{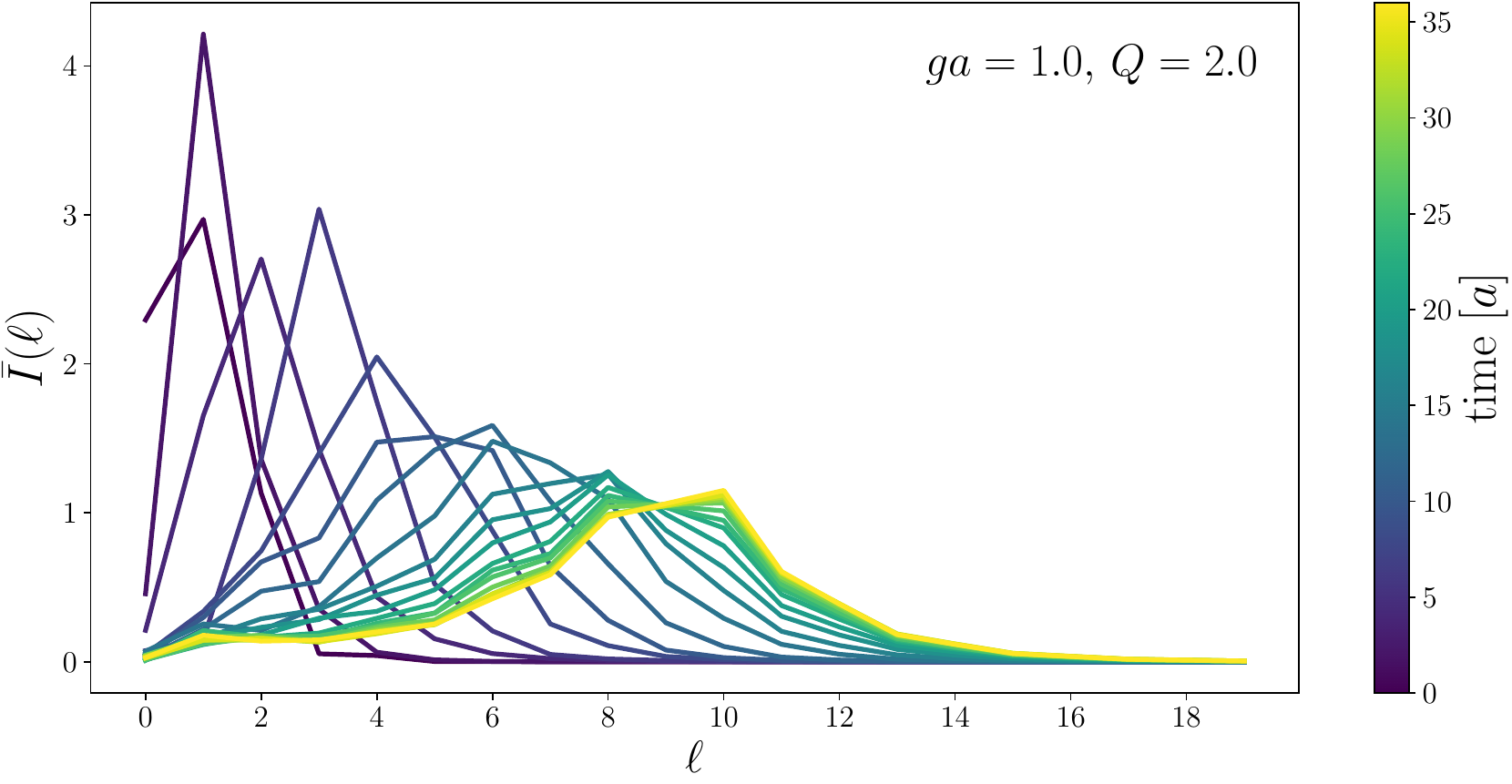}
    \includegraphics[width=.48\columnwidth]{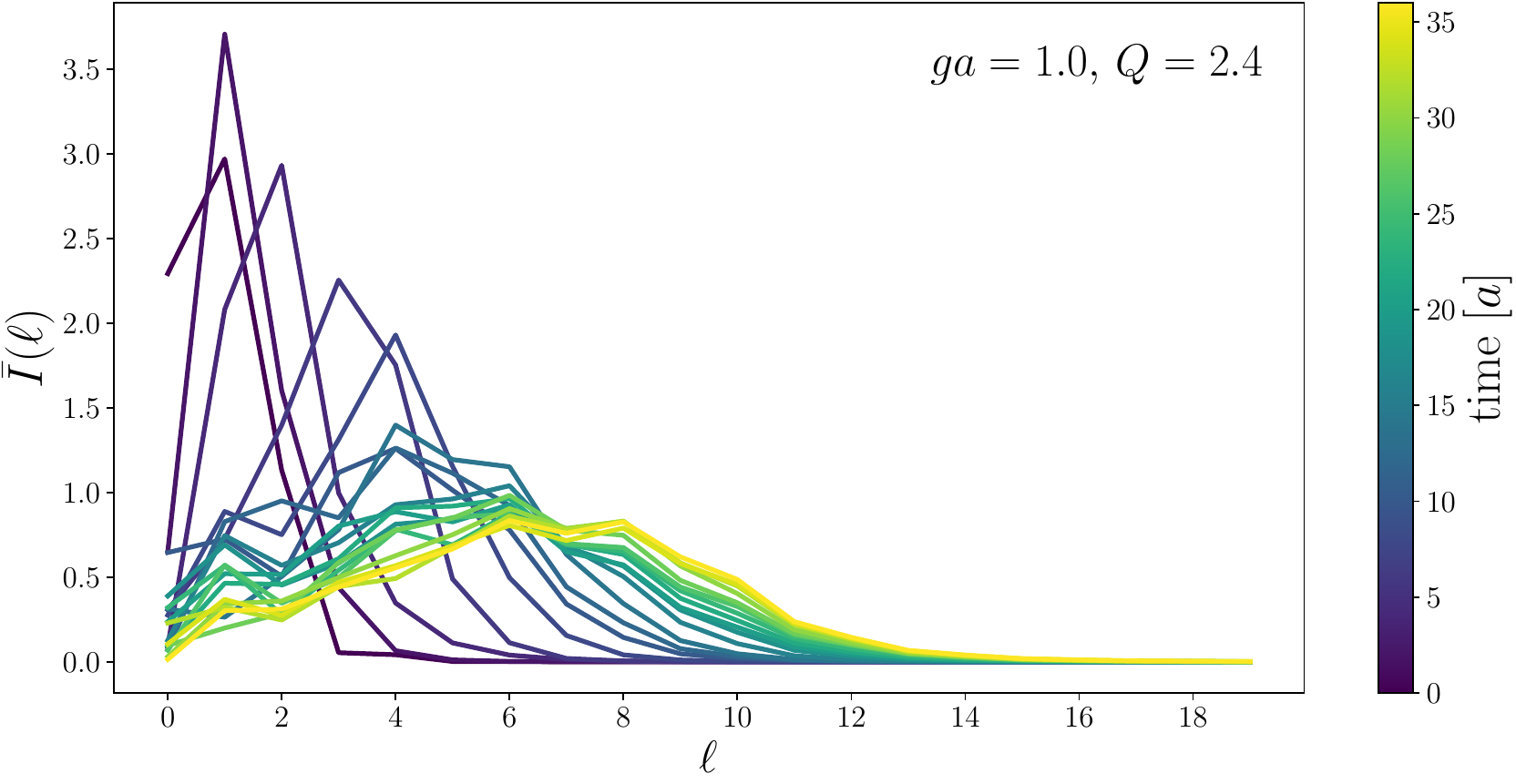}
    \includegraphics[width=.48\columnwidth]{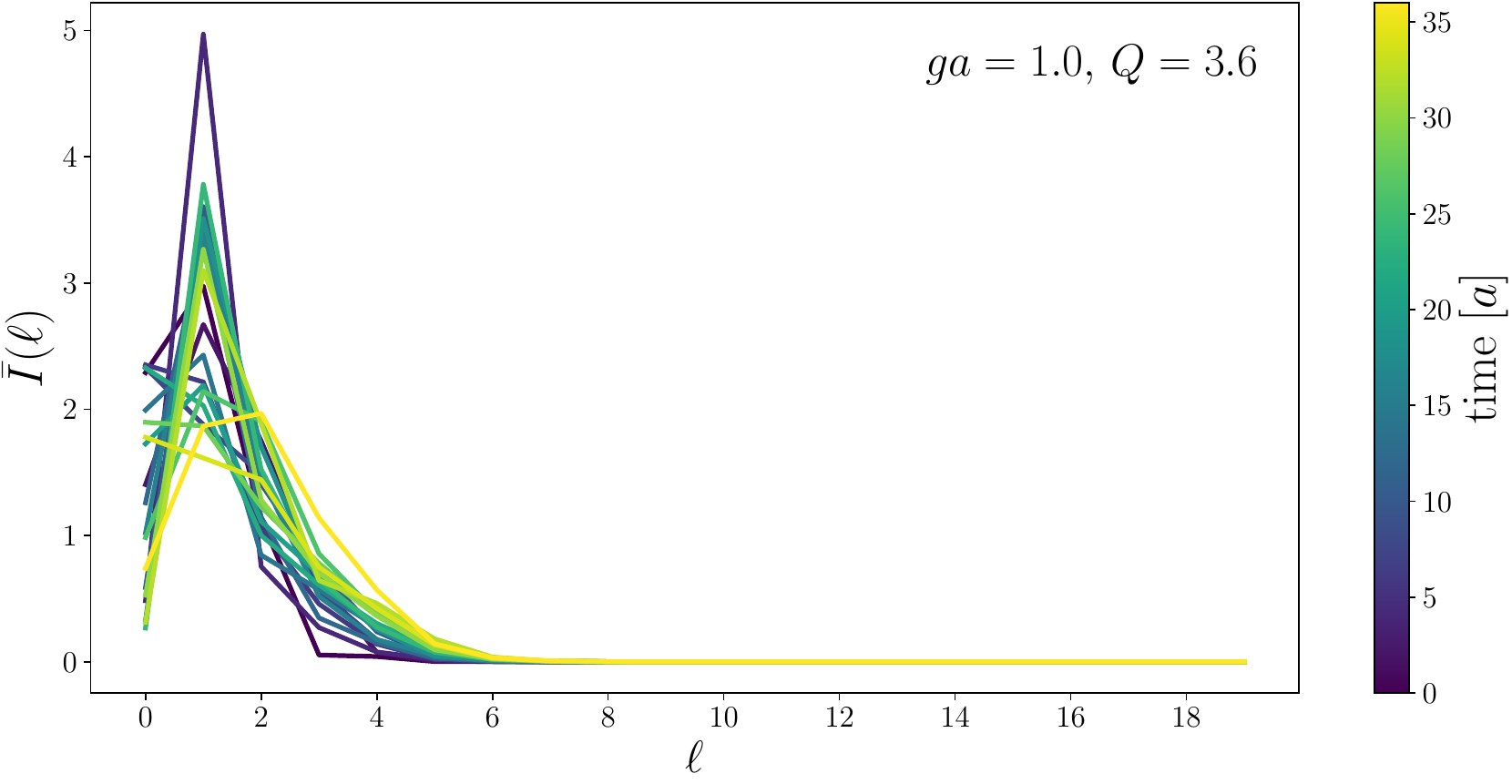}
    \caption{Partially integrated information per scale $\bar{I}(\ell)$ (see main text) for the time evolution illustrated in Fig.~\ref{fig:string_full_iln_2}.}
    \label{fig:partially_integrated_information_2}
\end{figure}

In Figs.~\ref{fig:string_full_iln_2} and \ref{fig:partially_integrated_information_2}, we show the local-information time evolution for $ga=1$, i.e., when string breaking can occur, setting the other parameters as in Figs.~\ref{fig:string_full_iln} and \ref{fig:partially_integrated_information}. For weak quenches, when the string is not broken, one again observes the formation of a translationally invariant and static state at the center of the chain with a peak of information at finite $\ell \approx 10$. This is clearly visible at the level of the partially integrated information per scale $\bar{I}(\ell)$ in Fig.~\ref{fig:partially_integrated_information_2}. However, when the external charge $Q$ increases and the applied external field is above the critical field, $\bar{I}(\ell)$ remains concentrated at $\ell \approx 1$ as in the initial state. This transition is apparent when comparing $Q=2.4$, where string breaking starts taking place, see Fig.~\ref{fig:11_full_charges}, with $Q=3.6$, where multiple string emerges. Considering the partially integrated information per scale in Fig.~\ref{fig:partially_integrated_information_2}, one can cleanly see the rise of correlations at $\ell \approx 1$ for nearly all times, indicating the dominance of the same state over the entire evolution.

\begin{figure}[h!]
    \centering
    \includegraphics[width=0.6\columnwidth]{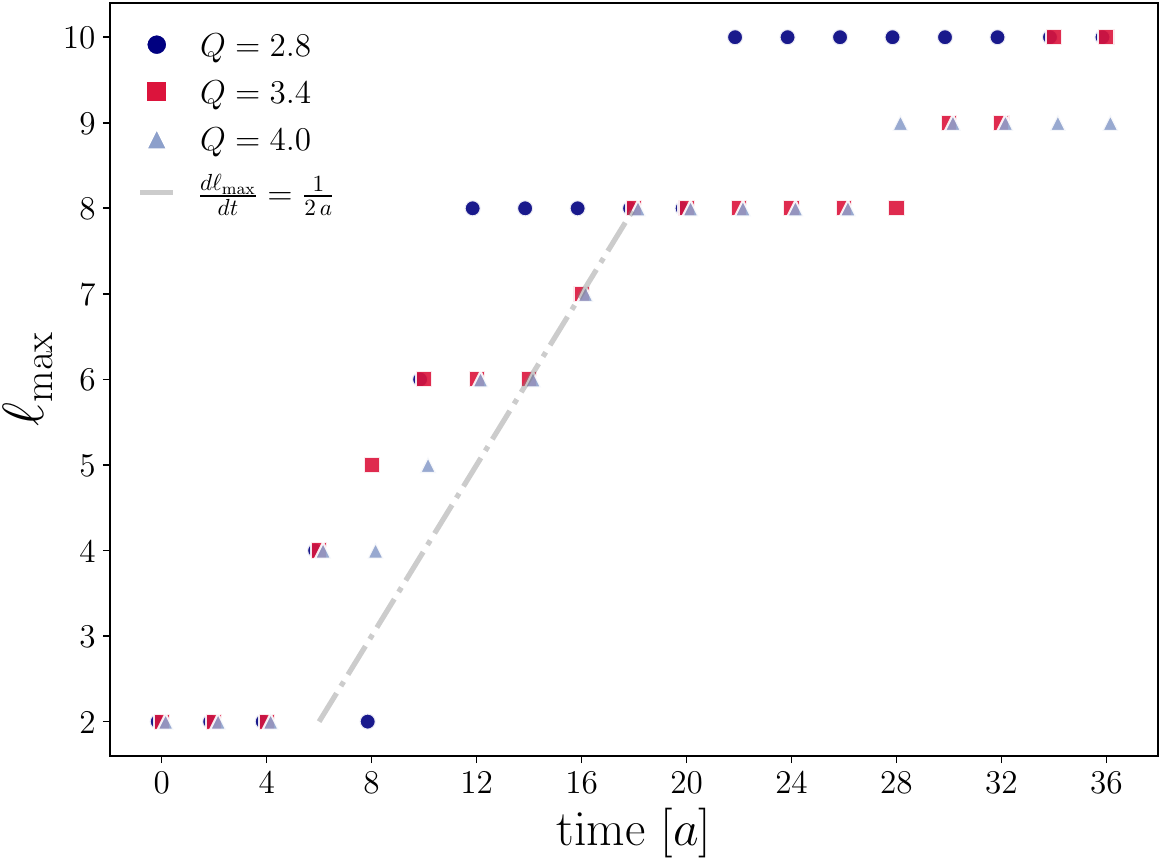}
    \caption{Peak position $\ell_{\mathrm{max}}$ for the $\bar{I}(\ell)$ distribution (see main text) for $ga=0.5$ and several $Q$ values as a function of time. The dotted dashed gray line indicates the slope for a ballistic propagation at half the light speed on the lattice. We set $ma=0.25$ and $N=100$.}
    \label{fig:max_path}
\end{figure}

Finally, we investigate how the information distribution evolves towards the long-time static state in the no-string-breaking regime. To that end, we extract the position of the maximum $\ell_{\rm max}$ of the $\bar I(\ell)$ distribution for several values of $Q$ over time, as depicted in Fig.~\ref{fig:max_path}. Notice that we ignore the peak at $\ell \approx 1$ and only consider that at finite $\ell$. Here we observe that, for intermediate times, the peak of the distribution moves roughly ballistically, up until the information distribution saturates to the long-time static one observed above. A ballistic flow of local information toward larger scales is generally expected during time evolution under Hamiltonians with local interactions. For instance, applying one cycle of a brickwork random unitary circuit to a product state generates correlations up to $\ell=2$, two cycles up to $\ell=4$, and so on; see Ref.~\cite{artiaco2024efficient}. This flow corresponds to the linear growth of entanglement entropy in generic local interacting Hamiltonians.

\begin{figure}[h!]
    \centering
    \includegraphics[width=0.32\linewidth]{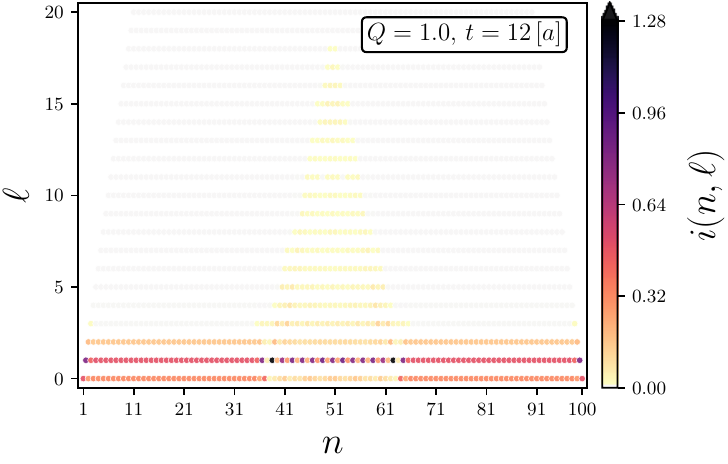}
    \includegraphics[width=0.32\linewidth]{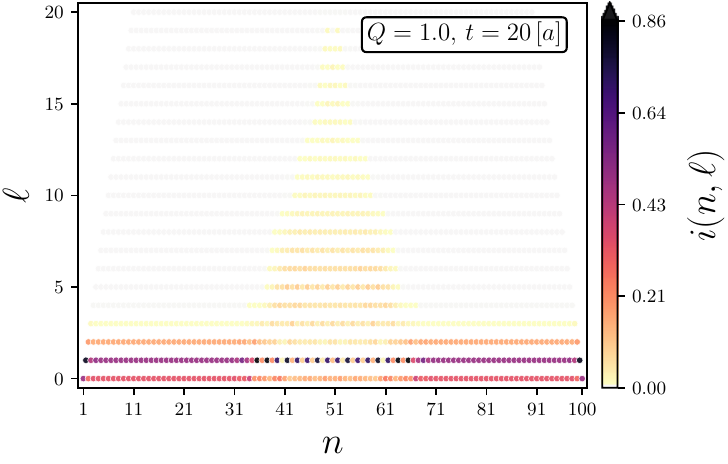}
    \includegraphics[width=0.32\linewidth]{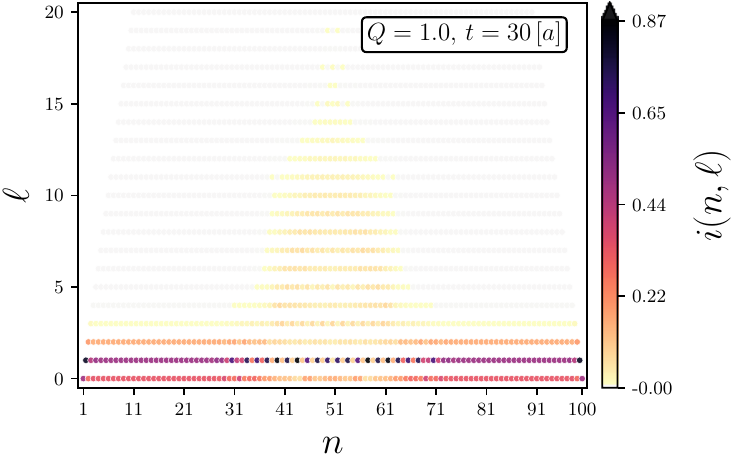}
    \includegraphics[width=0.32\linewidth]{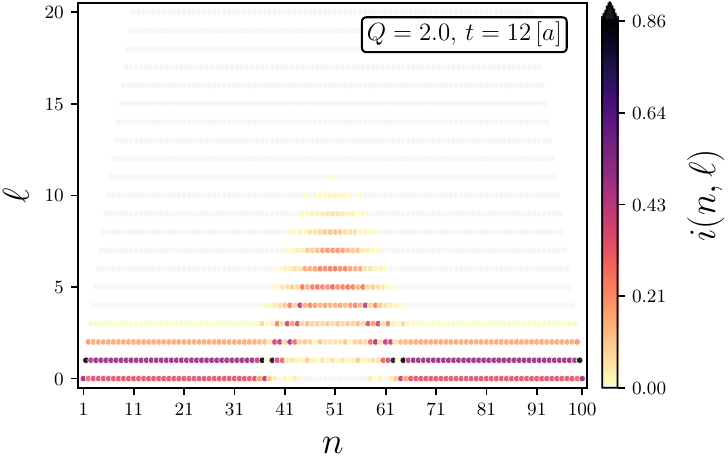}
    \includegraphics[width=0.32\linewidth]{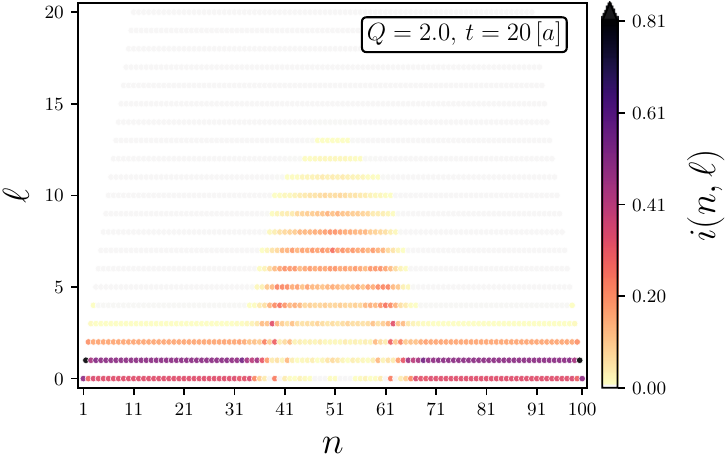}
    \includegraphics[width=0.32\linewidth]{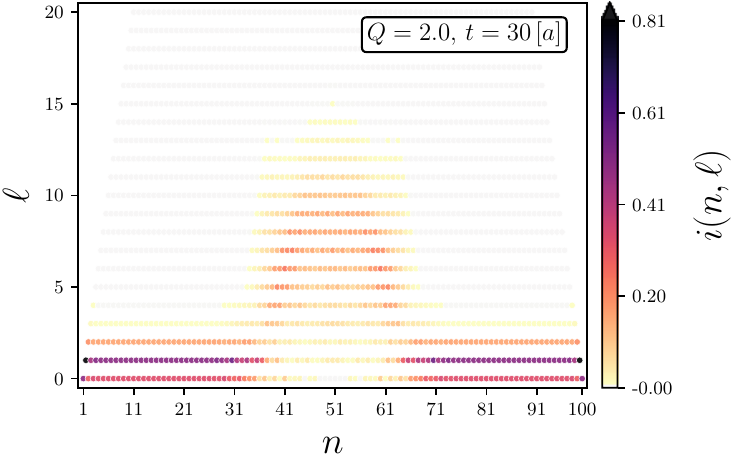}
    \includegraphics[width=0.32\linewidth]{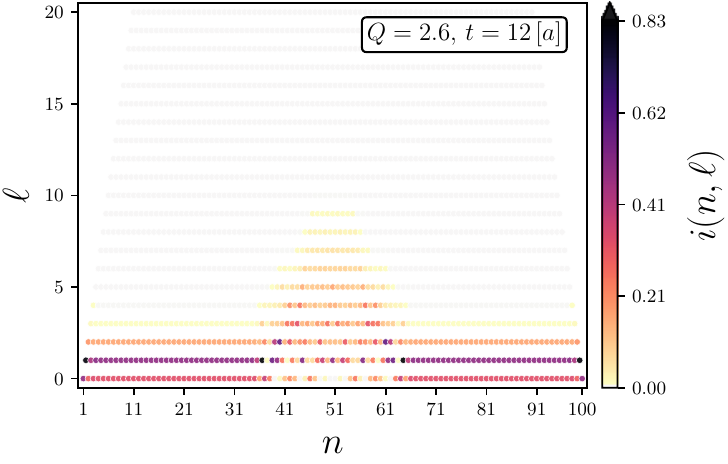}
    \includegraphics[width=0.32\linewidth]{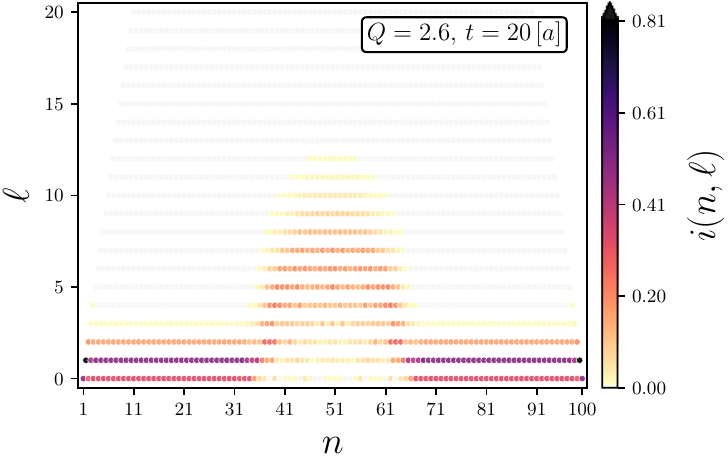}
    \includegraphics[width=0.32\linewidth]{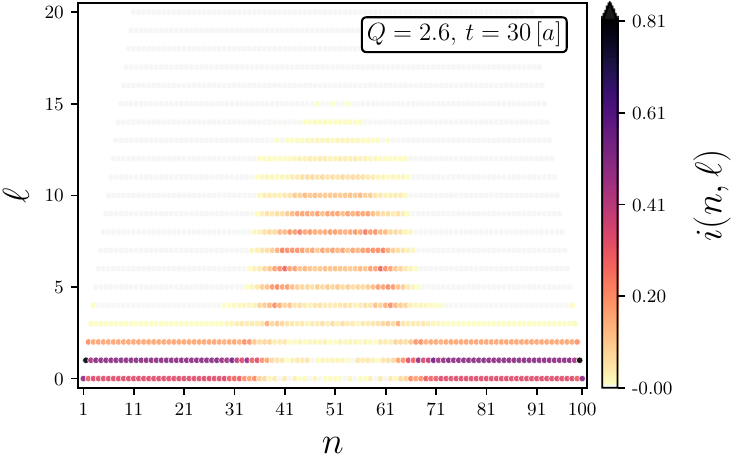}
    \includegraphics[width=0.32\linewidth]{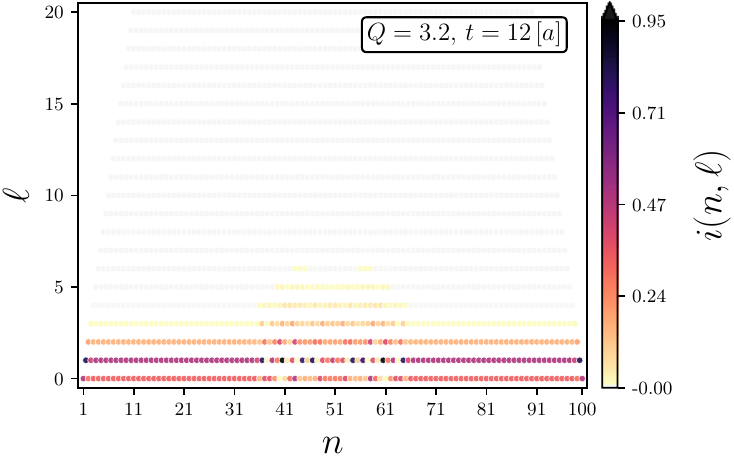}
    \includegraphics[width=0.32\linewidth]{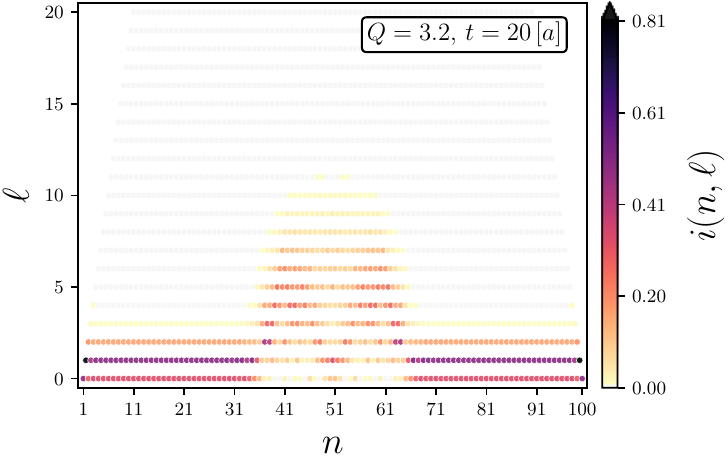}
    \includegraphics[width=0.32\linewidth]{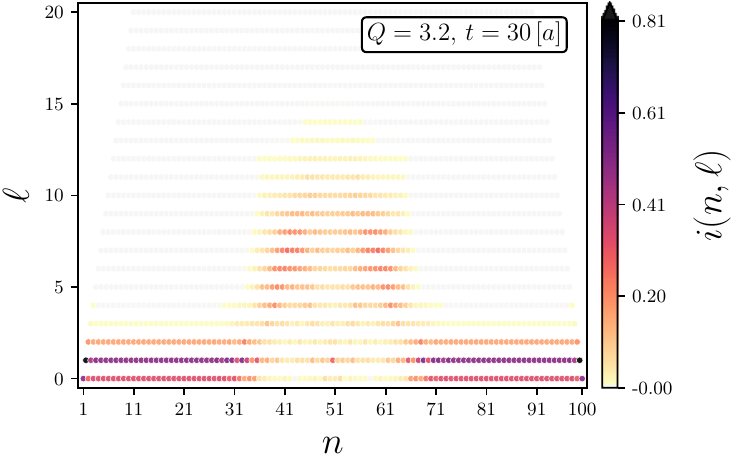}
    \caption{Snapshots of the information lattice for the quench in which the charges are removed at $t=12 \, a$ as in Fig.~\ref{fig:E_field_Entropy_stopped} for $Q=\{1,2,2.6,3.2\}$, $ga=1$, $ma=0.25$, and $N=100$.}
    \label{fig:stopped_charges_iln}
\end{figure}

In Fig.~\ref{fig:stopped_charges_iln}, we show the information lattice at different times and for different $Q$ values when the external charges are removed at $t=12 \, a$. Here, we again observe that this quench has different characteristics compared to the one where the system is always driven. At a smaller $Q$, we observe that, after the quench, the system exhibits oscillations, consistent with a back reaction to the applied electric field. As a result, the characteristic $\ell$ does not increase. For larger $Q$, we observe that the local-information distribution becomes static and peaked at finite $\ell$. Interestingly, for $Q=3.2$, we see that a two-short string configuration is generated during the evolution with the external charge ($t<12 \, a$), which then survives at late times and manifests as a two-peaked information distribution. Comparing with the results in Fig.~\ref{fig:E_field_Entropy_stopped}, these findings support the interpretation that this quench features a (partial) survival of the multi-string configuration generated before the charges are removed.

\section{Conclusion and discussion}\label{sec:conclusion}

In this work, we characterized the out-of-equilibrium dynamics of a lattice gauge theory using the information lattice. Focusing on the $(1+1)$D U(1) Schwinger model, we showed that the time evolution of local information serves as a clear and intuitive diagnostic of non-equilibrium dynamics. The two quench protocols studied---particle scattering and string breaking---exhibit distinct information-flow patterns that directly reflect their underlying physical processes.

In near-threshold particle scattering processes, the information lattice enables a clear identification of the production of new states through changes in the distribution of local information during time evolution. For example, the production of a heavy scalar meson from the collision of two lighter vector mesons manifests as the emergence of correlations at a larger length scale, directly illustrating the conversion of kinetic energy into mass.
For string dynamics, the information-lattice framework clearly distinguishes between confining and string-breaking regimes. In the confining regime, stable strings evolve into nonthermal steady states with static correlation profiles, whereas in the string-breaking regime, unstable strings display a recurring cycle of information buildup and decay, providing a direct view of real-time particle-antiparticle pair creation.

The key advantage of the information-lattice approach lies in its ability to render complex many-body correlations in a local, scale-resolved, and physically intuitive manner. Unlike standard local observables, which are often insensitive to the global correlation structure, or multi-point correlators, which can be difficult to interpret, the information lattice offers a comprehensive view of how quantum information organizes and flows during time evolution. In doing so, it provides a clear bridge between the abstract quantum state and its emergent physical properties.

Despite the strengths of the information lattice as a diagnostic tool, some aspects of the present analysis require further refinement. First, our results were obtained with a relatively small bond dimension $D$. Although varying $D$ did not lead to qualitative changes, a more systematic extrapolation to 
$D \to \infty$ would be desirable to confirm the conclusions. Second, while our approach enhances interpretability, distinguishing states with commensurate characteristic scales $\ell$ remains challenging. Addressing this will require a more detailed study of the full information lattice, including partially integrated information quantities---a direction we leave for future work.

Looking ahead, several promising directions emerge from this work. While our analysis was restricted to $(1+1)$D, ongoing efforts aim to extend the information-lattice framework to higher dimensions~\cite{flor2025workinprogress}. Its application to $(2+1)$D and $(3+1)$D gauge theories could shed light on complex phenomena such as confinement dynamics in non-Abelian gauge theories. In particular, the information lattice may provide a novel perspective on deeply nonperturbative features via local information flows---an aspect of growing relevance for quantum-simulation approaches to high-energy physics~\cite{Barata:2023clv, Barata:2024apg, Qian:2024xnr, Banuls:2025wiq, Lamm:2019uyc, Shao:2025ygy, Grieninger:2024axp, Li:2021kcs, Farrell:2024fit, Jha:2024jan, Chen:2025zeh,Davoudi:2025rdv,Bauer:2021gup,Florio:2025hoc}.

Perhaps most importantly, the information-lattice framework is particularly well-suited to the emerging era of quantum simulation. As experimental platforms gain the capability to prepare and evolve gauge-theory states, the key challenge becomes extracting physically meaningful insights from highly complex wavefunctions. The information lattice provides a practical and powerful method for experimentalists to characterize the states they create. It can be used to verify particle production, probe for thermalization, and identify novel non-equilibrium phases of matter---thereby providing a natural bridge between theory and experiment.

\section*{Acknowledgments}
We are grateful to Mari Carmen Bañuls, Jens H. Bardarson, and Wenyang Qian for helpful discussions.
We thank Ian Matthias Flór for kindly granting permission to adapt material from their work.
C.A. acknowledges funding from the European Research Council (ERC) under the European Union’s Horizon 2020 research and innovation program (Grant Agreement No.~101001902).
E.R. acknowledges the financial support received from the IKUR Strategy under the collaboration agreement between the Ikerbasque Foundation and UPV/EHU on behalf of the Department of Education of the Basque Government. E.R. acknowledges support from the BasQ strategy of the Department of Science, Universities, and Innovation of the Basque Government. E.R. is supported by the grant PID2021-126273NB-I00 funded by MCIN/AEI/10.13039/501100011033 and by “ERDF A way of making Europe” and the Basque Government through Grant No. IT1470-22. This work was supported by the EU via QuantERA project T-NiSQ grant PCI2022-132984 funded by MCIN/AEI/10.13039/501100011033 and by the European Union “NextGenerationEU”/PRTR. This work has been financially supported by the Ministry of Economic Affairs and Digital Transformation of the Spanish Government through the QUANTUM ENIA project, called Quantum Spain project, and by the European Union through the Recovery, Transformation, and Resilience Plan – NextGenerationEU within the framework of the Digital Spain 2026 Agenda. This work has been partially funded by the Eric \& Wendy Schmidt Fund for Strategic Innovation through the CERN Next Generation Triggers project under grant agreement number SIF-2023-004.

\bibliographystyle{JHEP-2modlong.bst}

\bibliography{Lib.bib}

@article{Calabrese:2006rx,
    author = "Calabrese, Pasquale and Cardy, John L.",
    title = "{Time-dependence of correlation functions following a quantum quench}",
    eprint = "cond-mat/0601225",
    archivePrefix = "arXiv",
    doi = "10.1103/PhysRevLett.96.136801",
    journal = "Phys. Rev. Lett.",
    volume = "96",
    pages = "136801",
    year = "2006"
}

@article{Kormos_2016,
   title={Real-time confinement following a quantum quench to a non-integrable model},
   volume={13},
   ISSN={1745-2481},
   DOI={10.1038/nphys3934},
   number={3},
   journal={Nature Phys.},
   publisher={Springer Science and Business Media LLC},
   author={Kormos, Marton and Collura, Mario and Takács, Gabor and Calabrese, Pasquale},
   year={2016}, 
   pages={246},
   eprint = "1604.03571",
   archivePrefix = "arXiv",
   primaryClass = "cond-mat"
}

@article{klein2022time,
  author        = {Klein Kvorning, Thomas and Herviou, Lo{\"\i}c and Bardarson, Jens H},
  title         = {Time-evolution of local information: {T}hermalization dynamics of local observables},
  journal       = {SciPost Phys.},
  year          = {2022},
  volume        = {13},
  number        = {4},
  pages         = {080},
  doi={10.21468/SciPostPhys.13.4.080},
  eprint = "2105.11206",
  archivePrefix = "arXiv",
  primaryClass = "quant-ph"
}

@article{artiaco2024efficient,
  title = {Efficient Large-Scale Many-Body Quantum Dynamics via Local-Information Time Evolution},
  author = {Artiaco, Claudia and Fleckenstein, Christoph and Aceituno Ch\'avez, David and Kvorning, Thomas Klein and Bardarson, Jens H.},
  journal = {PRX Quantum},
  volume = {5},
  issue = {2},
  pages = {020352},
  numpages = {28},
  year = {2024},
  publisher = {American Physical Society},
  doi = {10.1103/PRXQuantum.5.020352},
  eprint = "2310.06036",
  archivePrefix = "arXiv",
  primaryClass = "quant-ph"
  
}

@article{harkins2025nanoscale,
    author = {Kieren Harkins  and Christoph Fleckenstein  and Noella D’Souza  and Paul M. Schindler  and David Marchiori  and Claudia Artiaco  and Quentin Reynard-Feytis  and Ushoshi Basumallick  and William Beatrez  and Arjun Pillai  and Matthias Hagn  and Aniruddha Nayak  and Samantha Breuer  and Xudong Lv  and Maxwell McAllister  and Paul Reshetikhin  and Emanuel Druga  and Marin Bukov  and Ashok Ajoy },
    title = {Nanoscale engineering and dynamic stabilization of mesoscopic spin textures},
    journal = {Sci. Adv.},
    volume = {11},
    number = {13},
    pages = {eadn9021},
    year = {2025},
    doi = {10.1126/sciadv.adn9021},
    eprint = "2310.05635",
    archivePrefix = "arXiv",
    primaryClass = "quant-ph"
}

@article{ryu2006holographic,
  title = {Holographic Derivation of Entanglement Entropy from the anti--de {S}itter Space/Conformal Field Theory Correspondence},
  author = {Ryu, Shinsei and Takayanagi, Tadashi},
  journal = {Phys. Rev. Lett.},
  volume = {96},
  issue = {18},
  pages = {181602},
  numpages = {4},
  year = {2006},
  publisher = {American Physical Society},
  doi = {10.1103/PhysRevLett.96.181602},
  eprint = "hep-th/0603001",
  archivePrefix = "arXiv",
  primaryClass = "hep-ph"
}

@article{nozaki2013holographic,
  title={Holographic local quenches and entanglement density},
  author={Nozaki, Masahiro and Numasawa, Tokiro and Takayanagi, Tadashi},
  journal={JHEP},
  volume={2013},
  number={5},
  pages={1--40},
  year={2013},
  publisher={Springer},
  doi={https://doi.org/10.1007/JHEP05(2013)080},
  eprint = "1302.5703",
  archivePrefix = "arXiv",
  primaryClass = "hep-ph"
}

@article{artiaco2025universal,
  title = {Universal Characterization of Quantum Many-Body States through Local Information},
  author = {Artiaco, Claudia and Klein Kvorning, Thomas and Aceituno Ch\'avez, David and Herviou, Lo\"{\i}c and Bardarson, Jens H.},
  journal = {Phys. Rev. Lett.},
  volume = {134},
  issue = {19},
  pages = {190401},
  numpages = {8},
  year = {2025},
  publisher = {American Physical Society},
  doi = {10.1103/PhysRevLett.134.190401},
  eprint = "2410.10971",
  archivePrefix = "arXiv",
  primaryClass = "quant-ph"
  
}

@article{flor2025workinprogress,
    title={Higher-dimensional information lattice},
    author={Flór, Ian Matthias and Artiaco, Claudia and Klein Kvorning, Thomas and Bardarson, Jens H.},
    journal={In preparation},
    year={2025}
}

@misc{aceituno2024thermalization,
  title = {{Thermalization and Localization: Novel Perspectives from Random Circuits and the Information Lattice (Ph.D. Thesis)}},
  author = {Aceituno Ch\'avez, David},
  year = {2024},
  url={https://urn.kb.se/resolve?urn=urn:nbn:se:kth:diva-356733}
}

@book{wilde2013quantum,
  title={Quantum information theory},
  author={Wilde, Mark M},
  year={2013},
  publisher={Cambridge University Press}
}

@article{Banuls:2019bmf,
    author = "Ba\~nuls, M. C. and others",
    title = "{Simulating Lattice Gauge Theories within Quantum Technologies}",
    eprint = "1911.00003",
    archivePrefix = "arXiv",
    primaryClass = "quant-ph",
    doi = "10.1140/epjd/e2020-100571-8",
    journal = "Eur. Phys. J. D",
    volume = "74",
    number = "8",
    pages = "165",
    year = "2020"
}

@article{Jordan:2012xnu,
    author = "Jordan, Stephen P. and Lee, Keith S. M. and Preskill, John",
    title = "{Quantum Algorithms for Quantum Field Theories}",
    eprint = "1111.3633",
    archivePrefix = "arXiv",
    primaryClass = "quant-ph",
    doi = "10.1126/science.1217069",
    journal = "Science",
    volume = "336",
    pages = "1130--1133",
    year = "2012"
}

@article{Chen:2025zeh,
    author = "Chen, Jiunn-Wei and Chen, Yu-Ting and Meher, Ghanashyam",
    title = "{Parton Distributions on a Quantum Computer}",
    eprint = "2506.16829",
    archivePrefix = "arXiv",
    primaryClass = "hep-lat",
    month = "6",
    year = "2025"
}

@article{Banuls:2018jag,
    author = {Ba\~nuls, Mari Carmen and Cichy, Krzysztof and Cirac, J. Ignacio and Jansen, Karl and K\"uhn, Stefan},
    title = "{Tensor Networks and their use for Lattice Gauge Theories}",
    eprint = "1810.12838",
    archivePrefix = "arXiv",
    primaryClass = "hep-lat",
    doi = "10.22323/1.334.0022",
    journal = "PoS",
    volume = "LATTICE2018",
    pages = "022",
    year = "2018"
}

@article{Pichler:2015yqa,
    author = "Pichler, T. and Dalmonte, M. and Rico, E. and Zoller, P. and Montangero, S.",
    title = "{Real-time Dynamics in U(1) Lattice Gauge Theories with Tensor Networks}",
    eprint = "1505.04440",
    archivePrefix = "arXiv",
    primaryClass = "cond-mat.quant-gas",
    reportNumber = "INT-PUB-15-013",
    doi = "10.1103/PhysRevX.6.011023",
    journal = "Phys. Rev. X",
    volume = "6",
    pages = "011023",
    year = "2016"
}

@article{Papaefstathiou:2024zsu,
    author = "Papaefstathiou, Irene and Knolle, Johannes and Ba\~nuls, Mari Carmen",
    title = "{Real-time scattering in the lattice {S}chwinger model}",
    eprint = "2402.18429",
    archivePrefix = "arXiv",
    primaryClass = "hep-lat",
    doi = "10.1103/PhysRevD.111.014504",
    journal = "Phys. Rev. D",
    volume = "111",
    pages = "014504",
    year = "2025"
}

@article{Casher:1974vf,
    author = "Casher, A. and Kogut, John B. and Susskind, Leonard",
    title = "{Vacuum polarization and the absence of free quarks}",
    doi = "10.1103/PhysRevD.10.732",
    journal = "Phys. Rev. D",
    volume = "10",
    pages = "732--745",
    year = "1974"
}

@article{Schwinger:1951nm,
    author = "Schwinger, Julian S.",
    editor = "Milton, K. A.",
    title = "{On gauge invariance and vacuum polarization}",
    doi = "10.1103/PhysRev.82.664",
    journal = "Phys. Rev.",
    volume = "82",
    pages = "664--679",
    year = "1951"
}

@article{bauer2025local,
  title = {Local information flow in quantum quench dynamics},
  author = {Bauer, Nicolas P. and Trauzettel, Bj\"orn and Klein Kvorning, Thomas and Bardarson, Jens H. and Artiaco, Claudia},
  journal = {Phys. Rev. A},
  volume = {112},
  issue = {2},
  pages = {022221},
  numpages = {14},
  year = {2025},
  publisher = {American Physical Society},
  doi = {10.1103/v7gb-5gq8},
  eprint = "2505.00537",
  archivePrefix = "arXiv",
  primaryClass = "quant-ph"
}

@article{Chanda:2019fiu,
    author = "Chanda, Titas and Zakrzewski, Jakub and Lewenstein, Maciej and Tagliacozzo, Luca",
    title = "{Confinement and lack of thermalization after quenches in the bosonic Schwinger model}",
    eprint = "1909.12657",
    archivePrefix = "arXiv",
    primaryClass = "cond-mat.stat-mech",
    doi = "10.1103/PhysRevLett.124.180602",
    journal = "Phys. Rev. Lett.",
    volume = "124",
    pages = "180602",
    year = "2020"
}

@article{Cheneau_2012,
   title={Light-cone-like spreading of correlations in a quantum many-body system},
   volume={481},
   ISSN={1476-4687},
   DOI={10.1038/nature10748},
   number={7382},
   journal={Nature},
   publisher={Springer Science and Business Media LLC},
   author={Cheneau, Marc and Barmettler, Peter and Poletti, Dario and Endres, Manuel and Schauß, Peter and Fukuhara, Takeshi and Gross, Christian and Bloch, Immanuel and Kollath, Corinna and Kuhr, Stefan},
   year={2012},
   pages={484},
   eprint = "1111.0776",
   archivePrefix = "arXiv",
   primaryClass = "cond-mat",}

@article{James_2019,
  title = {Nonthermal States Arising from Confinement in One and Two Dimensions},
  author = {James, Andrew J. A. and Konik, Robert M. and Robinson, Neil J.},
  journal = {Phys. Rev. Lett.},
  volume = {122},
  issue = {13},
  pages = {130603},
  numpages = {8},
  year = {2019},
  publisher = {American Physical Society},
  doi = {10.1103/PhysRevLett.122.130603},
  eprint = "1804.09990",
  archivePrefix = "arXiv",
  primaryClass = "cond-mat"
}

@article{Robinson_2019,
  title = {Signatures of rare states and thermalization in a theory with confinement},
  author = {Robinson, Neil J. and James, Andrew J. A. and Konik, Robert M.},
  journal = {Phys. Rev. B},
  volume = {99},
  issue = {19},
  pages = {195108},
  numpages = {29},
  year = {2019},
  publisher = {American Physical Society},
  doi = {10.1103/PhysRevB.99.195108},
  eprint = "1808.10782",
  archivePrefix = "arXiv",
  primaryClass = "cond-mat"
}

@article{Alba:2017lvc,
    author = "Alba, Vincenzo and Calabrese, Pasquale",
    title = "{Entanglement dynamics after quantum quenches in generic integrable systems}",
    eprint = "1712.07529",
    archivePrefix = "arXiv",
    primaryClass = "cond-mat.stat-mech",
    doi = "10.21468/SciPostPhys.4.3.017",
    journal = "SciPost Phys.",
    volume = "4",
    pages = "017",
    year = "2018"
}

@article{Susskind:1976jm,
    author = "Susskind, Leonard",
    title = "{Lattice Fermions}",
    reportNumber = "PTENS-76-1",
    doi = "10.1103/PhysRevD.16.3031",
    journal = "Phys. Rev. D",
    volume = "16",
    pages = "3031--3039",
    year = "1977"
}

@article{Banks:1975gq,
    author = "Banks, Tom and Susskind, Leonard and Kogut, John B.",
    title = "{Strong Coupling Calculations of Lattice Gauge Theories: (1+1)-Dimensional Exercises}",
    reportNumber = "CLNS-318",
    doi = "10.1103/PhysRevD.13.1043",
    journal = "Phys. Rev. D",
    volume = "13",
    pages = "1043",
    year = "1976"
}

@article{Coleman:1976uz,
    author = "Coleman, Sidney R.",
    title = "{More About the Massive Schwinger Model}",
    reportNumber = "Print-76-0357 (HARVARD)",
    doi = "10.1016/0003-4916(76)90280-3",
    journal = "Annals Phys.",
    volume = "101",
    pages = "239",
    year = "1976"
}

@article{Mandelstam:1975hb,
    author = "Mandelstam, S.",
    editor = "Stone, M.",
    title = "{Soliton Operators for the Quantized Sine-Gordon Equation}",
    reportNumber = "Print-75-0198 (UC,BERKELEY)",
    doi = "10.1103/PhysRevD.11.3026",
    journal = "Phys. Rev. D",
    volume = "11",
    pages = "3026",
    year = "1975"
}

@article{Hebenstreit:2014rha,
    author = {Hebenstreit, Florian and Berges, J\"urgen},
    title = "{Connecting real-time properties of the massless Schwinger model to the massive case}",
    eprint = "1406.4273",
    archivePrefix = "arXiv",
    primaryClass = "hep-ph",
    doi = "10.1103/PhysRevD.90.045034",
    journal = "Phys. Rev. D",
    volume = "90",
    pages = "045034",
    year = "2014"
}

@article{Kogut:1974ag,
    author = "Kogut, John B. and Susskind, Leonard",
    title = "{Hamiltonian Formulation of Wilson's Lattice Gauge Theories}",
    reportNumber = "Print-74-1186 (CORNELL)",
    doi = "10.1103/PhysRevD.11.395",
    journal = "Phys. Rev. D",
    volume = "11",
    pages = "395--408",
    year = "1975"
}

@article{Hamer:1997dx,
    author = "Hamer, C. J. and Zheng, Wei-hong and Oitmaa, J.",
    title = "{Series expansions for the massive Schwinger model in Hamiltonian lattice theory}",
    eprint = "hep-lat/9701015",
    archivePrefix = "arXiv",
    doi = "10.1103/PhysRevD.56.55",
    journal = "Phys. Rev. D",
    volume = "56",
    pages = "55--67",
    year = "1997"
}

@article{PhysRevLett.69.2863,
  title = {Density matrix formulation for quantum renormalization groups},
  author = {White, Steven R.},
  journal = {Phys. Rev. Lett.},
  volume = {69},
  issue = {19},
  pages = {2863--2866},
  numpages = {0},
  year = {1992},
  publisher = {American Physical Society},
  doi = {10.1103/PhysRevLett.69.2863}
  }

@article{PhysRevB.48.10345,
  title = {Density-matrix algorithms for quantum renormalization groups},
  author = {White, Steven R.},
  journal = {Phys. Rev. B},
  volume = {48},
  issue = {14},
  pages = {10345--10356},
  numpages = {0},
  year = {1993},
  publisher = {American Physical Society},
  doi = {10.1103/PhysRevB.48.10345},
  
}

@article{itensor,
	title={{The ITensor Software Library for Tensor Network Calculations}},
	author={Matthew Fishman and Steven R. White and E. Miles Stoudenmire},
	journal={SciPost Phys. Codebases},
	pages={4},
	year={2022},
	publisher={SciPost},
	doi={10.21468/SciPostPhysCodeb.4},
	eprint = "2007.14822",
    archivePrefix = "arXiv",
    primaryClass = "cs.MS",
}

@article{Haegeman:2011zz,
    author = "Haegeman, Jutho and Cirac, J. Ignacio and Osborne, Tobias J. and Pizorn, Iztok and Verschelde, Henri and Verstraete, Frank",
    title = "{Time-Dependent Variational Principle for Quantum Lattices}",
    eprint = "1103.0936",
    archivePrefix = "arXiv",
    primaryClass = "cond-mat.str-el",
    doi = "10.1103/PhysRevLett.107.070601",
    journal = "Phys. Rev. Lett.",
    volume = "107",
    pages = "070601",
    year = "2011"
}

@article{PhysRevB.94.165116,
  title = {Unifying time evolution and optimization with matrix product states},
  author = {Haegeman, Jutho and Lubich, Christian and Oseledets, Ivan and Vandereycken, Bart and Verstraete, Frank},
  journal = {Phys. Rev. B},
  volume = {94},
  issue = {16},
  pages = {165116},
  numpages = {10},
  year = {2016},
  publisher = {American Physical Society},
  doi = {10.1103/PhysRevB.94.165116},
  eprint = "1408.5056",
  archivePrefix = "arXiv",
  primaryClass = "quant-ph",
}

@article{Barata:2025jhd,
    author = "Barata, Jo{\~a}o and Frenklakh, David and Mukherjee, Swagato",
    title = "{Thermal modifications of mesons and energy-energy correlators from real-time simulations of a $U(1)$ lattice gauge theory}",
    eprint = "2507.16890",
    archivePrefix = "arXiv",
    primaryClass = "hep-ph",
    reportNumber = "CERN-TH-2025-133",
    month = "7",
    year = "2025"
}

@article{Barata:2024apg,
    author = "Barata, Jo{\~a}o and Mukherjee, Swagato",
    title = "{Probing celestial energy and charge correlations through real-time quantum simulations: Insights from the Schwinger model}",
    eprint = "2409.13816",
    archivePrefix = "arXiv",
    primaryClass = "hep-ph",
    doi = "10.1103/PhysRevD.111.L031901",
    journal = "Phys. Rev. D",
    volume = "111",
    pages = "L031901",
    year = "2025"
}

@article{Mo:1992sv,
    author = "Mo, Yi-Zhang and Perry, Robert J.",
    title = "{Basis function calculations for the massive Schwinger model in the light front Tamm-Dancoff approximation}",
    reportNumber = "OSU-NT-92-128",
    doi = "10.1006/jcph.1993.1171",
    journal = "J. Comput. Phys.",
    volume = "108",
    pages = "159--174",
    year = "1993"
}

@article{Harada:1995tv,
    author = "Harada, Koji and Okazaki, Atsushi and Taniguchi, Masa-aki",
    title = "{Six body LFTD and wave functions for the massive Schwinger model}",
    eprint = "hep-th/9502102",
    archivePrefix = "arXiv",
    reportNumber = "KYUSHU-HET-22",
    doi = "10.1103/PhysRevD.52.2429",
    journal = "Phys. Rev. D",
    volume = "52",
    pages = "2429--2438",
    year = "1995"
}

@article{Ferreres-Sole:2018vgo,
    author = {Ferreres-Sol{\'e}, Silvia and Sj{\"o}strand, Torbj{\"o}rn},
    title = "{The space{\textendash}time structure of hadronization in the Lund model}",
    eprint = "1808.04619",
    archivePrefix = "arXiv",
    primaryClass = "hep-ph",
    reportNumber = "LU TP 18-18, MCnet-18-20",
    doi = "10.1140/epjc/s10052-018-6459-8",
    journal = "Eur. Phys. J. C",
    volume = "78",
    number = "11",
    pages = "983",
    year = "2018"
}

@book{Andersson:1997xwk,
    author = "Andersson, Bo",
    title = "{The Lund Model}",
    doi = "10.1017/9781009401296",
    isbn = "978-1-009-40129-6, 978-1-009-40125-8, 978-1-009-40128-9, 978-0-521-01734-3, 978-0-521-42094-5, 978-0-511-88149-7",
    publisher = "Cambridge University Press",
    volume = "7",
    year = "1998"
}

@article{Winter:2003tt,
    author = "Winter, Jan-Christopher and Krauss, Frank and Soff, Gerhard",
    title = "{A Modified cluster hadronization model}",
    eprint = "hep-ph/0311085",
    archivePrefix = "arXiv",
    reportNumber = "CERN-TH-2003-272",
    doi = "10.1140/epjc/s2004-01960-8",
    journal = "Eur. Phys. J. C",
    volume = "36",
    pages = "381--395",
    year = "2004"
}

@article{Coleman:1975pw,
    author = "Coleman, Sidney R. and Jackiw, R. and Susskind, Leonard",
    title = "{Charge Shielding and Quark Confinement in the Massive Schwinger Model}",
    reportNumber = "MIT-CTP-470",
    doi = "10.1016/0003-4916(75)90212-2",
    journal = "Annals Phys.",
    volume = "93",
    pages = "267",
    year = "1975"
}

@article{Deutsch:1991msp,
    author = "Deutsch, J. M.",
    title = "{Quantum statistical mechanics in a closed system}",
    doi = "10.1103/PhysRevA.43.2046",
    journal = "Phys. Rev. A",
    volume = "43",
    number = "4",
    pages = "2046",
    year = "1991"
}

@article{Srednicki:1994mfb,
    author = "Srednicki, Mark",
    title = "{Chaos and Quantum Thermalization}",
    eprint = "cond-mat/9403051",
    archivePrefix = "arXiv",
    doi = "10.1103/PhysRevE.50.888",
    journal = "Phys. Rev. E",
    volume = "50",
    month = "3",
    year = "1994"
}

@article{Cobos:2025krn,
    author = {Cobos, Jes{\'u}s and Fraxanet, Joana and Benito, C{\'e}sar and di Marcantonio, Francesco and Rivero, Pedro and Kap{\'a}s, Korn{\'e}l and Werner, Mikl{\'o}s Antal and Legeza, {\"O}rs and Bermudez, Alejandro and Rico, Enrique},
    title = "{Real-Time Dynamics in a (2+1)-D Gauge Theory: The Stringy Nature on a Superconducting Quantum Simulator}",
    eprint = "2507.08088",
    archivePrefix = "arXiv",
    primaryClass = "quant-ph",
    reportNumber = "CERN-TH-2025-111",
    month = "7",
    year = "2025"
}

@article{DiMarcantonio:2025cmf,
    author = "Di Marcantonio, Francesco and Pradhan, Sunny and Vallecorsa, Sofia and Ba{\~n}uls, Mari Carmen and Ortega, Enrique Rico",
    title = "{Roughening and dynamics of an electric flux string in a (2+1)D lattice gauge theory}",
    eprint = "2505.23853",
    archivePrefix = "arXiv",
    primaryClass = "hep-lat",
    reportNumber = "CERN-TH-2025-105",
    month = "5",
    year = "2025"
}

@article{Gonzalez-Cuadra:2024xul,
    author = "Gonzalez-Cuadra, Daniel and others",
    title = "{Observation of string breaking on a (2 + 1)D Rydberg quantum simulator}",
    eprint = "2410.16558",
    archivePrefix = "arXiv",
    primaryClass = "quant-ph",
    doi = "10.1038/s41586-025-09051-6",
    journal = "Nature",
    volume = "642",
    number = "8067",
    pages = "321--326",
    year = "2025"
}

@article{Mallick:2024slg,
    author = "Mallick, Arindam and Lewenstein, Maciej and Zakrzewski, Jakub and Plodzie{\'n}, Marcin",
    title = "{String-breaking dynamics in an Ising chain with local vibrations}",
    eprint = "2501.00604",
    archivePrefix = "arXiv",
    primaryClass = "quant-ph",
    doi = "10.1103/mdcm-5w9k",
    journal = "Phys. Rev. B",
    volume = "112",
    pages = "024311",
    year = "2025"
}

@article{Liu:2024lut,
    author = "Liu, Ying and Zhang, Wei-Yong and Zhu, Zi-Hang and He, Ming-Gen and Yuan, Zhen-Sheng and Pan, Jian-Wei",
    title = "{String-Breaking Mechanism in a Lattice Schwinger Model Simulator}",
    eprint = "2411.15443",
    archivePrefix = "arXiv",
    primaryClass = "cond-mat.quant-gas",
    doi = "10.1103/mwy1-v9hk",
    journal = "Phys. Rev. Lett.",
    volume = "135",
    pages = "101902",
    year = "2025"
}

@article{Buyens:2015tea,
    author = "Buyens, Boye and Haegeman, Jutho and Verschelde, Henri and Verstraete, Frank and Van Acoleyen, Karel",
    title = "{Confinement and string breaking for QED$_2$ in the Hamiltonian picture}",
    eprint = "1509.00246",
    archivePrefix = "arXiv",
    primaryClass = "hep-lat",
    doi = "10.1103/PhysRevX.6.041040",
    journal = "Phys. Rev. X",
    volume = "6",
    pages = "041040",
    year = "2016"
}

@article{Qian:2024xnr,
    author = "Qian, Wenyang and Wu, Bin",
    title = "{Quantum computation in fermionic thermal field theories}",
    eprint = "2404.07912",
    archivePrefix = "arXiv",
    primaryClass = "hep-ph",
    doi = "10.1007/JHEP07(2024)166",
    journal = "JHEP",
    volume = "07",
    pages = "166",
    year = "2024"
}

@article{Banuls:2025wiq,
    author = "Ba{\~n}uls, Mari Carmen and Cichy, Krzysztof and Lin, C. -J. David and Schneider, Manuel",
    title = "{Parton Distribution Functions in the {S}chwinger model from Tensor Network States}",
    eprint = "2504.07508",
    archivePrefix = "arXiv",
    primaryClass = "hep-lat",
    month = "4",
    year = "2025"
}

@article{Lamm:2019uyc,
    author = "Lamm, Henry and Lawrence, Scott and Yamauchi, Yukari",
    collaboration = "NuQS",
    title = "{Parton physics on a quantum computer}",
    eprint = "1908.10439",
    archivePrefix = "arXiv",
    primaryClass = "hep-lat",
    doi = "10.1103/PhysRevResearch.2.013272",
    journal = "Phys. Rev. Res.",
    volume = "2",
    pages = "013272",
    year = "2020"
}

@article{Shao:2025ygy,
    author = "Shao, Haiyang and Chen, Shile and Shi, Shuzhe",
    title = "{Onset of Bjorken Flow in Quantum Evolution of the Massive Schwinger Model}",
    eprint = "2509.10855",
    archivePrefix = "arXiv",
    primaryClass = "hep-ph",
    month = "9",
    year = "2025"
}

@article{Grieninger:2024axp,
    author = "Grieninger, Sebastian and Zahed, Ismail",
    title = "{Quasifragmentation functions in the massive Schwinger model}",
    eprint = "2406.01891",
    archivePrefix = "arXiv",
    primaryClass = "hep-ph",
    doi = "10.1103/PhysRevD.110.116009",
    journal = "Phys. Rev. D",
    volume = "110",
    pages = "116009",
    year = "2024"
}

@article{Li:2021kcs,
    author = "Li, Tianyin and Guo, Xingyu and Lai, Wai Kin and Liu, Xiaohui and Wang, Enke and Xing, Hongxi and Zhang, Dan-Bo and Zhu, Shi-Liang",
    collaboration = "QuNu",
    title = "{Partonic collinear structure by quantum computing}",
    eprint = "2106.03865",
    archivePrefix = "arXiv",
    primaryClass = "hep-ph",
    doi = "10.1103/PhysRevD.105.L111502",
    journal = "Phys. Rev. D",
    volume = "105",
    pages = "L111502",
    year = "2022"
}

@article{Barata:2023clv,
    author = "Barata, Jo{\~a}o and Du, Xiaojian and Li, Meijian and Qian, Wenyang and Salgado, Carlos A.",
    title = "{Quantum simulation of in-medium QCD jets: Momentum broadening, gluon production, and entropy growth}",
    eprint = "2307.01792",
    archivePrefix = "arXiv",
    primaryClass = "hep-ph",
    doi = "10.1103/PhysRevD.108.056023",
    journal = "Phys. Rev. D",
    volume = "108",
    pages = "056023",
    year = "2023"
}

@article{Farrell:2024fit,
    author = "Farrell, Roland C. and Illa, Marc and Ciavarella, Anthony N. and Savage, Martin J.",
    title = "{Quantum simulations of hadron dynamics in the Schwinger model using 112 qubits}",
    eprint = "2401.08044",
    archivePrefix = "arXiv",
    primaryClass = "quant-ph",
    reportNumber = "IQuS@UW-21-069, NT@UW-24-1",
    doi = "10.1103/PhysRevD.109.114510",
    journal = "Phys. Rev. D",
    volume = "109",
    pages = "114510",
    year = "2024"
}

@article{Jha:2024jan,
    author = "Jha, Raghav G. and Milsted, Ashley and Neuenfeld, Dominik and Preskill, John and Vieira, Pedro",
    title = "{Real-time scattering in Ising field theory using matrix product states}",
    eprint = "2411.13645",
    archivePrefix = "arXiv",
    primaryClass = "hep-th",
    doi = "10.1103/9dxz-k5wb",
    journal = "Phys. Rev. Res.",
    volume = "7",
    pages = "023266",
    year = "2025"
}

@article{PhysRevD.89.074011,
  title = {Turbulent thermalization process in heavy-ion collisions at ultrarelativistic energies},
  author = {Berges, J. and Boguslavski, K. and Schlichting, S. and Venugopalan, R.},
  journal = {Phys. Rev. D},
  volume = {89},
  issue = {7},
  pages = {074011},
  numpages = {10},
  year = {2014},
  publisher = {American Physical Society},
  doi = {10.1103/PhysRevD.89.074011},
  eprint = "1303.5650",
  archivePrefix = "arXiv",
  primaryClass = "hep-ph"
}

@article{banuls2013mass,
  title={The mass spectrum of the Schwinger model with matrix product states},
  author={Ba{\~n}uls, Mari Carmen and Cichy, K and Cirac, J Ignacio and Jansen, Karl},
  journal={JHEP},
  volume={2013},
  number={11},
  pages={1--21},
  year={2013},
  publisher={Springer},
  doi={10.1007/JHEP11(2013)158},
  eprint = "1305.3765",
  archivePrefix = "arXiv",
  primaryClass = "hep-lat"
}

@article{PhysRevX.6.041040,
  title = {Confinement and String Breaking for ${\mathrm{QED}}_{2}$ in the Hamiltonian Picture},
  author = {Buyens, Boye and Haegeman, Jutho and Verschelde, Henri and Verstraete, Frank and Van Acoleyen, Karel},
  journal = {Phys. Rev. X},
  volume = {6},
  issue = {4},
  pages = {041040},
  numpages = {30},
  year = {2016},
  publisher = {American Physical Society},
  doi = {10.1103/PhysRevX.6.041040},
  eprint = "1509.00246",
  archivePrefix = "arXiv",
  primaryClass = "hep-lat"
}

@article{PhysRevD.96.114501,
  title = {Real-time simulation of the {S}chwinger effect with matrix product states},
  author = {Buyens, Boye and Haegeman, Jutho and Hebenstreit, Florian and Verstraete, Frank and Van Acoleyen, Karel},
  journal = {Phys. Rev. D},
  volume = {96},
  issue = {11},
  pages = {114501},
  numpages = {15},
  year = {2017},
  publisher = {American Physical Society},
  doi = {10.1103/PhysRevD.96.114501},
  eprint = "1612.00739",
  archivePrefix = "arXiv",
  primaryClass = "hep-lat"
}

@article{PhysRevD.101.054507,
  title = {Topological vacuum structure of the {S}chwinger model with matrix product states},
  author = {Funcke, Lena and Jansen, Karl and K\"uhn, Stefan},
  journal = {Phys. Rev. D},
  volume = {101},
  issue = {5},
  pages = {054507},
  numpages = {16},
  year = {2020},
  publisher = {American Physical Society},
  doi = {10.1103/PhysRevD.101.054507},
  eprint = "1908.00551",
  archivePrefix = "arXiv",
  primaryClass = "hep-lat"
}

@article{KASPER2016742,
title = {Schwinger pair production with ultracold atoms},
journal = {Phys. Lett. B},
volume = {760},
pages = {742-746},
year = {2016},
issn = {0370-2693},
doi = {https://doi.org/10.1016/j.physletb.2016.07.036},
author = {V. Kasper and F. Hebenstreit and M.K. Oberthaler and J. Berges},
eprint = "1506.01238",
archivePrefix = "arXiv",
primaryClass = "cond-mat"
}

@article{yang2020observation,
  title={Observation of gauge invariance in a 71-site {B}ose--{H}ubbard quantum simulator},
  author={Yang, Bing and Sun, Hui and Ott, Robert and Wang, Han-Yi and Zache, Torsten V and Halimeh, Jad C and Yuan, Zhen-Sheng and Hauke, Philipp and Pan, Jian-Wei},
  journal={Nature},
  volume={587},
  number={7834},
  pages={392--396},
  year={2020},
  publisher={Nature Publishing Group UK London},
  doi={10.1038/s41586-020-2910-8},
  eprint = "2003.08945",
  archivePrefix = "arXiv",
  primaryClass = "cond-mat"
}

@article{martinez2016real,
  title={Real-time dynamics of lattice gauge theories with a few-qubit quantum computer},
  author={Martinez, Esteban A and Muschik, Christine A and Schindler, Philipp and Nigg, Daniel and Erhard, Alexander and Heyl, Markus and Hauke, Philipp and Dalmonte, Marcello and Monz, Thomas and Zoller, Peter and others},
  journal={Nature},
  volume={534},
  number={7608},
  pages={516--519},
  year={2016},
  publisher={Nature Publishing Group UK London},
  doi={10.1038/nature18318},
  eprint = "1605.04570",
  archivePrefix = "arXiv",
  primaryClass = "quant-ph"
}

@article{kokail2019self,
  title={Self-verifying variational quantum simulation of lattice models},
  author={Kokail, Christian and Maier, Christine and van Bijnen, Rick and Brydges, Tiff and Joshi, Manoj K and Jurcevic, Petar and Muschik, Christine A and Silvi, Pietro and Blatt, Rainer and Roos, Christian F and others},
  journal={Nature},
  volume={569},
  number={7756},
  pages={355--360},
  year={2019},
  publisher={Nature Publishing Group UK London},
  doi={10.1038/s41586-019-1177-4},
  eprint = "1810.03421",
  archivePrefix = "arXiv",
  primaryClass = "quant-ph"
}

@article{PRXQuantum.3.020324,
  title = {Digital Quantum Simulation of the {S}chwinger Model and Symmetry Protection with Trapped Ions},
  author = {Nguyen, Nhung H. and Tran, Minh C. and Zhu, Yingyue and Green, Alaina M. and Alderete, C. Huerta and Davoudi, Zohreh and Linke, Norbert M.},
  journal = {PRX Quantum},
  volume = {3},
  issue = {2},
  pages = {020324},
  numpages = {15},
  year = {2022},
  publisher = {American Physical Society},
  doi = {10.1103/PRXQuantum.3.020324},
  eprint = "2112.14262",
  archivePrefix = "arXiv",
  primaryClass = "quant-ph"
}

@article{PhysRevB.109.134304,
  title = {Probing off-diagonal eigenstate thermalization with tensor networks},
  author = {Luo, Maxine and Trivedi, Rahul and Ba\~nuls, Mari Carmen and Cirac, J. Ignacio},
  journal = {Phys. Rev. B},
  volume = {109},
  issue = {13},
  pages = {134304},
  numpages = {13},
  year = {2024},
  publisher = {American Physical Society},
  doi = {10.1103/PhysRevB.109.134304},
  eprint = "2312.00736",
  archivePrefix = "arXiv",
  primaryClass = "quant-ph"
}

@article{Davoudi:2025rdv,
    author = "Davoudi, Zohreh and Hsieh, Chung-Chun and Kadam, Saurabh V.",
    title = "{Quantum computation of hadron scattering in a lattice gauge theory}",
    eprint = "2505.20408",
    archivePrefix = "arXiv",
    primaryClass = "quant-ph",
    reportNumber = "UMD-PP-025-02, IQuS@UW-21-102",
    month = "5",
    year = "2025"
}

@article{Bauer:2021gup,
    author = "Bauer, Christian W. and Freytsis, Marat and Nachman, Benjamin",
    title = "{Simulating Collider Physics on Quantum Computers Using Effective Field Theories}",
    eprint = "2102.05044",
    archivePrefix = "arXiv",
    primaryClass = "hep-ph",
    doi = "10.1103/PhysRevLett.127.212001",
    journal = "Phys. Rev. Lett.",
    volume = "127",
    number = "21",
    pages = "212001",
    year = "2021"
}

@article{Florio:2025hoc,
    author = "Florio, Adrien and Frenklakh, David and Grieninger, Sebastian and Kharzeev, Dmitri E. and Palermo, Andrea and Shi, Shuzhe",
    title = "{Thermalization from quantum entanglement: jet simulations in the massive Schwinger model}",
    eprint = "2506.14983",
    archivePrefix = "arXiv",
    primaryClass = "hep-ph",
    month = "6",
    year = "2025"
}

@article{Xu:2025abo,
    author = "Xu, Kaidi and Borla, Umberto and Moroz, Sergej and Halimeh, Jad C.",
    title = "{String Breaking Dynamics and Glueball Formation in a $2+1$D Lattice Gauge Theory}",
    eprint = "2507.01950",
    archivePrefix = "arXiv",
    primaryClass = "hep-lat",
    month = "7",
    year = "2025"
}

@article{Tian:2025mbv,
    author = "Tian, Yizhuo and Srivatsa, N. S. and Xu, Kaidi and Osborne, Jesse J. and Borla, Umberto and Halimeh, Jad C.",
    title = "{Role of Plaquette Term in Genuine $2+1$D String Dynamics on Quantum Simulators}",
    eprint = "2508.05736",
    archivePrefix = "arXiv",
    primaryClass = "quant-ph",
    month = "8",
    year = "2025"
}

@article{Cataldi:2025cyo,
    author = "Cataldi, Giovanni and Orlando, Simone and Halimeh, Jad C.",
    title = "{Real-Time String Dynamics in a $2+1$D Non-Abelian Lattice Gauge Theory: String Breaking, Glueball Formation, Baryon Blockade, and Tension Reduction}",
    eprint = "2509.08868",
    archivePrefix = "arXiv",
    primaryClass = "hep-lat",
    month = "9",
    year = "2025"
}

\end{document}